\documentclass[traditabstract]{aa}
\usepackage{amssymb,amsmath}
\usepackage{subfigure}
\usepackage{graphicx}%for EPS, PS files
\usepackage{natbib}
\usepackage{url}

\footnotesize
\newdimen\digitwidth    %define ! a one digit width for tables
\setbox0=\hbox{\rm0}
\digitwidth=\wd0
\catcode`!=\active
\def!{\kern\digitwidth}
\normalsize

\title{Observing pulsars and fast transients with LOFAR}
\author{B.~W.~Stappers\inst{\ref{jod}} \and J.~W.~T.~Hessels\inst{\ref{astron} \and \ref{uva}}
A. ~Alexov\inst{\ref{uva}}\and
K.~Anderson\inst{\ref{uva}}\and
T.~Coenen\inst{\ref{uva}} \and
T.~Hassall\inst{\ref{jod}}\and
A. ~Karastergiou\inst{\ref{ox}}
\and
V. ~I.~Kondratiev\inst{\ref{astron}}
\and
M. ~Kramer\inst{\ref{mpifr} \and \ref{jod}}
\and
J.~van Leeuwen\inst{\ref{astron} \and \ref{uva}} 
\and
J.~D.~Mol\inst{\ref{astron}}
\and
A.~Noutsos\inst{\ref{mpifr}}
\and
J.~W.~Romein\inst{\ref{astron}}
\and
P.~Weltevrede\inst{\ref{jod}}
\and
R.~Fender\inst{\ref{soton}}
\and
R.~A.~M.~J.~Wijers\inst{\ref{uva}}
\and
L.~B\"ahren\inst{\ref{uva}}\and
M.~E.~Bell\inst{\ref{soton}} \and 
J.~Broderick\inst{\ref{soton}} \and
E.~J.~Daw\inst{\ref{sheff}}\and
V.~S.~Dhillon\inst{\ref{sheff}}\and
J.~Eisl\"offel\inst{\ref{tls}} \and 
H.~Falcke\inst{\ref{nijmegen} \and \ref{astron}} \and 
J.~Griessmeier\inst{\ref{astron} \and \ref{cnrs}} \and 
C.~Law\inst{\ref{berkley} \and \ref{uva}} \and           
S.~Markoff\inst{\ref{uva}}
J.~C.~A.~Miller-Jones\inst{\ref{curtin} \and \ref{uva}} \and 
B.~Scheers\inst{\ref{uva}} \and 
H.~Spreeuw\inst{\ref{uva}} \and 
J.~Swinbank\inst{\ref{uva}} \and 
S.~ter Veen\inst{\ref{nijmegen}}
M.~W.~Wise\inst{\ref{astron} \and \ref{uva}} \and 
O.~Wucknitz\inst{\ref{ubonn}} \and 
P.~Zarka\inst{\ref{meudon}} \and 
J.~Anderson\inst{\ref{mpifr}} \and 
A.~Asgekar\inst{\ref{astron}} \and 
I.~M.~Avruch\inst{\ref{astron} \and \ref{kapteyn}} \and 
R.~Beck\inst{\ref{mpifr}} \and 
P.~Bennema\inst{\ref{astron}} \and 
M.~J.~Bentum\inst{\ref{astron}} \and 
P.~Best\inst{\ref{roe}} \and 
J.~Bregman\inst{\ref{astron}} \and 
M.~Brentjens\inst{\ref{astron}} \and 
R.~H.~van de Brink\inst{\ref{astron}} \and 
P.~C.~Broekema\inst{\ref{astron}} \and 
W.~N.~Brouw\inst{\ref{kapteyn}} \and 
M.~Br\"uggen\inst{\ref{bremen}} \and 
A.~G.~de Bruyn\inst{\ref{astron} \and \ref{kapteyn}} \and 
H.~R.~Butcher\inst{\ref{astron} \and \ref{anu}} \and      
B.~Ciardi\inst{\ref{mpifa}} \and 
J.~Conway\inst{\ref{oso}} \and 
R.-J. ~Dettmar\inst{\ref{raiub}} \and 
A.~van Duin\inst{\ref{astron}} \and 
J.~van Enst\inst{\ref{astron}} \and                            
M.~Garrett\inst{\ref{astron} \and \ref{leiden}} \and 
M.~Gerbers\inst{\ref{astron}} \and 
T.~Grit\inst{\ref{astron}} \and 
A.~Gunst\inst{\ref{astron}} \and 
M.~P.~van Haarlem\inst{\ref{astron}} \and 
J.~P.~Hamaker\inst{\ref{astron}}
G.~Heald\inst{\ref{astron}} \and 
M.~Hoeft\inst{\ref{tls}} \and 
H.~Holties\inst{\ref{astron}} \and 
A.~Horneffer\inst{\ref{mpifr} \and \ref{nijmegen}} \and
L.~V.~E.~Koopmans\inst{\ref{kapteyn}} \and 
G.~Kuper\inst{\ref{astron}} \and 
M.~Loose\inst{\ref{astron}} \and 
P.~Maat\inst{\ref{astron}} \and 
D.~McKay-Bukowski\inst{\ref{stfc}} \and 
J.~P.~McKean\inst{\ref{astron}} \and 
G.~Miley\inst{\ref{leiden}} \and 
R.~Morganti\inst{\ref{astron} \and \ref{kapteyn}} \and 
R.~Nijboer\inst{\ref{astron}} \and 
J.~E.~Noordam\inst{\ref{astron}} \and 
M.~Norden\inst{\ref{astron}} \and 
H.~Olofsson\inst{\ref{oso}} \and 
M.~Pandey-Pommier\inst{\ref{leiden} \and \ref{lyon}} \and 
A.~Polatidis\inst{\ref{astron}} \and 
W.~Reich\inst{\ref{mpifr}} \and 
H.~R\"ottgering\inst{\ref{leiden}} \and 
A.~Schoenmakers\inst{\ref{astron}} \and 
J.~Sluman\inst{\ref{astron}} \and 
O.~Smirnov\inst{\ref{astron}} \and 
M.~Steinmetz\inst{\ref{aip}} \and 
C.~G.~M.~Sterks\inst{\ref{groningen}} \and  
M.~Tagger\inst{\ref{cnrs}} \and 
Y.~Tang\inst{\ref{astron}} \and 
R.~Vermeulen\inst{\ref{astron}} \and 
N.~Vermaas\inst{\ref{astron}} \and 
C.~Vogt\inst{\ref{astron}} \and 
M.~de Vos\inst{\ref{astron}} \and 
S.~J.~Wijnholds\inst{\ref{astron}} \and 
S.~Yatawatta\inst{\ref{kapteyn}} \and 
A.~Zensus\inst{\ref{mpifr}} 
}

\institute{Jodrell Bank Center for Astrophysics,
School of Physics and Astronomy,
The University of Manchester,
Manchester M13 9PL,UK\label{jod}
\email{Ben.Stappers@manchester.ac.uk}
\and Netherlands Institute for Radio Astronomy (ASTRON), Postbus 2, 7990 AA Dwingeloo, The Netherlands\label{astron}
\and Astronomical Institute 'Anton Pannekoek', University of Amsterdam, Postbus 94249, 1090 GE Amsterdam, The Netherlands\label{uva}
\and Astrophysics, University of Oxford, Denys Wilkinson Building, Keble Road, Oxford OX1 3RH\label{ox}
\and
Max-Planck-Institut f\"ur Radioastronomie, Auf dem H\"ugel 69, 53121 Bonn, Germany\label{mpifr}
\and School of Physics and Astronomy, University of Southampton, Southampton, SO17 1BJ, UK\label{soton}
\and Max Planck Institute for Astrophysics, Karl Schwarzschild Str. 1, 85741 Garching, Germany\label{mpifa}
\and Department of Physics \& Astronomy, Hicks Building, Hounsfield Road, Sheffield S3 7RH, United Kingdom\label{sheff}
\and
Leiden Observatory, Leiden University, PO Box 9513, 2300 RA Leiden, The Netherlands\label{leiden}
\and
Kapteyn Astronomical Institute, PO Box 800, 9700 AV Groningen, The Netherlands\label{kapteyn}
\and
Onsala Space Observatory, Dept. of Earth and Space Sciences, Chalmers University of Technology, SE-43992 Onsala, Sweden\label{oso}
\and
Department of Astrophysics/IMAPP, Radboud University Nijmegen, P.O. Box 9010, 6500 GL Nijmegen, The Netherlands\label{nijmegen}
\and
International Centre for Radio Astronomy Research - Curtin University, GPO Box U1987, Perth, WA 6845, Australia\label{curtin}
\and
STFC Rutherford Appleton Laboratory,  Harwell Science and Innovation Campus,  Didcot  OX11 0QX, UK\label{stfc}
\and
Institute for Astronomy, University of Edinburgh, Royal Observatory of Edinburgh, Blackford Hill, Edinburgh EH9 
3HJ, UK\label{roe}
\and
LESIA, UMR CNRS 8109, Observatoire de Paris, 92195   Meudon, France\label{meudon}
\and
Argelander-Institut f\"ur Astronomie, University of Bonn, Auf dem H\"ugel 71, 53121, Bonn, Germany\label{ubonn}
\and
Leibniz-Institut fŸr Astrophysik Potsdam (AIP), An der Sternwarte 16, 14482 Potsdam, Germany\label{aip}
\and
Th\"uringer Landessternwarte, Sternwarte 5, D-07778 Tautenburg, Germany\label{tls}
\and
Astronomisches Institut der Ruhr-Universit\"at Bochum, Universitaetsstrasse 150, 44780 Bochum, Germany\label{raiub}
\and
Jacobs University Bremen, Campus Ring 1, 28759 Bremen, Germany\label{bremen}
\and
Laboratoire de Physique et Chimie de lÕEnvironnement et de lÕEspace, CNRS/UniversitŽ dÕOrlŽans, France
\label{cnrs}
\and
Center for Information Technology (CIT), University of Groningen, The Netherlands\label{groningen}
\and
Radio Astronomy Lab, UC Berkeley, CA, USA\label{berkley}
\and
Centre de Recherche Astrophysique de Lyon, Observatoire de Lyon, 9 av Charles Andr\'e, 69561 Saint Genis Laval Cedex, France\label{lyon}
\and
Mt Stromlo Observatory, Research School of Astronomy and Astrophysics, Australian National University, Weston, A.C.T. 2611, Australia\label{anu}
}

\date{}
\begin{document}

\abstract{Low frequency radio waves, while challenging to observe, are
  a rich source of information about pulsars. The LOw Frequency ARray
  (LOFAR) is a new radio interferometer operating in the lowest 4
  octaves of the ionospheric ''radio window": 10-240MHz, that will
  greatly facilitate observing pulsars at low radio frequencies.
  Through the huge collecting area, long baselines, and flexible
  digital hardware, it is expected that LOFAR will revolutionize radio
  astronomy at the lowest frequencies visible from Earth.  LOFAR is a
  next-generation radio telescope and a pathfinder to the Square
  Kilometre Array (SKA), in that it incorporates advanced
  multi-beaming techniques between thousands of individual elements.
  We discuss the motivation for low-frequency pulsar observations in
  general and the potential of LOFAR in addressing these science
  goals.  We present LOFAR as it is designed to perform
  high-time-resolution observations of pulsars and other fast
  transients, and outline the various relevant observing modes and
  data reduction pipelines that are already or will soon be
  implemented to facilitate these observations.  A number of results
  obtained from commissioning observations are presented to
  demonstrate the exciting potential of the telescope.  This paper
  outlines the case for low frequency pulsar observations and is also
  intended to serve as a reference for upcoming pulsar/fast transient
  science papers with LOFAR.}


\maketitle

\section{Introduction}
\label{sec:intro}

Pulsars are rapidly rotating, highly magnetised neutron stars that
were first identified via pulsed radio emission at the very low radio
observing frequency of 81~MHz \citep{hbp+68}. They have subsequently
been shown to emit pulsations across the electromagnetic spectrum, at
frequencies ranging from 17~MHz to above 87~GHz
(e.g. \citealt{bu76,bu77,mkt+97}) in the radio and at optical, X-ray and
$\gamma$-ray wavelengths (see \citealt{tho00b} and references
therein), although the vast majority are seen to emit only at radio
wavelengths.  These pulsations provide invaluable insights into the
nature of neutron star physics, and most neutron stars would be
otherwise undetectable with current telescopes.  Though radio pulsars
form over 85\% of the known neutron star population, they are
generally very weak radio sources with pulsed flux densities ranging
from 0.0001 to 5~Jy with a median of 0.01~Jy at a frequency of
400~MHz. The pulsed flux density at radio wavelengths exhibits a
steep spectrum ($S \propto \nu^{\alpha}$; $-4 < \alpha < 0$;
$\alpha_{\rm mean} = -1.8$, \citep{mkkw00a} that often peaks and turns over
at frequencies between 100 and 200~MHz \citep{kms+78,sab86,mgj+94}.

After their discovery, a lot of the early work on pulsars
(e.g. \citealt{col69,sr68,rcc+70}) continued at low radio frequencies
(defined here as $< 300$~MHz).  However, despite the fact that most
pulsars are intrinsically brightest in this frequency range, since
then the vast majority of pulsars have been discovered and studied at
frequencies in the range $300-2000$~MHz; much of our knowledge of the
properties of the radio emission mechanism stems from studies at these
frequencies and above.  There are three main reasons for this (see
Sect. \ref{sec:challenges}): the deleterious effects of the interstellar
medium (ISM) on pulsed signals; the effective background sky
temperature of the Galactic synchrotron emission; and ionospheric
effects.  All three of these effects have steep power law dependencies
on frequency and therefore become worse towards lower frequencies.
Combined with the generally steep spectra of pulsars, these effects
conspire to make observing frequencies of $\sim 300-2000$~MHz the
range of choice for most pulsar studies and searches.

\begin{figure*}
\includegraphics[width=\linewidth]{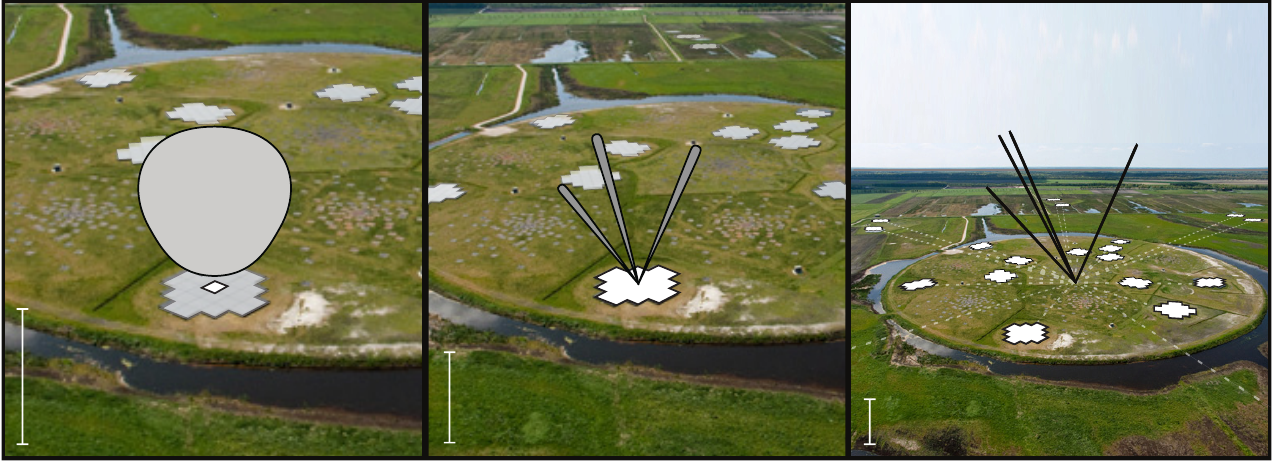}

\caption{Three successive zoom-outs showing the stations in the LOFAR
  core. The different scales of the hierarchically organised HBA
  elements are highlighted and their respective beam sizes are
  shown. The large circular area marks the edge of the Superterp,
  which contains the inner-most 6 stations (i.e. 12 HBA sub-stations:
  where there are 2 sub-stations, each of 24 tiles, in each HBA core
  station); other core stations can be seen highlighted beyond the
  Superterp in the third panel. Left: a single HBA tile and associated
  beam.  Middle: A single HBA sub-station with three simultaneous
  station beams.  Right: The 6 stations of the Superterp plus 3 core
  stations in the background are highlighted. Four independent beams
  formed from the coherent combination of all 24 core HBA stations,
  most of which are outside this photo, are shown. For the LBA
  stations, a similar scheme applies except that each LBA dipole can
  effectively see the whole sky. Fields of the relatively sparsely
  distributed LBA antennas are visible in between the highlighted HBA
  stations in all three panels.}

\label{fig:lofar}
\end{figure*}

However, despite these challenges, there are many reasons why it is
important and interesting to observe pulsars in a significantly lower
frequency regime than now commonly used; these are discussed in detail
in Sect. \ref{sec:science}.  In recent years some excellent studies have
continued at frequencies between 20$-$110~MHz mainly using the
Pushchino, Gauribidanur and UTR-2 telescopes (e.g. \citealt{mm10,
  mms00, ad05, pku+06b, uzk+06}).  These studies have begun to map,
e.g., the low-frequency spectra, pulse morphologies, and pulse energy
distributions of pulsars, but have in some cases been limited by the
available bandwidths and/or polarisation and tracking capabilities of
these telescopes (see Sect. 2).

The Low Frequency Array (LOFAR) was designed and constructed by
ASTRON, the Netherlands Institute for Radio Astronomy, and has
facilities in several countries, that are owned by various parties
(each with their own funding sources), and that are collectively
operated by the International LOFAR Telescope (ILT) foundation under a
joint scientific policy. LOFAR provides a great leap forward in
low-frequency radio observations by providing large fractional
bandwidths and sophisticated multi-beaming capabilities.  In this
paper we present the LOFAR telescope as it will be used for pulsar and
other high-time-resolution beamformed observations; this will serve as
a reference for future science papers that use these LOFAR modes.  We
also describe the varied pulsar and fast transient science LOFAR will
enable and present commissioning results showing how that potential is
already being realised.  LOFAR is well suited for the study
of known sources, and its huge field of view (FoV) makes it a powerful
survey telescope for finding new pulsars and other ``fast-transients".
In Sect. \ref{sec:LOFAR} we present the basic design parameters of the
LOFAR telescope. The challenges associated with observing at low radio
frequencies and how they can be mitigated with LOFAR will be discussed
in Sect. \ref{sec:challenges}. A detailed description of the science that
will be possible with LOFAR is presented in Sect. \ref{sec:science}. The
flexible nature of LOFAR means that there are many possible observing
modes; these are introduced in Sect. \ref{sec:obsmodes}. In
Sect. \ref{sec:pipelines} we discuss the different pulsar pipelines that
are being implemented.  Commissioning results, which demonstrate that
LOFAR is {\it already} performing pulsar and fast transients
observations of high quality, are presented in
Sect. \ref{sec:commissioning}. We summarise the potential of LOFAR for
future pulsar observations in Sect. \ref{conclusions}.
\section{LOFAR}
\label{sec:LOFAR}

Instrumentation in radio astronomy is undergoing a revolution that
will exploit massive computing, clever antenna design, and digital
signal processing to greatly increase the instantaneous FoV and
bandwidth of observations.  This work is part of the international
effort to create the ``Square Kilometre Array" \citep{cr04}, a radio
telescope orders of magnitude better than its predecessors.

One of the first ``next generation" radio telescopes to implement
these techniques is LOFAR, which operates in the frequency range
$10-240$~MHz.  The large collecting area of LOFAR is comprised of many
thousands of dipole antennas, hierarchically arranged in stations
which come in three different configurations (Table
\ref{tab:stations}). These stations are distributed in a sparse array
with a denser core region near Exloo, the Netherlands, extending out
to remote stations in the Netherlands and then on further to stations
in France, Germany, Sweden and the United Kingdom.  There are a total
of 40 stations in the Netherlands and 8 international stations, with
the prospect of more to come.  A schematic diagram of some of the
LOFAR stations in the inner core of LOFAR $-$ the ``Superterp" as it
is known $-$ is shown in Figure \ref{fig:lofar}. As will be discussed
in more detail later, pulsar observations can utilise all of these
stations to achieve a variety of diverse science goals. Details of
system architecture and signal processing can be found in \cite{dgn09}
and a full description of LOFAR will soon be published (van Haarlem et
al. in prep.) we limit the discussion only to the most important points
related to pulsar observations.

LOFAR has two different types of antennas to cover the frequency range
10--240~MHz. The low band antennas, LBAs, cover the frequency range
10--90~MHz, although they are optimised for frequencies above
30~MHz. The lower limit of 10~MHz is defined by transmission of radio
waves through the Earth's ionosphere. There are 48/96 active LBA
dipoles in each Dutch/international station (Table
\ref{tab:stations}). The high band antennas, HBAs, cover the frequency
range 110--240~MHz, and consist of 16 folded dipoles grouped into
tiles of $4 \times 4$ dipoles each, which are phased together using an
analogue beamformer within the tile itself. There are 48/96 tiles in
each Dutch/international station, with a separation into two sub-stations
of 24 HBA tiles each in the case of core stations\footnote{We note
  that when we refer to dipoles and tiles we are generally referring
  to both the X and Y polarisations together, that is dipole pairs,
  and we draw a distinction between the two polarisations only when
  necessary.}.

The received radio waves are sampled at either 160 or 200~MHz in one
of three different Nyquist zones to access the frequency ranges
0--100, 100--200 and 160--240~MHz. There are filters in place to
optionally remove frequencies below 30~MHz and the FM band
approximately encompassing 90--110~MHz. The 80 or 100~MHz wide bands
are filtered at the stations into 512 subbands of exactly
156.25/195.3125~kHz using a poly-phase filter. Up to 244 of these
subbands can be transported back to the Central Processor, CEP, giving
a maximum instantaneous bandwidth of about 39/48~MHz\footnote{This
  number may further increase if the number of bits used to describe
  each sample is reduced.}. There is no restriction
on which 244 of the 512 subbands can be selected to be processed at
CEP, therefore this bandwidth can be distributed throughout the entire
available 80/100~MHz.  Alternatively it is possible to portion out the
bandwidth into multiple beams. Previously there was a limit of eight
per station, however a recent new implementation of the beam server
software has enabled each of the 244 subbands to be pointed in a
different direction. These subbands can be further divided into
narrower frequency channels in CEP as will be discussed below. The
degree of flexibility afforded by these choices of frequency and beams
allows a wide range of high-time-resolution pulsar-like observations
with LOFAR; these different modes are described in detail in
Sect. \ref{sec:obsmodes}.

In Table \ref{tab:compare} we compare the properties of LOFAR with
those of other telescopes currently operating in (part of) the same
frequency range.  LOFAR is the {\it only} existing or planned
telescope capable of covering the entire lowest 4 octaves of the radio
window ($10-240$~MHz, above the Earth's ionospheric cut-off).  In some
modes this entire range can be observed simultaneously (see Sect. \ref{sec:obsmodes}).
LOFAR's total effective collecting area and instantaneous sensitivity
places it at the forefront of existing low-frequency radio telescopes,
especially in the range $100-240$~MHz, but collecting area is only
one aspect of LOFAR's capabilities.  As will be described in more
detail later, LOFAR offers many advantages over current telescopes
through its multi-beaming capabilities, flexible backend, high spatial
resolution, large instantaneous bandwidth, wide total available
frequency range, ability to track, and ability to observe a large
fraction of the sky (i.e. declinations greater than $-$30 degrees, see
Sect. 5).

%Table comparing LOFAR stations
\begin{table}
\caption{Arrangement of elements in LOFAR stations.} 
\centering
\label{tab:stations}
\begin{tabular}{l c c c}
\hline \hline
Station Type &  LBA (no.) &  HBA tiles (no.) & Baseline (km) \\
\hline
Core               & 2$\times$48 & 2$\times$24 & $0.1-1$ \\ 
Remote          & 2$\times$48 &  48                 & $1-10$s \\
International  &   96                &  96                 & $\sim 100$s \\
\hline
\end{tabular}
\tablefoot{Arrangement of elements in the three types of LOFAR stations,
  along with their typical distance from the center of the array
  (baseline). In the Core and Remote stations there are 96 LBA
  dipoles but only 48 can be beamformed at any one time.  For these
  stations, one can select either the inner circle or the outer ring
  of 48 LBA dipoles depending on the science requirements. The HBA
  sub-stations can be correlated, or used in beamforming,
  independently.}
\end{table}

%Table comparing LOFAR to other low frequency telescopes.
\begin{table*}
\caption{Comparison of some telescopes that have operated at frequencies below 200~MHz.}
\label{tab:compare}
\centering
\begin{tabular}{l c c c c c c c c }
\hline \hline
Telescope &  Area  &  T$_{rec}$ & T$_{sky}$ & Frequencies & Bandwidth & Nbits & Beams & Transit\\
 & (m$^2$) & (K) & (K) & (MHz) & (MHz) & & & \\
\hline
Arecibo &  30000 & & 5700--5000 & 47--50 & & & 1 & Restricted\\
Cambridge 3.6 ha &  36000 & & 1400  & 81.5 & 1 & 1 & 16 & Yes \\
DKR-1000 & 5000-8000 & & 18000--500 & 30--120 & 1--2 & & & Yes \\
Gauribidanur &  12000 & 500 & 30000--12000 & 25-35 & 2 & 2 & 1 & Yes \\
GMRT   &  48000 & 295 & 276 & 151 &  16/32 & 8/4 & 2 & No\\ 
LOFAR  (LBA) &  75200  &  500 &  320000--1000 & 10--90 & 48 & 16 & $<$244 & No \\
LOFAR (HBA)  &  57000 &  140--180 & 630--80  & 110--240 & 48 & 16 & $<$244 & No \\
LWA & 1000000 & & 320000--1100 & 10-88 & 19.2 & 8 & 4 & No\\
MRT & 41000 & 500 & 276 & $\sim$150 & 2 & 2 & 1 & Yes\\
MWA & 8000 & 150 & 1400--45  & 80--300 & 32 & 5 & 16(32) & No\\ 
Nan\c{c}ay Decametric Array & 4000 & & 320000--800 & 10-100 & 32 & 14 & 1 & No \\
LPA/BSA - Pushchino & 20000 & 110 & 900 & 109--113 & 2.5 & 12 & 16(32) & Yes\\
UTR-2   &  150000 & & 550000--9000 & 8--40 & 24 & 16 & 5(8) & No\\
VLA & 13000  &  & 1800 & 74 & 1.56 & 2 & 1 & No\\
WSRT & 7000 & 400 & 650--175 & 110--180 & 8$\times$2.5 & 8 & 1 & No\\
\hline
\end{tabular}
\tablefoot{Telescopes that are observing in (parts of) the same
  frequency range or have done so in the past are considered.  The
  bandwidths presented are those that are used for pulsar observations
  and do not take into consideration what fraction may not be useful
  due to radio frequency interference.  In the case of LOFAR pulsar
  observations, only a small fraction of the bandwidth requires
  masking (see Sect. 6).  Note that the collecting area is the maximum and
  does not take into account efficiencies or hour angle dependent
  effects. The transit instruments such as the Mauritius Radio
  Telescope (MRT; \citealt{gss+98}), Gauribidanur \citep{dss89}, and
  the Cambridge 3.6 ha telescope \protect{\citep{stw98}} have some
  tracking ability but are usually limited to observing times of a few
  minutes. The collecting area of Arecibo is based on an illumination
  equivalent to a 200~m dish. The collecting area for the UTR-2
  \citep{abzk01} is quoted for a frequency of 20~MHz and the 5-beam
  and 8-beam modes are discussed in Ryabov et
  al. (2010)\nocite{rvz+10} and Abranin et al. (2001)\nocite{abzk01}
  respectively. The Murchison Widefield Array (MWA) can have 32 single
  polarisation beams and the area is quoted for a frequency of 200~MHz
  \citep{lcm+09}. LOFAR can have up to a total of 244 station beams
  which equals the number of subbands.  The Long Wavelength Array
  (LWA) \citep{ecc+09} collecting area is quoted for a frequency of
  20~MHz and decreases as $\lambda^2$. We note that the LWA and MWA
  collecting areas are those projected for the final system and are
  not yet in place. For the LOFAR LBA entry, the effective area is
  based on using the outermost antenna configuration at a freqeuncy of
  30~MHz, which maximizes collecting area. The HBA collecting area is
  quoted for a frequency of 150~MHz. The GMRT \citep{sak+91}, VLA
  \citep{clc+07} and WSRT \citep{kss10} are multiple dish
  interferometers. In some cases, not all parameters are available in
  the literature and so these entries were left blank. T$_{sky}$ is an
  approximate value calculated on a cold piece of sky and is quoted
  for the full range of available frequencies at that telescope.}
\end{table*}

\section{Challenges of observing at low frequencies}
\label{sec:challenges}

As previously mentioned, currently the majority of radio pulsar observations are
done between $300-2000$~MHz because this frequency range greatly
reduces the severity of several low-frequency observational
impediments.  Here we describe these challenges in more detail and
explain how LOFAR is capable of mitigating them.

\subsection{Interstellar medium}

Simply put, effects in the ISM between the observer
and the pulsar degrade the effective time resolution of the data and,
in severe cases, completely mask short time-scale variations like
the typically milliseconds-wide pulses of pulsars.  Free electrons in
the ISM between the pulsar and the Earth cause the pulsed signal to be
both dispersed and scattered. Dispersion results in the pulses at
lower frequencies being delayed relative to those at higher
frequencies. This effect is directly proportional to the number of
free electrons along the line of sight, which is expressed as the
dispersion measure (DM); the quadratic frequency dependence of the
delay means the effect is particularly severe when observing with even
1\% fractional bandwidths at frequencies below 200~MHz. For example,
at a central frequency of 100~MHz and using a bandwidth of 2~MHz the
pulses from a pulsar with a relatively low DM of 20~pc~cm$^{-3}$
will be smeared out by $\sim$~0.33~s, roughly 10 times the pulse
width of most pulsars.  Even with this modest bandwidth and DM the
smearing is sufficient to make the majority of pulsars undetectable
and certainly prevents detailed study of the emission physics or pulse
morphology.

Traditionally this problem has been solved by dividing the available
bandwidth into narrower frequency channels, which are then delayed
relative to each other by an amount dictated by the DM of the
source. Because the maximum rate at which the data can be sampled is
the inverse of the channel width, there is a limit to the number of
channels one can divide a given bandwidth into, and thus at some point
the resulting sampling time itself becomes too long.  As pulsars are
weak radio sources, large bandwidths are nonetheless required to
improve the signal-to-noise ratio (S/N) of the detections. However
until the advent of digital systems it was not possible to form the
large number of channels required to use large bandwidths.  A
technique to fully correct for the effects of dispersion in the ISM
was developed by \citealt{han71} but because it is computationally
very expensive, this so-called coherent dedispersion has only seen
regular use in the last 15 years or so. The strong dependence of
dispersion on frequency has meant that coherent dedispersion has only
seen limited use at low frequencies
(e.g. \citealt{pku+06b,kss10}). Until now the effect of dispersion has
therefore been a strongly limiting factor on the number and types of
pulsars which can be observed at low frequencies.

Furthermore, the ionised ISM is not distributed evenly
and the inhomogeneities along a given line of sight between the Earth
and a pulsar will cause the pulsed signal to take multiple paths,
resulting in the pulse being scattered. Depending on the scattering
regime, either strong or weak \citep{ric90}, along a particular sight
line, the degree of temporal scattering of the pulse profile can scale
as steeply as $\nu^{-4.4}$.  Along a given line of sight it also shows
a dependence on the total electron content, i.e. DM, but there are
deviations from this relation of at least an order of magnitude
\citep{bcc+04}. Considering the same pulsar and observing frequecy and
system as discussed in the first paragraph of this section, and using
the relation between dispersion measure and scattering from
\citet{bcc+04} we find that the scattering delay would lie between
0.01 and 1 seconds. The stochastic nature of scattering means that it
is difficult to uniquely correct for it without an underlying
assumption for the intrinsic pulse shape; thus, scattering becomes the
limiting factor on the distance out to which low-frequency
observations can be used for detailed study, or even detection, of
pulsars of given rotational periods.

\subsection{Galactic background}

Diffuse radio continuum emission in our Galaxy at frequencies below a
few GHz is predominantly due to synchrotron emission from cosmic rays
moving in the Galactic magnetic field. This emission has a strong
frequency dependence ($\nu^{-2.6}$; \citealt{lmop87,rr88}) and can
thus be a significant component, or even dominate, the system
temperature, $T_{\rm sys}$, at low frequencies. We note however that
this spectral index does vary over the sky and especially in the low
band it may turn over \citep{rcls99}.  Moreover, as the effect of the
sky temperature on sensitivity can have a frequency dependence that is
steeper than that of the flux density of some pulsars, pulsars located
in regions of bright synchrotron emission, such as along the Galactic
plane, become more difficult to detect at low radio frequencies.  This
is especially relevant as two-thirds of all known pulsars are found
within $5^{\circ}$ of the Galactic plane.

\subsection{Ionosphere}

LOFAR operates at frequencies just above the Earth's ionospheric
cutoff, below which radio waves from space are
reflected\footnote{Low-frequency radio frequency interference is also
  reflected back to Earth by the ionosphere and could potentially be
  detected from well below the local horizon.}.  In this regime, the
ionosphere still plays an important role in observations by
contributing an additional time and frequency dependent phase delay to
incoming signals.  In particular, separate ionospheric cells can cause
differential delays across the extent of an interferometric array like
LOFAR, greatly complicating the calibration needed to add the signals
from multiple stations together in phase. For a more detailed
discussion of the ionosphere and LOFAR see \cite{int09} and
\cite{wtnv10} and references therein. The specifics of the challenge
for beam formed observations is described in greater detail in
Sect. \ref{sec:pipelines}.

\subsection{Addressing these challenges with LOFAR}

The flexibility afforded by the almost fully digital nature of the
signal path and the associated processing power of LOFAR mean that it
can address these observational challenges better than any previous
low frequency radio telescope. As will be discussed in
Sect. \ref{sec:pipelines} it is possible to take the complex channelised
data coming from the LOFAR stations and either further channelise to
achieve the required frequency resolution to correct for dispersion,
or, when higher time resolution is required, it is possible to perform
coherent dedispersion on the complex channels. While it is not
presently possible to correct for the effects of scattering, we note
that variations in the magnitude of scattering are so large that some
pulsars with high DMs will still be accessible, as evidenced by our
detection of PSR~B1920+21 with a DM of 217~pc~cm$^{-3}$
(Figure~\ref{fig:profiles}).

The system temperature of LOFAR is sky dominated at nearly all
frequencies and when combined with the very large collecting area this
makes LOFAR very sensitive despite the contribution from the Galactic
synchrotron emission. The Galactic plane is clearly the hottest
region, but that is only a small fraction of the sky\footnote{Note
  however that LOFAR's complex sidelobe pattern means that the
  Galactic plane can still contribute somewhat to the total sky
  temperature even for observations far from the plane itself.} and as
we are most sensitive to the nearby population it will not greatly
affect the number of new pulsars LOFAR will find (see
Sect. \ref{sec:science}).  Our detection of PSR B1749$-$28
(Figure~\ref{fig:profiles}), which is only $\sim 1^{\circ}$ from the
direction to the Galactic Centre shows that observations of bright
pulsars are still possible even with the large background temperature
of the Galactic plane.  Moreover, for observations of known pulsars in
the direction of the Galactic plane, the narrow tied-array beams (see
Sect. \ref{sec:obsmodes}) will reduce the contribution from discrete
extended sources, such as supernova remnants (SNRs), to the system
temperature and thus further improve the sensitivity over single dish
or wide beam telescopes.

In terms of calibrating the station/time/frequency dependent
ionospheric phase delays, we will exploit LOFAR's multi-beaming
capability and its ability to simultaneously image and record
high-time-resolution pulsar data (Sect. \ref{sec:obsmodes}).  The
envisioned scheme is to use a separate station beam to track a
calibration source during an observation, use this to calculate the
required phase adjustments per station, and to implement these online
while observing the main science target with another beam.  This is
admittedly a difficult problem and the solution has yet to be
implemented.  In the case of the innermost core stations on the
Superterp however, differential ionospheric delays are unlikely to be
a major problem.  This means that using the static phase solutions
these stations can be combined coherently without ionospheric
calibration.

\section{Pulsar science at low frequencies}
\label{sec:science}

Here we give an overview of the pulsar science that can be done at
low radio frequencies with LOFAR. This is not intended as an
exhaustive list of envisioned studies, but rather as a general
scientific motivation for LOFAR's pulsar modes.

\subsection{Pulsar and fast transient searches}

%Galactic population, need for new surveys
There are estimated to be approximately 100,000 actively
radio-emitting neutron stars in the Milky Way
\citep{vml+04,lfl+06,fk06}, of which at least 20,000 are visible as
radio pulsars due to fortuitous geometrical alignment of the radio
beam with the direction towards Earth.  With a sample of close to
2,000 known radio pulsars to study, we still have only a rough idea of
this population's overall properties and of the detailed physics of
pulsars \citep{lfl+06}.  This uncertainty is partly a product of the
difficulty in disentangling the intrinsic properties of the population
from the observational biases inherent to past surveys, most of which
were conducted at $\sim 350$~MHz or $\sim
1.4$~GHz.\footnote{Reconciling the total number of neutron stars and
  the supernova rate \citep{kk08} also remains an outstanding, related
  issue, with many fundamental questions still unanswered.}
Discovering a large, nearby sample of pulsars with LOFAR will allow
the determination of the distribution of pulsar luminosities in the
low-luminosity regime, crucial for extrapolating to the total Galactic
population, with the potential for detecting a cut-off in that
distribution.  Such a survey will also quantify the beaming fraction,
that is what fraction of the sky is illuminated by the pulsar beams,
at low frequencies, showing how this evolves from the more commonly
observed 350/1400-MHz bands.  Though a LOFAR pulsar survey will of
course also be observationally biased towards a particular subset of
the total Galactic pulsar population, these biases are in many ways
complementary to those of past surveys, providing the opportunity to
fully characterise the known population.

\begin{figure*}
\centering
     \subfigure{
           \label{fig:pulse-hba}
           \includegraphics[width=0.48\textwidth]{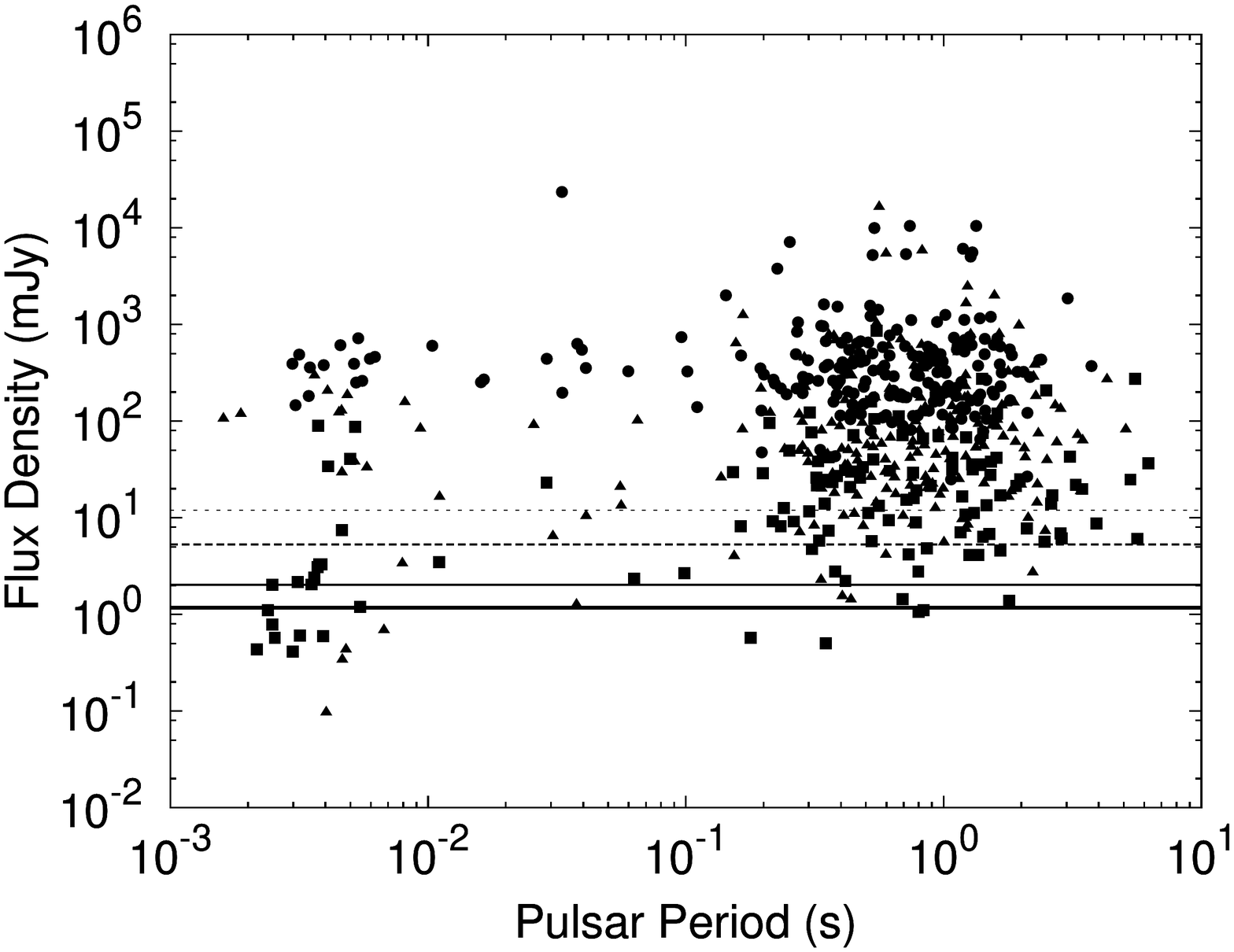}}
     \subfigure{
          \label{fig:pulse-lba}
          \includegraphics[width=0.48\textwidth]{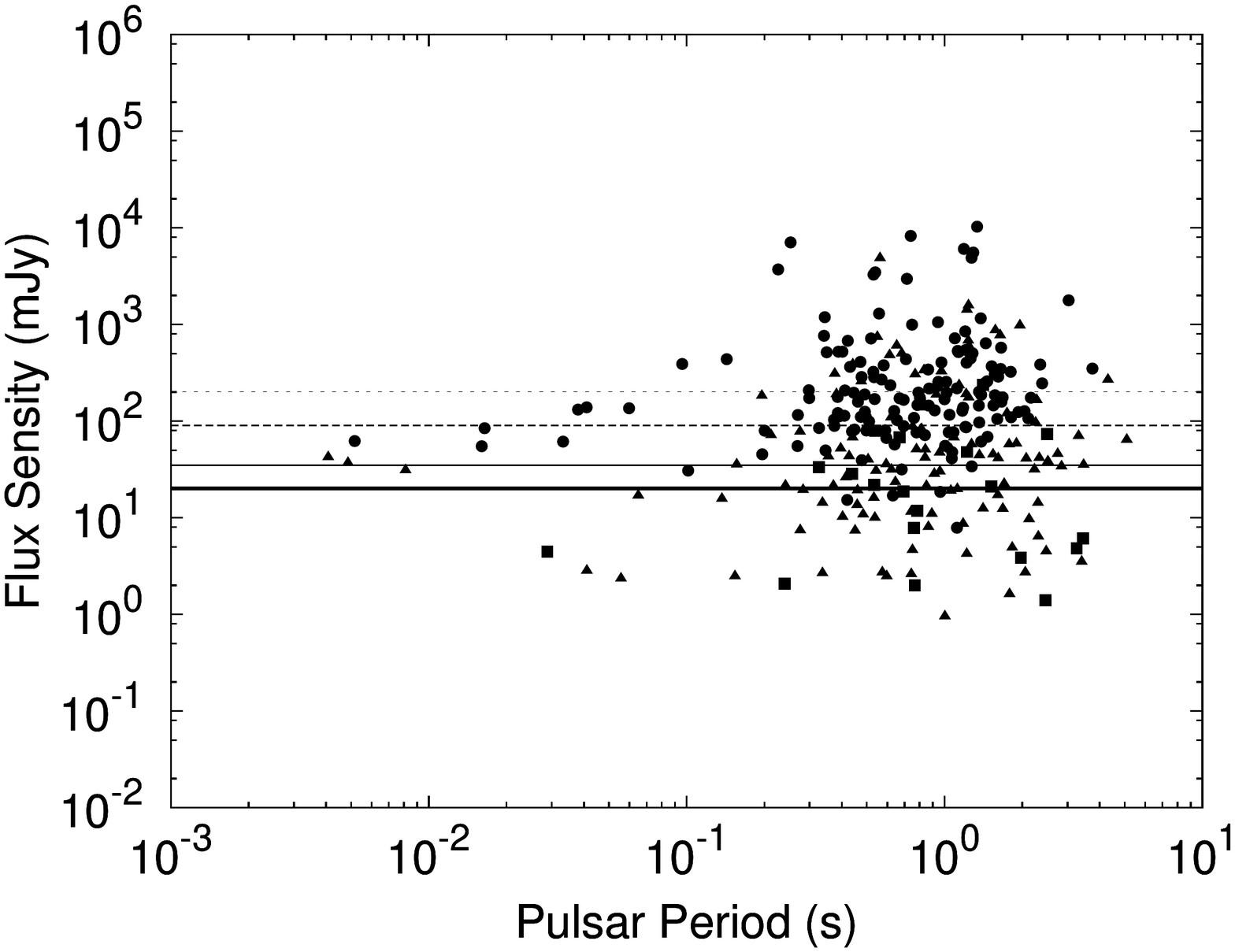}}\\

     \caption{Sensitivity curves for 600-s observations using the HBAs
       (left) and the LBAs (right) with a bandwidth of 48~MHz,
       compared to the extrapolated flux densities of 787 known
       pulsars at 150 and 50~MHz respectively. Flux densities were
       determined using, if known, 100 MHz fluxes from \cite{mms00}
       (circles) otherwise fluxes at either 400 MHz (triangles) or
       1400 MHz (diamonds) \citep{mhth05}, whichever were available,
       were used and scaled to 150~MHz using a known spectral index or
       a typical spectral index of $-1.8$ (squares). All of these flux
       densities were scaled by $\sqrt{W/(P-W)}$, where $W$ is the
       effective pulse width and $P$ is the period of the pulsar, to
       incorporate broadening of the profile by scattering in the
       ISM. $W$ was determined by combining the pulse width at high
       frequencies with the broadening due to scattering in the ISM,
       based on the model of Bhat et al (2004). If $W > 0.75P$ then
       the pulsar was deemed undetectable. It was assumed that
       broadening of the profile due to uncorrected dispersive
       smearing was negligible. MSPs are shown by the open symbols and
       ``normal'' pulsars by filled symbols. The lines in both cases
       correspond to the sensitivity of a single international station
       (fine dashed line), the incoherent sum of 20 core stations
       (heavy dashed line), incoherent sum of all LOFAR stations
       (filled line) and the coherent sum of 20 core stations (bold
       line).}

\label{fig:sensitivity} 
\end{figure*}

%Characteristics and advantages of LOFAR survey
Detailed simulations of potential LOFAR surveys were done by van
Leeuwen \& Stappers (2010)\nocite{ls10} and show that an
all-Northern-sky survey with LOFAR will find about 1000 new pulsars
and will provide a nearly complete census of all radio-emitting
neutron stars within $\sim$ 2~kpc. An indication of the sensitivity of
LOFAR for pulsar surveys can be seen in Figure
\ref{fig:sensitivity}. The limiting distance out to which pulsars can
be detected is governed predominantly by scattering in the ISM. There
are however both pulsar and telescope characteristics that make low
frequency surveys an attractive prospect. The pulsar beam broadens at
low frequencies (Cordes 1978\nocite{cor78} find that the width
changes with frequency $\nu$ as $\nu^{-0.25}$), which nearly doubles
the beaming fraction compared to 1.4~GHz surveys. Furthermore, the
large FoV and high sensitivity of LOFAR mean that such a survey can be
carried out far more quickly and efficiently than any other pulsar
survey, past or present.  Even in the Galactic plane, where the
diffuse background temperature will reduce the sensitivity, the
ability to form narrow tied-array beams (see Sect. \ref{sec:obsmodes})
will reduce the contribution from sources which are large enough to be
resolved.

The large FoV also affords relatively long dwell times, improving the
sensitivity to rare, but repeating events like the pulses from RRATs
(rotating neutron stars which emit approximately once in every 1000
rotations; \citealt{mll+06}), while the rapid survey speed will allow
multiple passes over the sky.  Thus, such a survey is guaranteed to
have a large product of total observing time and sky coverage
($\Omega_{\rm tot} T_{\rm tot}$).  This factor is important for
finding neutron stars that, for intrinsic or extrinsic reasons, have
variable detectability, such as the RRATs and intermittent pulsars
(the latter can be inactive for days: \citealt{klo+06}), those in
relativistic binaries (where the acceleration near periastron can
alter the periods too rapidly to allow detection (e.g.,
\citealt{jk91,rce03}), and those millisecond pulsars,
MSPs,\footnote{MSPs are believed to be formed in binary systems, where
  the neutron star accretes matter from the companion star and is spun
  up to millisecond periods, sometimes referred to as recycling.}
which are found in eclipsing systems (e.g.,
\citealt{fst88,sbl+96}). It has recently become apparent that to
understand the pulsar and neutron star populations we need to know
what fraction of pulsars are relatively steady emitters compared with
those that pulse erratically like the RRATs.

This complete, volume-limited sample can be extrapolated for modelling
the entire neutron star population of the Galaxy, which then constrains
the population of massive stars and the supernova rate, the velocities
and spatial distribution of neutron stars, and the physics of neutron
stars in general.  Chances are that this largely unexplored population of
faint or intermittent radio-emitting neutron stars will also contain
exotic systems -- double neutron stars, double pulsars and possibly
even a black-hole pulsar binary. Such systems provide the best testing
ground for fundamental physical theories, ranging from solid-state to
gravitational physics \citep{ckl+04}.

%Finding MSPs
LOFAR is highly sensitive to MSPs and as they are generally much older
than other neutron stars, they have had more time to leave their
birth-place in the Galactic plane and to become equally distributed in
the halo as well as in the plane. Thus, despite being more easily
affected by scattering, the lower number of free electrons along the
lines of sight out of the Galactic plane means that MSPs are still
prime targets for LOFAR.  Moreover, they are bright at LOFAR
frequencies with the flux density spectra of some MSPs remaining steep
down to 30 MHz \citep{kl01, kll+99}.  Recent studies have successfully
detected about half of the known Northern MSPs using the relatively
insensitive low-frequency ($110-180$~MHz) frontends on the WSRT
\citep{skh08}. Malov \& Malofeev (2010) \nocite{mm10} have also shown
a number of detections at frequencies near 100~ MHz using the Large
Phased Array BSA radio telescope of the Pushchino Radio Astronomy
Observatory despite the limited time resolution.  It is also
noteworthy that the recently discovered ``missing-link''
MSP~J1023+0038 \citep{asr+09}, detected with WSRT at 150~MHz, could
have easily been found by a LOFAR survey, assuming it was observed out
of eclipse. The number of MSPs uncovered at radio wavelengths, often
at 350~MHz, by observations of unidentified {\it Fermi} sources
(e.g. \citealt{rrc+11,hrm+11}) indicates that there are still many
more relatively bright sources to be found. The high sensitiviy of
LOFAR will allow it to detect nearby low-luminosity MSPs and steep
spectrum high-luminosity MSPs which are too far away to be detected at
high frequencies.

%Extra-galactic sources
LOFAR also has the sensitivity to discover the brightest sources in
local group galaxies \citep{ls10}.  If observed face-on and located
away from the Galactic disk, the scatter broadening to pulsars in an
external galaxy will be relatively low and thus a LOFAR survey will
have high sensitivity for even rapidly rotating extragalactic
pulsars.  For a relatively close galaxy like M33, LOFAR could detect
all pulsars more luminous than $\sim$50~Jy kpc$^2$. Ten of the
currently known pulsars in our own Galaxy have comparable luminosity
\citep{mhth05}. There are at least 20 (dwarf) galaxies for which LOFAR
will have good sensitivity to their pulsar population. Complementing
searches for periodic signals, detecting the bursts of either giant or
RRAT pulses can equally pin-point pulsars. In particular, the
ultra-bright giant pulses could be visible in even more remote
galaxies \citep{mc03}.  A survey for extragalactic pulsars would allow
us to investigate if the bright end of the pulsar distribution in
other galaxies differs from that of our Galaxy, and how that ties into
galaxy type and star formation history. Such pulsars can also feed the
understanding of the history of massive star formation in these
galaxies and also, if sufficient numbers can be found, they can be
used to probe the intergalactic medium and possibly constrain or
measure the intergalactic magnetic field.

In addition to the discovery of bright pulses from pulsars in nearby
galaxies, an all-sky LOFAR survey will also cast a wide net for other
extragalactic and cosmological bursts
(e.g. \citealt{fws+08,vkr+08,lbm+07}). The large FoV of LOFAR means
that only 1200/200 pointings of station beams are required to cover
the full Northern Hemisphere with the HBAs/LBAs at central frequencies
of 150 and 60~MHz respectively\footnote{Compare for instance with the
  1.4~GHz Parkes Multibeam system, which requires of the order of
  35000 pointings to cover a similar area of sky.}.  Through a
dedicated pulsar/fast transient survey, {\it and} regular
piggy-backing on the imaging observations of other projects, it will
be possible to obtain an unprecedented $\Omega_{\rm tot} \times T_{\rm
  tot}$ figure of merit.  For example, 8000 HBA observations of 1~hr
each (roughly two year's worth of observations, assuming 50\%
observing efficiency) would provide 4 hours of all-sky coverage,
meaning that events with rates of only 6/day over the whole sky could
be detected in such a data set.  The parameter space of such rare
bursts has never been probed with such high sensitivity. Moreover
there are many options for performing wide angle rapid shallow
searches for fast transients using either single stations, forming
sub-arrays or forming 244 LBA station beams all at once.

LOFAR will also respond to high-energy X-ray/$\gamma$-ray or
high-frequency radio triggers.  Most high-energy events peak later in
the radio, providing ample time to follow-up on such events.
Dispersive delay, which can easily be 100s of seconds in the LOFAR
low-band, may also aid in detecting the prompt emission of some
bursts.  Efforts are being made to reduce the setup time of a typical
observation so that target of opportunity observations can begin
quickly and automatically.  Since LOFAR has no moving parts,
repointing is only a matter of reconfiguring the delay corrections
used in beamforming.  The ultimate goal is to repoint to any location
of the sky and begin new observations in just a few seconds.

\subsection{The physics of pulsar radio emission}

%Intro
The sensitivity and frequency range of LOFAR opens up the
low-frequency window to new studies of pulsar emission. It is
precisely in the LOFAR frequency range where some of the most
interesting changes in pulsar radio emission can be observed,
including significant broadening of the pulse profile, presumably due
to a ``radius-to-frequency'' mapping and deviations from this expected
relation; changes in the shape of pulse profile components; and a
turn-over in the flux density spectrum. The frequency range accessible
by LOFAR is also where propagation effects in the pulsar magnetosphere
are expected to be largest
(e.g.~\citealt{pet06,pet08,wsve03}). Therefore, simultaneous
multi-frequency observations (see Figures \ref{fig:lbaprofiles} and
\ref{fig:simult}) are expected to reveal interesting characteristics
of the pulsar magnetosphere, such as the densities and birefringence
properties \citep{smi88}, that will ultimately lead to a better
understanding of the emission mechanisms of pulsars.

\subsubsection{Pulsar flux densities and spectra}

LOFAR's large fractional bandwidth is a big advantage for measuring
the low-frequency flux densities and spectra of pulsars.  As discussed
in Sect. \ref{sec:subsing} it will be possible to use multiple
stations to observe contiguously from 10~MHz to 240~MHz. With the
ability to easily repeat observations, we can be certain to remove
effects such as diffractive scintillation, which may have affected
previous flux density estimates. It is also the case that the
timescale for refractive scintillation becomes very long at these
frequencies and this will have a modulating effect on the determined
fluxes. However this is typically of smaller amplitude. Repeating
these observations will also allow one to determine whether the
spectral characteristics are fixed in time, or whether they vary, for
any given source, something which has rarely been done in the
past.\nocite{gl98}\nocite{hx97} It is also important to note the role
that variable scattering might play in flux density determination
(e.g. \citealt{kljs08}).

So far, there are relatively few pulsars for which spectral
information in the LOFAR frequency range is available \citep{mms00}.
Obtaining a large, reliable set of low-frequency flux density
measurements over the 4 lowest octaves of the radio window is a
valuable missing piece in the puzzle of the pulsar emission process.
For instance, the generally steep increase in pulsed flux density
towards lower frequencies is seen to turn over for a number of pulsars
at frequencies between 100 and 250~MHz.  However, the physical origin
for this turn over is not yet clear.  Conversely, there are a number
of pulsars for which no such break in the spectrum has yet been seen
and studies at lower frequencies are needed to locate this spectral
break \citep{kl01,mal00}.  Furthermore, there is evidence for
complexity in the shape of pulsar spectra (e.g.~\citealt{mkkw00a}), which
again is an important signature of the pulsar emission process that
LOFAR's wide-band data can probe.  In addition to their importance for
our understanding of the emission process itself
(e.g.~\citealt{gbi93,mel04}), pulsar spectra and flux densities are a
basic ingredient for estimating the total radio luminosity and
modelling the total Galactic population of radio pulsars.

In many ways, MSPs show very similar radio emission properties to
those of non-recycled pulsars (e.g.~\citealt{kxl+98}); yet, they
differ from these by many orders of magnitude in terms of their
rotation period, magnetic field strength, and the size of their
magnetosphere.  A curious and potentially important distinction is
that MSPs tend to show un-broken flux density spectra, continuing with
no detected turn-over down to frequencies of 100~MHz
\citep{kll+99,kl01} and lower (e.g. \citealt{nbf+95}).  Kuzmin \&
Losovsky (2001) presented spectra of some 30 MSPs using measurements
close to 100~MHz. Using the sensitivity, bandwidth and tracking
abilities of LOFAR, it will be possible to more than double that
number, and to provide measurements with much wider bandwidths. Such
flux density measurements will also be possible using LOFAR's imaging
ability, even if severe scattering prevents detailed profile
studies. In general, however, when scattering is not the dominant
  effect the improved time resolution enabled by the coherent
dedispersion mode of LOFAR should also allow us to resolve individual
profile components in MSP profiles, so that the average profile
spectrum can be compared to that of single pulse components.

Correlating the spectral properties with other properties of the
pulsar (such as pulse shape, geometrical parameters or pulse energy
distributions) may reveal important physical relationships. These can
then be used to further improve pulsar emission and geometry
models. This is particularly important as more and more sources are
being detected at high energies with {\it Fermi}, placing interesting
constraints on the emission sites (e.g.~\citealt{aaa+10a}).

\begin{figure*}
\centering
     \subfigure{
           \label{fig:single-hba}
           \includegraphics[width=0.48\textwidth]{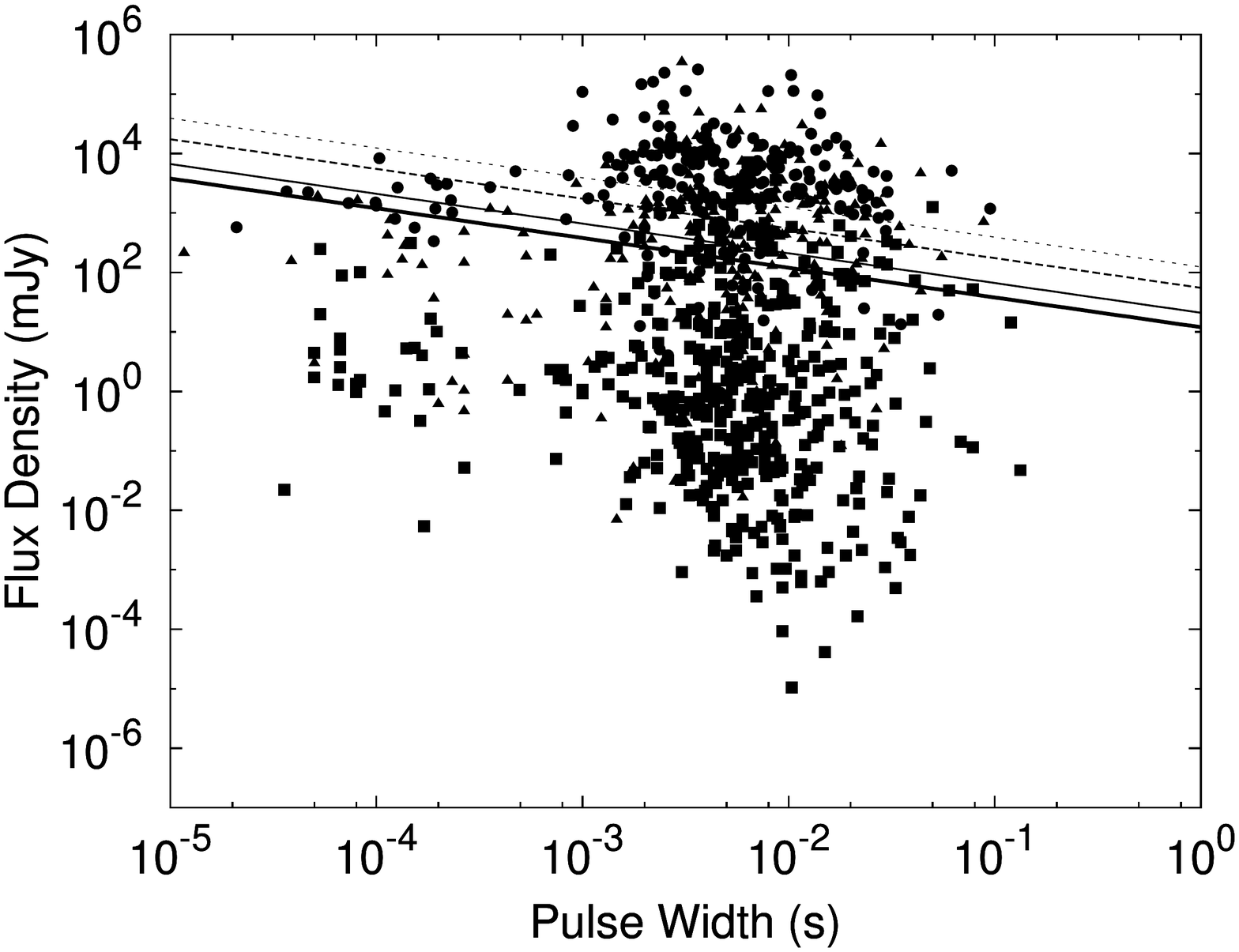}}
     \subfigure{
           \label{fig:single-lba}
          \includegraphics[width=0.48\textwidth]{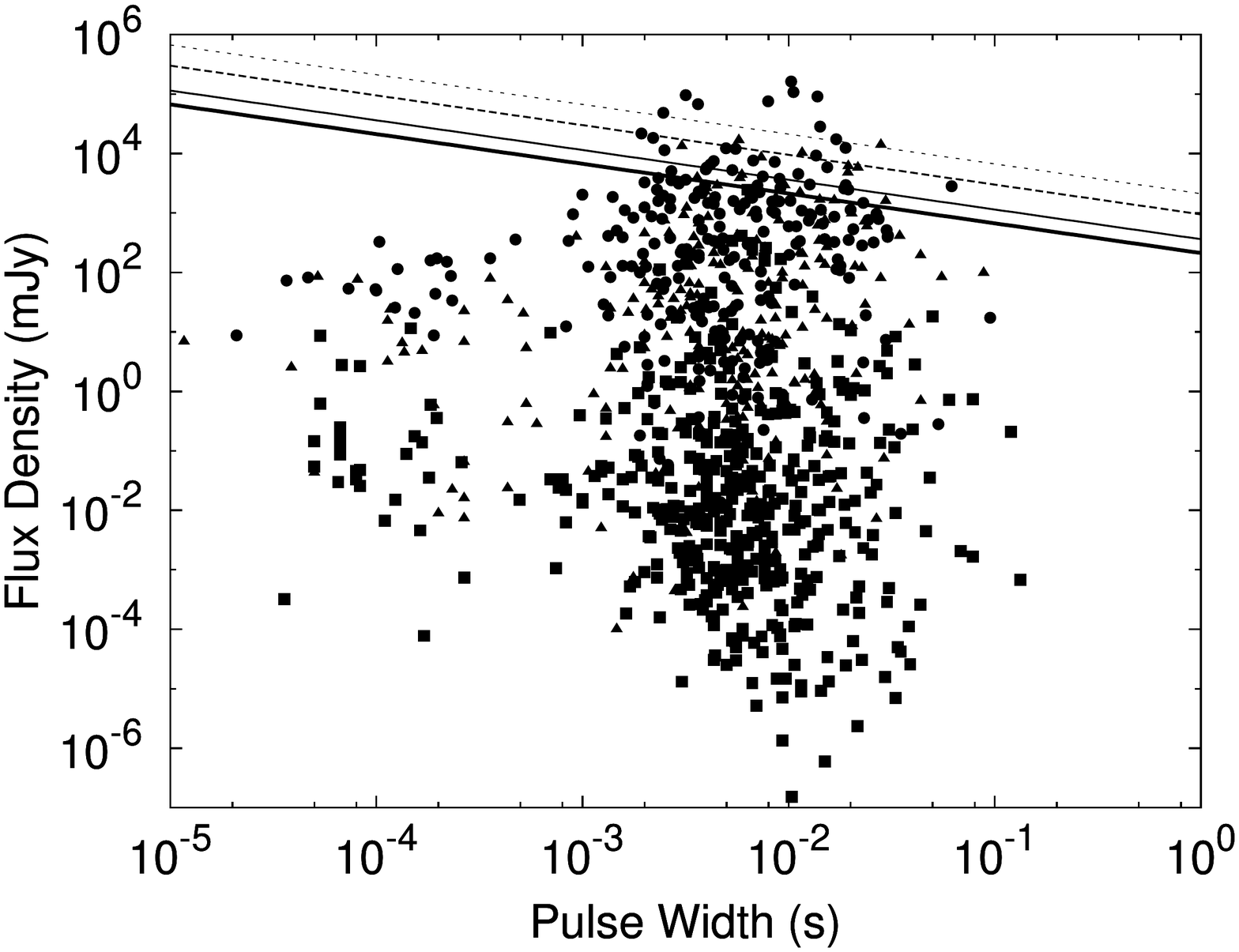}}

     \caption{Sensitivity curves for the HBAs (left panel) and LBAs
       (right panel) to single pulses as a function of pulse width at
       a central frequency of 150~MHz (HBAs) and 50~MHz (LBAs) and
       assuming a bandwidth of 48~MHz. Single pulse widths for 787
       pulsars were calculated by assuming them to be one third of the
       width of the average pulse and then scatter broadened assuming
       the relationship of Bhat et al (2004). It was assumed that
       broadening of the pulse due to uncorrected dispersive smearing was
       negligible. Single pulse fluxes were determined based on the
       assumed pulse width and, if known, 100~MHz fluxes from
       \cite{mms00} were used.  Otherwise, fluxes were extrapolated to
       150~MHz from either 400~MHz or 1400~MHz measurements (whichever
       was available, \citealt{mhth05}) using a known spectral index
       or a typical spectral index of $-1.8$. The symbols are the same
       as in Figure \ref{fig:sensitivity}. The lines in all cases
       correspond to the sensitivity of a single international station
       (fine dashed line), the incoherent sum of 20 core stations
       (heavy dashed line), incoherent sum of all the LOFAR stations
       (filled line) and the coherent sum of 20 core stations (bold
       line). }

\label{fig:single-sensitivity} 
\end{figure*}

\subsubsection{Pulse profile morphology}

\begin{figure*}
\begin{center}
\includegraphics[scale=0.6]{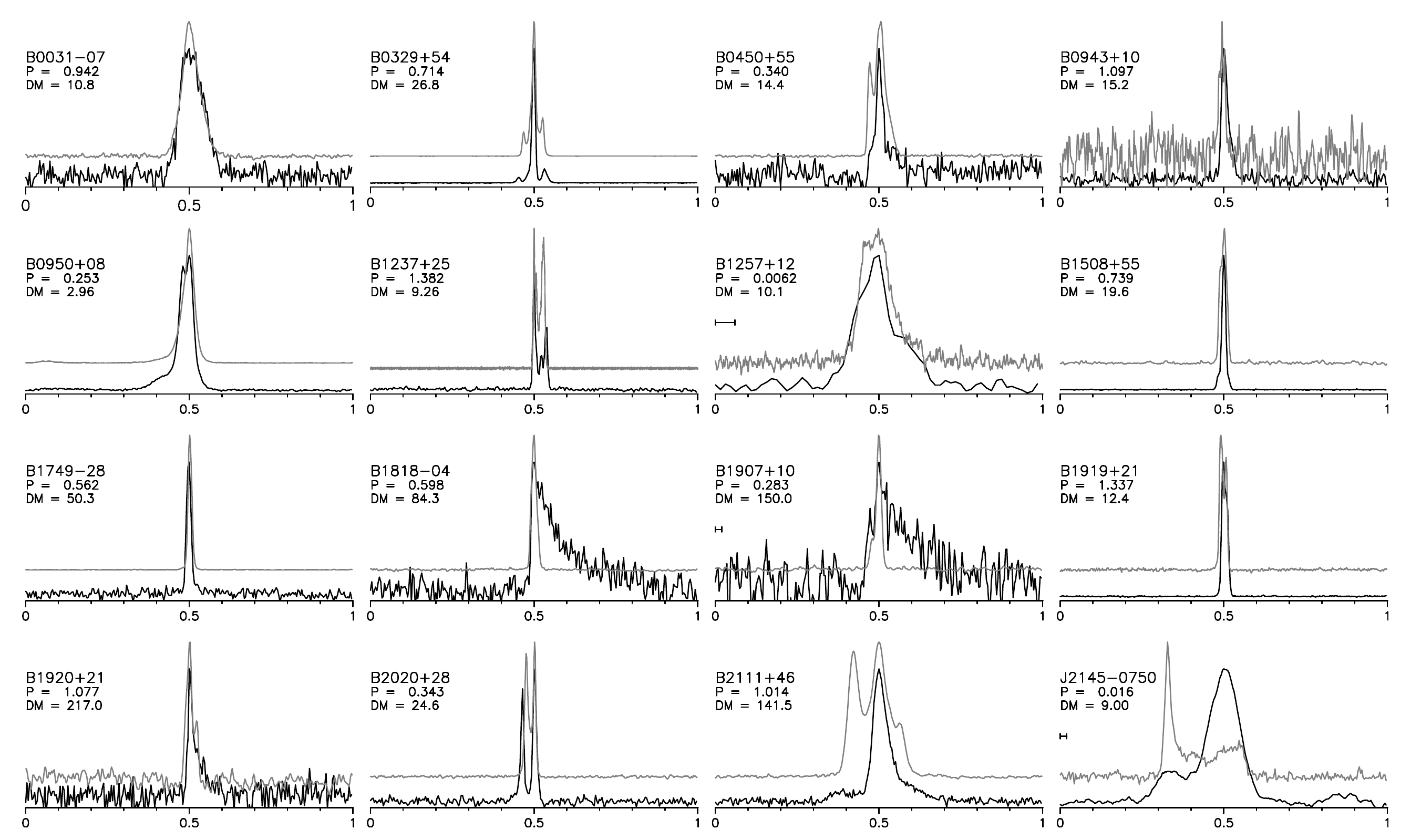}
\caption{A sample of average pulse profiles from LOFAR observations of
  various pulsars made during the commissioning period. The dark line
  corresponds to LOFAR observations made with the HBAs with 48~MHz of
  bandwidth, divided into 3968 channels, at a central frequency of
  163~MHz. Typically about 12 stations were used and they were
  combined incoherently. At present there is no flux calibration for
  the LOFAR data, hence no flux scale is shown. The majority of the
  observations are $\sim 1000$~s in duration, except for PSR B1257+12,
  which was observed for 3~hr. The dispersion measure smearing is less
  than two bins for the majority of the pulsars shown; if the smearing
  is large, this is indicated by a bar on the left hand side. The
  lighter line corresponds to observations made at 1408~MHz by Gould
  \& Lyne (1998) except for the observations of PSR B0943+10 which was
  at 608~MHz and PSR B1237+25 which was at 1600~MHz (from van
  Hoensbroech \& Xilouris 1997). All archival profiles are from the
  EPN database (www.mpifr-bonn.mpg.de/div/pulsar/data/browser.html);
  they are shown for illustration purposes and are manually
  aligned. Both the LOFAR and EPN profiles have been normalised to the
  peak intensity. The periods are given in seconds and the dispersion
  measures in pc~cm$^{-3}$. }
\label{fig:profiles}
\end{center}
\end{figure*}

Observations of the average pulse profile morphology of about 50\% of
pulsars studied over a wide frequency range indicate substantial and
complex variations which are not easily understood in the standard
model of pulsar emission (e.g.~\citealt{ls04,lk05}). This simple
pulsar model has the plasma and emission properties dominated by the
dipolar magnetic field and is often combined with a model where the
emission obeys a ``radius-to-frequency mapping'', such that lower
frequencies are emitted further out in the magnetosphere resulting in
wider pulse profiles \citep{cor78}. However this simple picture is not
the full story as evidenced by the spectral behaviour of the different
pulse components (e.g. \citealt{mr02a,ran93}). The latter property is
in general supported by observations, and probably corresponds to
stratification of the density in the emission zone although plasma
propagation effects may play an important role here as well (see
e.g.~Lorimer \& Kramer 2005 for a review). The sensitivity and time
resolution possible with LOFAR, combined with the wide frequency
range, will allow us to obtain average profiles for the majority of
the known population of pulsars in the Northern sky (Figure
\ref{fig:sensitivity}). Additionally, hundreds of new pulsars,
expected to be discovered with LOFAR, can also be studied in detail;
as is demonstrated by the high quality data from our commissioning
observations (Figure \ref{fig:profiles}).

There is also observational evidence that profiles that are dominated
by the outer pulse components at high frequencies evolve to profiles
where the central component is comparatively much stronger at low
frequencies (e.g.~Lorimer \& Kramer 2005). The exact profile evolution
may depend on the distribution of plasma along the field lines or may
be the result of different emission zones. Indeed, the general picture is
much more complex: variations are seen in the pulse profile with
different components having different spectral indices and new
components becoming visible \citep{kwj+94}. By combining LOFAR data
over its wide frequency range with higher frequency observations we
can investigate these properties for a much larger group of pulsars
(see for an example Figure \ref{fig:simult}).

A further important probe of the emission mechanism, the geometry and
the plasma properties of the magnetosphere is polarisation. So far,
there have been only limited studies of the low-frequency polarisation
properties of pulsars.  In contrast, in simultaneous multiple
high-frequency observations strange polarisation variations have been
seen which point directly to the physics of the emission regions
(e.g.~\citealt{ijk+94,khk+01}). Polarisation studies with LOFAR have the
potential to provide insight into how the polarisation behaves at
these relatively unexplored low frequencies.  LOFAR polarisation data
can determine whether at low frequency the percentage of polarisation
evolves strongly with frequency and if there are even more severe
deviations of the polarisation position angle swing from the
expectation for a dipole field. It will also be possible to determine
if the polarisation properties change below the spectral break and whether
the sign changes of circular polarisation seen at higher frequencies show
similar properties at lower frequencies or exhibit no such sign changes.

\subsubsection{Single pulses}
\label{sec:singlepulses}

The average, or cumulative, pulse profiles of radio pulsars are formed
by adding at least hundreds of consecutive, individual pulses.  While
most pulsars have highly stable cumulative pulse profile morphologies,
their individual pulses tend to be highly variable in shape, and
require more complex interpretations than the average profiles. For
example, single pulse studies of some pulsars have revealed
quasi-periodic ``micropulses'' with periods and widths of the order of
microseconds \citep{cwh90,lk05}.  As such observed phenomena exhibit
properties and timescales closest to those expected from theoretical
studies of pulsar emission physics (e.g.~\citealt{mel04}) studying them can
strongly constrain models of the emission process. Single pulse
studies at low frequencies are particularly interesting, because
microstructure \citep{spk83,stb94} and subpulse modulation
\citep{wse07,uzb08} tend to be stronger there.  We also expect density
imbalances and plasma dynamics to be the most noticeable at low radio
frequencies \citep{pet06}.

In some pulsars, single sub-pulses are observed to drift in an
organised fashion through the pulse window (see Figure
\ref{fig:drift}), an effect which may either be related to the
creation of plasma columns near the polar cap or to the plasma
properties within the pulsar magnetosphere. Some pulsars show
significant frequency evolution in the properties of these drifting
subpulses and observations at low frequencies, which probe a very
different sight line across the typically enlarged emission beam, can
be used to reconstruct the distribution of emission within the
magnetosphere. Greater constraints can be obtained by simultaneous
observations at different frequencies, within LOFAR bands but also in
combination with high-frequency facilities. Such experiments will
reveal how the radius-to-frequency mapping manifests itself in the
sub-pulse modulation; for example, is the drifting more or less
organised at the low frequencies? There is some evidence that drifting
subpulses are more pronounced \citep{wse07} and perhaps easier to
study at low frequencies.

In order to gauge the number of pulsars whose single pulses can be
studied with LOFAR, we compare the estimated single-pulse flux
densities in Figure \ref{fig:single-sensitivity}.  We show all known
pulsars visible with LOFAR and scale them to LOFAR sensitivity in both
the high band and low band.  While there are a number of assumptions
(see caption of Figure \ref{fig:single-sensitivity}) that go into
these calculations, we see that on average we expect to be able to see
single pulses from about one third of visible pulsars in the high band
and ten percent of visible pulsars in the low band.  This is a
significant increase on what has previously been possible. When
combined with LOFAR's very wide bandwidth and the ability to track
sources, it is clear that LOFAR will allow for a very rich study of
the emission physics of radio pulsars (for example see Figure
\ref{fig:drift}). Based on the success of observing radio pulsars
using multiple telescopes (e.g.~\citealt{khk+01,kjk03,kkg+03}), we expect
that in particular the combined simultaneous multi-frequency
observations of single pulses at very low and high frequencies will be
extremely useful in confirming or refuting some of the models and
interpretations derived from previous data sets. The length and
quality of the available data will be much improved, both by
being able to follow sources for a much longer time than previously
possible with low-frequency transit telescopes, and by having
much better sensitivity, frequency coverage, and time resolution.  In
fact, LOFAR will also be able to detect single pulses from MSPs,
something which has so far been done for only very few sources
\citep{es03,jap01}. This will allow us to study for the first time if
pulsar phenomena discovered in normal, younger pulsars are also
present in their much older brethren. Such effects include
potential mode changing, nulling and drifting subpulse properties of
MSPs.  Understanding the occurrence of such effects will also be
important for high precision timing at higher frequencies, as profile
instabilities could, for example, hamper efforts to detect
gravitational waves \citep{cs10}.  We note, however, that the high time resolution
requirement of these observations compete with the effects of
interstellar scattering which will ultimately limit the number of
sources that can be studied.  Nonetheless, we expect that a few
hundred sources can be studied (Figure \ref{fig:single-sensitivity}),
making this a particularly interesting aspect of pulsar studies with
LOFAR.

\subsection{Pulsars as probes of the ISM and the Galactic magnetic field}

Pulsars are ideal probes of the ionized component of the ISM,
through the effects of scintillation, scattering, dispersion, and
Faraday rotation.  Most ISM effects become significantly stronger
towards low radio frequencies, meaning LOFAR is well placed to study
these phenomena in the direction of known pulsars.  Furthermore, by
greatly increasing the number of known pulsars, a LOFAR pulsar survey
will add a dense grid of new sight lines through the Galaxy.  As the
majority of these pulsars will be nearby or out of the Galactic plane,
the dispersion and scattering measures of this new sample will improve
our model of the distribution of the ionized ISM and its degree of
clumpiness \citep{cl02}.

Scintillation and scattering are related phenomena which result in
variations of the pulse intensity as a function of frequency and time,
and in broadening of the pulse profile respectively
(e.g. \citealt{ric70,ric90,nar92a}). They are caused by fluctuations
in the density of free electrons in the ISM.  Scintillation studies
have been revolutionised in the last few years by the discovery of
faint halos of scattered light, extending outward to $10-50$ times the
width of the core of the scattered image \citep{smc+01}.  This, in
turn, gives a wide-angle view of the scattering medium with
milliarcsecond resolution, and the illuminated patch scans rapidly
across the scattering material because of the high pulsar space
velocity.  Some of the most interesting effects are visible at low
frequencies. The dynamic nature of these phenomena also fits well with
LOFAR's  monitoring capabilities. In a sense, this
scintillation imaging will allow the monitoring of the range of
interstellar conditions encountered along a particular sight line.
The wide bandwidths and high sensitivities will also allow very
precise measurements of the DM of many pulsars; for MSPs this could be
as precise as $1:10^{4}$ or better for an {\it individual}
measurement.  Such precision, even for normal pulsars, will enable
detailed studies of variations in the DM which can be used with the
scintillation properties to study structures in the ISM
(e.g. \citealt{yhc+07}).

Scattering is a spatial effect, whereby the inhomogeneities in the ISM
cause ray paths to be redirected into the line of sight with a
resulting delay associated with the extra travel time. Scattering is
typically assumed to happen in a thin screen located midway between
the pulsar and the Earth, so that the turbulence obeys a Kolmogorov
law even though deviations are observed (e.g. \citealt{lmg+04}). There
is evidence that the scattering does not have the expected frequency
dependence for such a spectrum and that there is also an inner and
outer scale to the turbulence which can be measured
(e.g. \citealt{acr81,ars95,ss02}).  Low-frequency measurements with
LOFAR will allow us to measure (even subtle) scattering tails for many
more pulsars, affording better statistical studies. The study of
individual, moderately scattered but bright pulsars will provide a
probe of the distribution of electrons along the line of sight,
enabling one to distinguish between the different frequency
dependences of the dispersive and scattering delays that affect the
arrival time of the pulse (see for example Figure
\ref{fig:0329scatt}).  Identifying such effects has the potential to
extract distances and thicknesses of scattering screens
\citep{rjt+09,ss10}.

LOFAR can increase the number of known rotation measures to pulsars,
which will place important constraints on the overall magnetic field
structure of the Milky Way, something that is still not well
characterised (e.g. \citealt{njkk08,han09,wfl+10}). At LOFAR frequencies
it is possible to also measure the very small rotation measures of the
nearby population of pulsars which provides an unprecedented tool for
studying the local magnetic field structure. The large number of new
sight lines will also allow statistical studies of small-scale
fluctuations of electron density and magnetic field variations down to
the smallest scales that can be probed with pulsars ($\lesssim 1$~kpc).  These
same techniques will open up new vistas when applied to the first
(truly) extragalactic pulsars, which will be discovered by LOFAR in
the survey of local group galaxies that will be undertaken.

\subsection{Pulsar timing}

Ultra-high-precision pulsar timing (at the ns$-$sub-$\mu$s level) is not
possible at low radio frequencies because the timing precision is
strongly affected by ISM effects that, currently at least, cannot be
compensated for.  However LOFAR's large FoV, multi-beaming capability, and the
availability and sensitivity of single stations working in parallel to
the main array enable many pulsars to be timed at sufficient precision and
cadence to be of scientific interest. Regular monitoring of the
rotational behaviour of radio pulsars plays an important role in
understanding the internal structure and spin evolution of neutron
stars and how they emit. It is also an important probe of the ISM and
potentially for the emission of gravitational waves.  Measuring DM
variations and modelling the scattering of pulse profiles at LOFAR
frequencies has the potential to help correct pulse shape variations
at high observing frequencies and thus reduce systematics in high
precision timing.

Rotational irregularities such as glitches are attributed to the
physics of the super-dense superfluid present in the neutron star
core, and so their study allows us to probe a physical regime far
beyond those that can be reached in laboratories on Earth
(e.g. \citealt{dal96,vm10}). These glitches might also trigger heating
of the neutron star surface or magnetic reconnections which can be
studied in either or both gamma- and X-rays \citep{tc01,rem91}. They
may also cause sufficient deformation of the neutron star that it
might be a gravitational wave emitter (e.g. \citealt{bvm10}). To allow
these multi-messenger follow-up observations requires the accurate
determination of glitch occurence times which can be achieved through
regular monitoring with LOFAR. Moreover, the studies of high-energy
emission from pulsars, which is presently done in detail by the {\it
  Fermi} and {\it Agile} satellites and other telescopes like MAGIC
and H.E.S.S.  (e.g. \citealt{aaa+09a,tba+09,aaa+08,aab+07}), as well as the
search for non-burst-like gravitational waves from pulsars
\citep{aaa+10b} can only be performed with the provision of accurate
pulsar rotational histories.

Three relatively recently discovered manifestations of radio emitting
neutron stars, the RRATs, the intermittent pulsars, and the radio
emitting magnetars \citep{mll+06,klo+06,lyu02,crh+06}, can all greatly
benefit from the monitoring capabilities of LOFAR.  All of these
source types show only sporadic radio activity, which in some cases is
only present for less than a second per day.  As such, a large amount
of on-sky observing is often preferable to raw sensitivity, making
such studies particularly well-suited for individual LOFAR stations
observing independently from the rest of the array.  It is possible
that there are neutron stars whose radio emission is even more
sporadic; discovering these will require potentially several days of
cummulative integration time per sky position, which is achievable as
part of the LOFAR Radio Sky Monitor \citep{fws+08}.

So far some 30 RRATs have been discovered, though periods and period
derivatives have only been determined for slightly more than half of these
\citep{mlk+09}.  Detailed studies of these sources are hampered by the
fact that they pulse so infrequently as to make proper timing
follow-up prohibitively expensive for most telescopes.  LOFAR's
multi-beaming capabilities and the separate use of single stations
make such studies feasible.  These can determine key parameters like
the spin period $P_{\rm spin}$ and its derivative $\dot{P}_{\rm spin}$
as well as better characterising the nature of the pulses themselve.
These are all essential for understanding the relationship of RRATs to
the ``normal'' pulsars.

Unlike the RRATs, the intermittent pulsars appear as normal pulsars
for timescales of days to months, but then suddenly turn off for
similar periods \citep{klo+06}.  Remarkably, when they turn off they
spin down more slowly than when they are on, providing a unique and
exciting link with the physics of the magnetosphere. Observing more
on/off transitions will improve our understanding of the transition
process.  Furthermore, recent work by \cite{lhk+10} has shown similar
behaviour in normal pulsars, relating timing noise, pulse shape
variations and changes in spin-down rate. Moreover they demonstrate
that with sufficiently regular monitoring, that can be achieved with
telescopes like LOFAR, it may be possible to correct the effects these
variations cause on the pulsar timing. This will allow for the
possibility of improving their usefulness as clocks.

Another manifestation of radio pulsars which show highly variable
radio emission are the magnetars, extremely magnetised neutron stars
($B_{\rm surf} \sim 10^{14-15}$~G), which were thought to emit only in
gamma- and X-rays.  Recently, however, two such sources have been
shown to emit in the radio, albeit in a highly variable way
\citep{crh+06,crhr07}.  This emission may have been triggered by an
X-ray burst, and was subsequently observed to fade dramatically.  Even
more recently, a radio pulsar was discovered which has a rotation
period and magnetic field strength within the observed magnetar range.
This pulsar also shows similarly variable radio emission properties,
as seen for the other two ``radio magnetars'', though no X-ray burst
has been observed \citep{lbb+10}. It is only by monitoring these
sources, and regularly searching the sky, in the radio that we can
understand the link between intermittent radio pulsations and
high-energy bursts, as well as the lifetime of these sources as radio
emitters.

\subsection{New populations}

The case for observing pulsars at low frequencies, as discussed so
far, is partly based on our knowledge of the typical spectral
behaviour of the known population.  However, there are pulsars like
PSR B0943+10, which have flux density spectra with spectral indices
steeper than $\alpha = -3.0$ \citep{dr94} and a number of MSPs which
have steep spectra and which do not show a turn-over even at
frequencies as low as 30~MHz (e.g. \citealt{nbf+95}). It is as yet
unclear whether these extreme spectral index sources represent a tail
of the spectral index distribution. There is a potential bias against
steep spectrum objects as they need to be very bright at low
frequencies in order to be detected in high-frequency surveys. For
example, the minimum flux density of the HTRU, High Time Resolution
Universe, survey with the Parkes telescope at 1400~MHz \citep{kjv+10}
is 0.2 mJy and for a spectral index $-3.0$ pulsar the corresponding
140~MHz flux density would be 200~mJy. In contrast, a source close to
the LOFAR search sensitivity of about 1~mJy would have to have a flux
density no steeper than $-$0.7 to be detected by HTRU, suggesting that
a large number of pulsars may only be detectable at low frequencies.

Beyond these regular pulsars with steep spectral indices, more exotic
neutron stars are seen to show transient radio emission. While the
anomalous X-ray pulsars XTE~J1810-197 and 1E~1547.0-5408
\citep{crh+06,crhr07} have shallow spectral indices and are very dim
in the LOFAR regime, several other AXPs and magnetars may appear to be
only detectable at frequencies near 100~MHz (e.g. \citealt{mmt06}),
offering an intriguing possibility to study more of these
high-magnetic-field objects in the radio band if confirmed. The
potential 100-MHz detection of Geminga, a prominent rotation-powered
pulsar visible at optical, X- ray and gamma-ray wavelengths
\citep{mm97,kl97,sp97b}, could exemplify a population of neutron stars
that may be only detectable at radio wavelengths with LOFAR. Also, the
non-detection of radio emission from X-ray dim isolated neutron stars
(XDINSs) thus far \citep{kml+09}, could be due to the beams of these
long-period sources being quite narrow at high frequencies and there
may be a better chance to detect them in radio at LOFAR frequencies.
In fact, weak radio emission from two XDINSs, RX J1308.6+2127 and RX
J2143.0+0654, was reported by Malofeev et al. (2005,
2007)\nocite{mmt+05,mmt07} at 111~MHz, hence it would be important to
confirm this detection with LOFAR.  This is also the case for the
pulsars discovered in blind searches of {\it Fermi} gamma-ray photons,
many of which do not exhibit detectable radio emission
\citep{aaa+09a,aaa+10a}, as well as the remaining unidentified
gamma-ray sources, which have characteristics of radio pulsars but
which have not yet been detected in the radio \citep{aaa+10c}.

\section{Observing modes}
\label{sec:obsmodes}

Here we describe LOFAR's various ``beamformed'' modes, both currently
available or envisioned, for observing pulsars and fast transients.
Though we concentrate on the pulsar and fast transient applications of
these modes, we note that LOFAR's beamformed modes are also
applicable to other high-time-resolution studies including dynamic
spectra of planets (both solar and extra-solar), flare stars, and the
Sun.  We also provide a brief description of LOFAR's standard
interferometric imaging mode, which can be run in parallel with some
beamformed modes.  For a detailed description of LOFAR imaging, see
Heald et al. (2010)\nocite{hmp+10} and van Haarlem et al. (2011).

LOFAR is an interferometer with sparsely spaced stations, distributed
in such a way as to produce reliable high-resolution images.  To
achieve this, the data from all stations are correlated\footnote{Note
  that the antennas within an individual station are first summed in
  phase to form one or multiple station beams on the sky.} with each
other, resulting in a significant increase in the amount of data. To
reduce the data rate to an acceptable level there is an averaging step
which reduces the time resolution of the data to typically about one
second, or longer (though somewhat shorter integrations are possible
in some cases). To sample the combined radio signal at significantly
higher time resolution than this ($t_{\rm samp} < 100$~ms), one has to
normally sacrifice spatial resolution, and/or the large FoV seen by
the individual elements, to form a single beam pointing in the
direction of the source of interest. It is these so-called beamformed
or pulsar-like modes that we will describe here.  The LOFAR Transients
Key Science Project \citep{fws+06} will use both the imaging and
beamformed modes to discover and study transient sources. The imaging
mode will probe flux changes on timescales of seconds to years, while
the beamformed modes will probe timescales from seconds down to
microseconds and will revisit the same sky locations over the course
of days to years. With the Transient Buffer Boards (TBBs;
Sect. \ref{sec:TBBs}) it will be possible to form images with high time
resolution, but limited observing durations.

There are many ways in which the various parts of LOFAR (antennas,
tiles, stations) can be combined to form beams (see Table
\ref{tab:modes}).  The almost completely digital nature of the LOFAR
signal processing chain means that it is highly flexible to suit a
particular observational goal. In the following sub-sections we will
discuss different options for combining these signals, to maximise
either the FoV, instantaneous sensitivity or to compromise between
these two factors.  For the sake of clarity however, we begin by
defining some related terms.  An {\it element beam} refers to the FoV
seen by a single element, a dipole in the case of the LBAs and a tile
of $4 \times 4$ dipoles in the case of the HBAs (recall that these
dipoles are combined into a tile beam using an analog
beamformer). The term {\it station beam} corresponds to the beam
formed by the sum of all of the elements of a station. For any given
observation there may be more than one station beam and they can be
pointed at any location within the wider element beam.  A {\it
  tied-array beam} is formed by coherently combining all the station
beams, one for each station, which are looking in a particular
direction.  There may be more than one tied-array beam for each
station beam.  Station beams can also be combined incoherently in
order to form {\it incoherent array beams}.  These retain the FoV of
the individual station beams and have increased sensitivity compared
with a single station.

\begin{table*}
\caption{Comparison of the LOFAR beam-forming modes}
\begin{center}
\begin{footnotesize}
\begin{tabular}{l c c c c c}
\hline \hline
Mode                            & Sensitivity & FoV        & Resolution & Data Rate & FoM     \\
                                & (Norm.)     & (sq. deg.) & (deg)      & (TB/hr)   & (Norm.) \\
\hline
\multicolumn{6}{c}{High-Band Antennas (HBAs)} \\
\hline
Single HBA sub-station         & 1 / 0.35          & 18 / 147    & 4.8   & 0.3 & 1   \\
Single Rem. Station            & 2 / 0.7           & 10 / 82     & 3.6   & 0.3 & 3   \\
Single Intl. Station           & 4 / 1.4           & 6  / 45     & 2.7   & 0.3 & 9   \\
Fly's Eye                      & 1 / 0.35          & 1050 / 8400 & 4.8   & 20  & 56   \\
Dutch Inc. Sum                 & 11  / 4           & 10 / 82     & 3.6   & 0.3 & 77   \\
Intl. Inc. Sum                 & 11  / 4           & 6  / 45     & 2.7   & 0.3 & 73   \\
Coherent Superterp (94 beams)  & 12  / 4           & 18 / 147    & 0.5   & 29  & 1382  \\
Coherent Sum Core (100 beams)  & 48 / 17           & 0.4 / 3     & 0.075 & 31  & 3206  \\
Constrained Coherent Core (29 beams) & 10 / 3.5    & 18 / 147    & 0.9   & 9   & 512  \\
\hline
\multicolumn{6}{c}{Low-Band Antennas (LBAs)} \\
\hline
Single Core Station Outer      & 1  / 0.35         & 17 / 132     & 4.6   & 0.3 & 1  \\
Single Core Station Inner      & $< 1$ / $< 0.35$  & 105 / 840    & 11.6  & 0.3 & $< 1$ \\
Single Rem. Station            & 1 / 0.35          & 17 / 132     & 4.6   & 0.3 & 1  \\
Single Intl. Station           & 2 / 0.7           & 26 / 211     & 5.8   & 0.3 & 5  \\ 
Fly's Eye                      & 1 / 0.35          & 660  / 5300  & 4.6   & 12  & 40 \\
Dutch Inc. Sum                 & 6 / 2             & 17 / 132     & 4.6   & 0.3 & 40 \\
Intl. Inc. Sum                 & 6 / 2             & 26 / 211     & 5.8   & 0.3 & 44 \\
Coherent Superterp (15 beams)  & 6 / 2             & 17 / 132     & 1.2   & 4.5 & 138 \\
Coherent Sum Core (100 beams)  & 24 / 8.5          & 3 / 23       & 0.19  & 30  & 2460 \\
\hline \hline
\end{tabular}
\end{footnotesize}
\end{center}
\label{tab:modes}
\tablefoot{LOFAR beam-formed modes and their (approximate)
  associated sensitivity, FoV, resolution (i.e. $\Delta \Omega$),
  data-rate, and survey FoM (see text).  High-band (HBA) and low-band
  (LBA) sensitivities and FoMs have been normalized to that of a
  single 24-tile HBA sub-station or a 48-dipole Dutch LBA field
  respectively (Recall that each Dutch LBA field contains 96
    dipoles, only 48 of which are used in any particular observation.
    Unless otherwise stated, we assume the LBA Outer mode is being
    used.  This mode gives somewhat higher gain, but reduced FoV
    compared with the LBA Inner mode.).  Quantities are quoted
  assuming one beam per station (48\,MHz bandwidth) and 8 beams per
  station (6\,MHz bandwidth per beam) respectively.  FoV ($\propto
  \lambda^2_{\rm obs}$) and resolution (i.e. FWHM of the beam,
  $\propto \lambda_{\rm obs}$) are quoted for a central observing
  frequency of 150\,MHz (HBA, $\lambda_{\rm obs} = 2$\,m) and 60\,MHz
  (LBA, $\lambda_{\rm obs} = 5$\,m).  Note that FWHM is taken to be
  $\alpha \times \lambda_{\rm obs} / D$, where $\alpha = 1.3$ and $D$
  is the size of a station or the maximum baseline between combined
  stations where applicable.  As LOFAR stations consist of several
  square tiles, they are not perfectly circular; thus, the product of
  FoV and sensitivity is not constant when station size increases.  We
  have used LBA (Inner) / LBA (Outer) / HBA station sizes of
  32.3\,m / 81.3\,m / 30.8\,m (core), 32.3\,m / 81.3\,m / 41.1\,m
  (remote), and 65\,m / 56\,m (international, Inner/Outer mode does
  not apply here).  Further empirical beam modeling will likely refine
  the value of $\alpha$, and will somewhat effect the rough values
  quoted here.  Where applicable, we assume that 24 core stations of
  $2 \times 24$ HBA tiles / 48 LBA dipoles, 16 Dutch remote stations
  of 48 HBA tiles / 48 active LBA dipoles, and 8 international
  stations of 96 HBA tiles / 96 LBA dipoles are available and can be
  recorded separately if desired.  Fly's Eye mode assumes all Dutch
  stations - i.e. 48 HBA core sub-stations plus 16 remote HBA stations
  or 40 LBA fields of 48-dipoles each are used.  For the ``Coherent''
  modes, we assume the maximum number of tied-array beams required to
  cover the station beam, up to a maximum of 100 (per station beam),
  can be synthesized, and that the maximum baseline between stations
  is 300\,m for the Superterp and 2000\,m for the entire Core.  The
  ``Coherent Sum Core" mode assumes that all 48 Core sub-stations are
  combined coherently.  The ``Dutch Incoherent Sum" mode assumes that
  all 40 Dutch stations (24 core / 16 remote) are combined
  incoherently.  The ``Intl. Incoherent Sum" mode assumes that all 8
  international stations are combined incoherently.  The ``Constrained
  Coherent Core'' mode is a hybrid coherent/incoherent summation in
  which the two HBA sub-stations of each core station are first summed
  coherently at station level before these stations are in turn summed
  incoherently.  The integration time used in each mode is assumed to
  be the same, though this would likely differ in practice, especially
  in the case of wide-field surveys.  The data rates assume 16-bit
  samples (this could be reduced if desired), summed to form Stokes I,
  at the maximum possible spectral/time resolution, which for certain
  applications can be downgraded by a factor of a few in order to save
  on disk space and processing load.}
\end{table*}

\subsection{Coherent and incoherent station addition}
\label{obsmodes:coh_incoh}

\subsubsection{Coherent addition}

%Types of phase delays
To achieve the full sensitivity of the LOFAR array it is necessary to
combine the signals from each of the station beams coherently, meaning
that the phase relationship between the station signals from a
particular direction must be precisely determined.  Phase delays
between the signals are generated by a combination of geometric,
instrumental, and environmental effects. The geometric term is simply
related to the relative locations of the stations and the source and
is easily calculated. Instrumental contributions are related to the
observing system itself, such as the length of cables connecting the
various elements.  Consequently, the components in the signal chain
are designed to be stable on timescales that are long compared to the
observing duration.  The environmental terms can be much more variable
and are dominated by ionospheric delays at low radio frequencies.

%FoV of tied-array beams
Combining the station signals coherently into tied-array beams gives
the equivalent sensitivity to the sum of the collecting area of all
the stations being combined.  However, the resulting FoV is
significantly smaller than that of an individual station beam because
it is determined by the distance between stations, which for LOFAR is
significantly larger than the size of the stations themselves.  The
FoVs for various beam types are given in Table \ref{tab:modes} and
examples are shown in Figure \ref{fig:lofar}. The small FoV of
tied-array beams is not a problem for observing known point sources
$-$ in fact, it can be advantageous by reducing the contribution of
background sources $-$ but increases the data rate and processing
requirements when undertaking surveys (see below).

The coherent combination of telescopes for pulsar observations is
regularly used at the Westerbork Synthesis Radio Telescope
\citep{ksv08} and the Giant Meter-Wave Radio Telescope
\citep{ggj+00}. In both cases the phase delays between the telescopes
are determined by regular observations of known and unresolved
calibration sources in a process which is often called
``phasing-up''. The timescale between when observations of calibration
sources need to be made depends strongly on the observing frequency,
the largest separation between telescopes and the ionospheric
conditions, but range from a few hours to days.

At the low frequencies of LOFAR, phase delays between stations are
very susceptible to the influence of ionospheric disturbances
(e.g. \citealt{nn07}).  This directly affects how often phasing-up
needs to be done, as well as the maximum separation between stations
that can be coherently combined.  Fortunately, LOFAR's flexible signal
processing provides a number of options to rapidly phase-up (at least)
those stations located in the core (the innermost 2~km of the array). The first option makes use of
LOFAR's ability to produce images in near real time.  As part of the
imaging calibration, the phase delays between stations are determined.
As it is possible to operate imaging and beamformed modes
simultaneously (see also Sect. \ref{sec:simult}) the phases from the
imaging pipeline can be used to coherently combine the stations, and
to update the calibration continuously.  If for some reason this isn't
sufficient, a second option is available in which it is possible to
trade a small amount of observing bandwidth to form a station beam
which points at a strong calibrator somewhere in the wide FoV of the
element beam.  As long as this probes a similar ionospheric patch,
this can be used to monitor the phases.

An important aspect of keeping the signals from the stations in phase
is the time stamp given to each sample. In the original design, LOFAR
used separate rubidium clocks at each of the stations.  These clocks
in turn are governed by their own Global Positioning System receiver,
which steers the rubidium clock drifts to maintain long-term timing
stability. However it was recognised that while the instantaneous
accuracy was sufficient, there were medium term drifts which needed to
be corrected for. To alleviate the need for such corrections on the
LOFAR Superterp, a new single clock system has been implemented: the
clock signal from one GPS-governed rubidium clock is distributed to
these 6 stations.  The rest of the core stations will be kept in phase
for the coherent addition using either of the two methods described
above.

Once the phase delays have been determined between the stations, it is
possible to use different geometrical delays to form additional
tied-array beams. These additional tied-array beams can be used to
tesselate sections of a single station beam, in order to greatly
increase the total FoV for survey-like observations. They can also be
formed inside different station beams to point at many different known
point sources simultaneously, thereby greatly improving observing
efficiency (see Figure \ref{fig:lofar} for examples of beams).  The
processing capability of CEP will allow us to form at least 200
tied-array beams simultaneously, depending on the observing bandwidth
and time resolution used.  This allows LOFAR to achieve rapid
survey speed despite the decrease in FoV caused by the wide
distribution of the stations.  Though the resulting data rate is very
large, an advantage of surveying with tens to hundreds of tied-array
beams is that the position of newly found sources is immediately known
to an accuracy of $\sim 5^{\prime}$ (when all core stations are used).
This can save significant follow-up observation time that would
otherwise be spent refining the source position.

\subsubsection{Incoherent addition}

It is also possible to combine the stations without correcting for the
phase differences between them. If the signals are added after
detecting them (i.e. computing the power and thereby losing the phase
information) this results in a so-called incoherent sum. As there is
no longer any phase relation between the signals from the stations,
the signals do not constructively nor destructively interfere. The
combined signal is therefore sensitive to signals anywhere within the
FoV of the station beam, however the sensitivity increases only as the
square-root of the number of stations combined.

As it is not necessary to keep track of the phase relationship between
the stations, the incoherent addition can incorporate stations which
are more widely spread than those that can be coherently combined.
Determining which mode to use, for any given observing goal, depends
on the amount of collecting area which can be combined coherently, the
number of tied-array beams that can be formed and the science goals.

As previously discussed in \cite{hsl09} the observational
requirements, e.g. sensitivity and FoV, can be distilled into a simple
figure of merit (FoM, see Table \ref{tab:modes}), which is generally
applicable to a transient survey\footnote{We choose to give extra
  weighting to $A_{\rm eff}$.  One may also consider a FoM that scales
  linearly with $A_{\rm eff}$.} \citep[see also][and references therein
  for a deeper discussion of survey metrics]{clm04,cor08}:

\begin{equation}
FoM \propto A^2_{\rm eff} \frac{\Omega}{\Delta \Omega} \frac{T}{\Delta T}.
\end{equation}

To be sensitive to transients over a wide range of source parameter
space, including faint and rare events, this FoM should be maximized.
This means maximizing: $A_{\rm eff}$, the effective collecting area being 
used; $\Omega$, the instantaneous field of view (FoV); and $T$, the 
total time spent observing the sky. At the same time, adequate spatial ($\Delta
\Omega$) and time ($\Delta T$) resolution are needed to provide reasonable source
localization, for multi-wavelength follow-up and
identification, and to resolve short timescale phenomena.
As it is not possible to maximize both raw, instantaneous sensitivity and FoV 
simultaneously, different modes naturally probe different areas of transient 
parameter space (see below).

\subsubsection{Coherent versus incoherent addition for different science cases}

Contingent on the particular science goals, pulsar and fast transient
surveys with LOFAR are likely to be performed using both the coherent
and incoherent addition modes, sometimes in parallel.  As discussed by
van Leeuwen \& Stappers (2010)\nocite{ls10}, a high sensitivity survey
of the entire Northern Sky is preferentially done using the coherent
addition of stations if one can form at least $\sim$200 tied-array
beams\footnote{Current system performance tests indicate that this
  should be achievable.}.  Though such a mode provides the
best-possible source localization and sensitivity, this comes at the
price of a much larger data rate (Table \ref{tab:modes}) and
consequently much greater processing requirements. In comparison, the
wide FoV afforded by the incoherent sum gives it a very competitive
survey speed.  However, there are several drawbacks compared with the
coherent mode: source localization is comparatively poor, as is the
rejection of astronomical backgrounds and radio frequency interference
(RFI).  Also, in the incoherent modes much longer observing times are
needed to achieve the same sensitivity and/or FoM (Table
\ref{tab:modes}), which increases the processing load significantly if
acceleration searches for binary systems are to be performed
(e.g. \citealt{jk91,rce03}).  On the other hand, longer dwell times
may be advantageous for detecting certain types of transients.
Nonetheless, the incoherent addition mode provides a way to rapidly
survey the whole sky\footnote{Using 7 station beams per pointing, only
  about 200 HBA pointings are required to cover the entire
  LOFAR-visible sky ($\delta > -35^{\circ}$).}  with a competitive
sensitivity, thereby making it very useful for detecting classes of
pulsars which can emit relatively brightly but more erratically,
e.g. RRATs, intermittent pulsars, radio emitting magnetars, and
eclipsing sources, as well as other transient sources like flare
stars, planets and potentially new classes of transient sources.

Observations of known point sources at high time resolution, where the
maximum sensitivity over a given bandwidth is required, are best
achieved by using the coherent sum mode.  As discussed earlier, even
in this mode multiple sources can be observed simultaneously if
multiple station beams are used at the expense of total bandwidth.
Each of the tied-array beams can be considered as an independent
entity in the LOFAR processing chain and so, e.g., pulsar timing
campaigns can be run efficiently by observing multiple sources at
once.  It is also possible to have beams pointing at a range of source
types, for example a pulsar, a planet, a flare star, and an X-ray binary all
at the same time, with different data products being produced for each
of the different sources.

\subsection{Observing in multiple modes simultaneously}
\label{obsmodes:simul}

\subsubsection{Coherent and incoherent sum modes}

LOFAR is flexible enough to produce coherently and incoherently added
data simultaneously.  This provides, at the incoherent summation
sensitivity, the full FoV of a single station while still having the
full coherent sensitivity in the centre of the FoV.  Though even 200
tied-array beams, generated using all the stations on the core, only
cover $\sim3$\% of the single station FoV, this parallel mode allows
one to both survey the sky at very high sensitivity (over a small FoV)
and over a large FoV (at lower sensitivity) simultaneously (Table
\ref{tab:modes}). We note that a greater fraction of the single
station FoV can be covered if we coherently sum the Superterp stations
only, and this may also prove a useful survey mode.  Alternatively, a
mix of coherent and incoherent addition can be used when adding the
stations.  For example, one might choose to coherently sum the
stations in the compact core of the array, e.g. on the Superterp
(Figure \ref{fig:lofar}), and then incoherently add the stations
outside of the core to this.  This results in a compromise between FoV
and sensitivity because the core stations have relatively short
baselines between them (Figure~\ref{fig:lofar}).  It also results in a
more complex combined beam shape.  In contrast, groups of stations can
also be used separately as part of sub-arrays, as discussed below.

\subsubsection{Imaging and beamformed modes}

As mentioned above, beamformed data can be taken in parallel with
imaging observations. This opens up the opportunity for commensal
observing where searches for fast transients can take place in
parallel with imaging-based searches for slow transients or other
observations.  This will be done as part of the Transient Key Science
Project's Radio Sky Monitor, which will image the sky down to 1 second
timescales; we will continuously record the incoherent sum of the
stations to probe variability on even faster timescales.  In 
Sect. \ref{sec:simult} (Figure~\ref{fig:img_psr}) we present the first
such simultaneous imaging/beamformed observation made with LOFAR taken
in 2010 April, when only 7 LOFAR stations were available for imaging.

\subsection{Observing with sub-arrays and single stations}
\label{sec:subsing}

In addition to the incoherent and coherent modes of observing
described above, the array can be split up in various flexible ways to
accommodate particular science goals.  There are certain pulsar and
fast transient science goals with observational requirements that are
best served by grouping the stations into sub-arrays, or using
individual stations independently. In this way, sensitivity can be
exchanged for a larger total FoV and/or broader total frequency
coverage.

For large sky coverage, LOFAR sub-arrays consisting of one or several
stations (depending on the required sensitivity) can be used to point
at different directions, covering a large fraction of the sky $-$
potentially the {\it entire} LOFAR-visible sky in the case of the LBAs
(Table \ref{tab:modes}).  To maximize FoV, each sub-array can simply
be a single station pointing in a unique direction.  This mode,
referred to as Fly's Eye, is very similar to that developed for the
Allen Telescope Array (ATA; \citealt{wbb+09}) and will be used to
monitor for rare, very bright, fast transient events.  In this mode
the localization of detected bursts can only be achieved to within the
single station FoV, which ranges from a few square degrees to many
hundred square degrees across the LOFAR band.  Much better position
determination could be achieved by triggering as the source is
detected and then using the TBBs to subsequently image the sky to try
to find the source (see Sect. \ref{sec:TBBs} for more details).

Another option involves using sub-arrays of stations tuned to
complementary frequencies across the entire LOFAR band, pointing at
the same target.  Covering the full band requires two 48-MHz
sub-arrays for the LBAs and three for the HBAs. Such broad and dense
spectral coverage is crucial for our understanding of the intrinsic
processes in pulsar magnetospheres and propagation effects through the
interstellar medium at low radio frequencies (see
Sect. \ref{sec:science}).

Individual stations can also be used independently to conduct
targeted observations of known bright objects, or wide field
monitoring of transient events. In this respect, the international
LOFAR stations have the highest sensitivity and may be available
during normal LOFAR observations when the longest baselines are not in
demand. In addition to the conventional mode where single stations are
controlled through the central LOFAR facilities and send data back to
CEP, the data will be recorded and analysed locally, so that each
station can operate as a stand-alone instrument, or a group of stations
could operate as a coordinated sub-array.

\subsection{Polarisation}
\label{sec:poln}

After careful calibration, LOFAR will enable polarisation studies of
pulsar emission physics and the interstellar medium.  A multi-stage
process is required to achieve sufficient polarisation purity. We only
summarise the most relevant issues here and refer the reader to the
Magnetism Key Science project \citep{bbb+07} for more details.  Both
the LBAs and the HBAs have dual linearly polarised feeds that are
stationary on the ground, meaning that for most observations the beams
will be formed off-axis, potentially increasing the contribution from
polarisation leakage. The dipoles are aligned South-East to North-West
and South-West to North-East for the x- and y-dipoles respectively and
this means that as a source is tracked across the sky the projected
length of the dipole will change for the majority of sources.  Thus,
it will be important to accurately calibrate the gains of the
individual dipoles. For all antennas and stations, Mueller and Jones
matrices \citep{tin96,hbs96} need to be determined to correct for
voltage and Stokes-parameter coupling.  It is important to note that
the beam patterns are time-variable and different for each station,
however as discussed below it will be possible to calibrate the system
in all but the worst ionospheric conditions.

To calibrate the individual LBA dipoles and the HBA tiles in a given
station, delay lookup tables that provide static polarisation
calibration will be used initially (see also Figure
\ref{fig:lba_calibration}). However, a multi-level calibration
procedure based on the individual station elements and the stations
themselves is required to ensure polarimetric fidelity as outlined in
\citet{wv09a,wv09b}. Combining the stations to form tied-array beams
when including stations beyond the superterp, will require using
dynamic delays provided by the real-time calibration described above.

The LOFAR real-time calibration will correct for the ionospheric phase
delays, but removing the ionospheric Faraday rotation is more
challenging. The limited size of individual stations means that the
ionosphere in most cases will not cause significant distortions across
a station, ionospheric Faraday rotation introduces a differential, and
frequency dependent, polarisation component in addition to the
differential delays. If the properties of the ionosphere above each
station are determined, the ionospheric Faraday rotation can be
removed by applying a frequency dependent phase term to the complex
polarisations at the level of each station. The polarisation data of
the array can then finally be used to form 4 independent Stokes
parameters, for both the incoherent sum mode as well as for the
phased-array polarimetric observations, and subsequently processed as
part of the pulsar pipeline using the methods described in van Straten (2004)
and van Straten et al. (2010)\nocite{van04d,vmj+10}.

\subsection{Transient buffer boards}
\label{sec:TBBs}

The TBBs (see \citealt{sbf+08}) are an exciting and unique aspect of
LOFAR. At each station there is sufficient memory, RAM, to store the
full-bandwidth raw voltage data from all of the active elements in
that station for up to 1.3~s\footnote{This number may increase in the
  near future with the addition of more memory.}. These data can then
be used to form an image anywhere in the visible sky, in the case of
the LBAs, or anywhere in the element beam of the HBAs if a trigger,
either external or internal, arrives within 1.3~s of an event
occurring, and causes the data in the buffer to be frozen. This mode
will be used predominantly by the Cosmic Ray Key Science Project
\citep{fal+07} but will also be an invaluable tool for the Transients
Key Science Project. It will be used to localise sources, but to also
search for new transient events at very high time resolution.

The TBBs can also be operated in a mode where data are saved that has
passed through the polyphase filter at the station. This allows one to
choose a reduced bandwidth, $B_r$,  and so significantly increase the amount of
time which can be stored in the buffer: $1.3~{\rm s} \times 100~{\rm
  MHz}/B_r$. This mode will be very useful when expecting triggers
related to short duration bursts at shorter wavelengths as the unknown
dispersion delay in the interstellar medium may be many seconds.

\begin{figure}
\begin{center}
\includegraphics[scale=0.5]{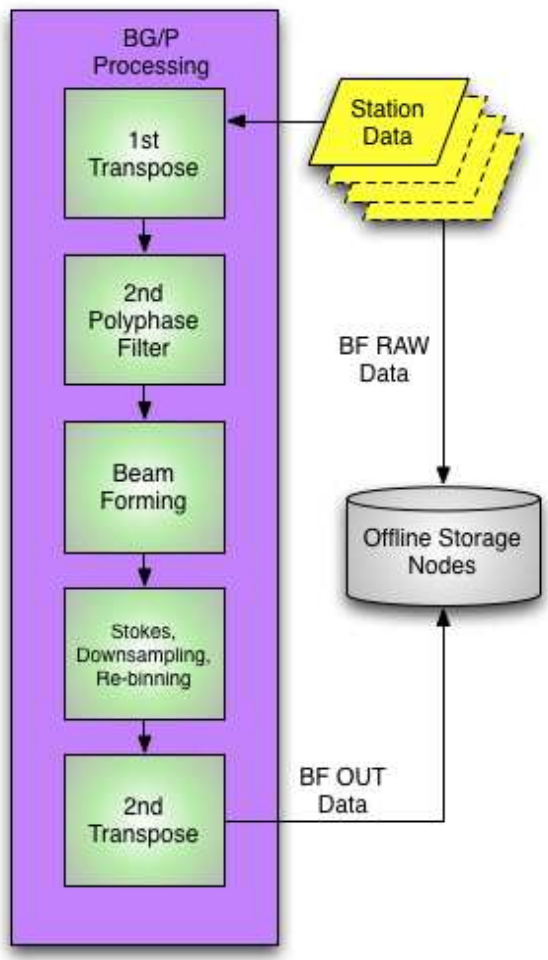}
\caption{Schematic diagram of the ``online" Tied-Array Beam (TAB)
  Pipeline, as it runs on the BG/P.  Streaming data from LOFAR
  stations are either written directly to disk per station (``BF RAW
  Data") or are read by the BG/P for pipeline processing in memory,
  combining station data into one or multiple beams, and then writing
  them to disk (``BF OUT Data").}
\label{fig:online_pipeline} 
\end{center}
\end{figure}

\section{LOFAR pulsar pipeline}
\label{sec:pipelines}

In this section we describe the standard and automated ``pipeline"
processing that is applied to the LOFAR beamformed data produced by
the various modes described in Sect. \ref{sec:obsmodes}.  We collectively refer
to this processing as the ``Pulsar Pipeline" though these data serve a
broad range of scientific interests, not only pulsar science.

The processing is split into two major, consecutive segments:
``online" processing, which is done by a Blue Gene P (BG/P)
supercomputer to streaming data from the LOFAR stations, and
``offline" processing which is further processing of these data on the
LOFAR offline cluster (or potentially elsewhere). This hardware is
located in Groningen in the Netherlands and the combination of the
BG/P and offline cluster is known as CEP. The online processing is
chiefly responsible for combining data from multiple stations into one
or multiple beams\footnote{As well as optionally producing
  correlations for imaging as described in Sect. \ref{sec:obsmodes}.},
while the offline processing is more science-specific and includes,
for example, dedispersion and folding of the data at a known pulsar
period.

Given that the processing of the raw station data is done entirely in
software, the pipeline is very flexible and extendable.  Here we
describe the current implementation and near future plans with the
expectation that this will continue to evolve to meet new scientific
requirements and to take advantage of increased computing resources in
the future.

\subsection{Online processing}
\label{onlineproc}

Here we briefly describe the online processing chain as it applies to
pulsar-like data.  For reference, Figure~\ref{fig:online_pipeline}
shows a simple block diagram of the processing chain.  For a more
detailed and general description of the LOFAR correlator see
\cite{rbmn10}.

\subsubsection{Station data}

Each one of the LOFAR stations is capable of sending up to 244
subbands, that are either 156 or 195~kHz wide, to CEP in
Groningen\footnote{We do not consider here the details of the
  processing that happens at the stations themselves but refer the
  interested reader to \cite{dgn09}.}. The subbands generated at the
stations are sent as complex numbers representing the amplitude and
phase, allowing division into smaller frequency channels if required,
and the coherent or incoherent combination of stations. As discussed
in Sect. \ref{sec:LOFAR} the subbands can be distributed over a
non-contiguous range of frequencies and divided into station beams
which can use different, or the same, set of subbands (see
van Haarlem et al. (in prep.) for further details of station capabilities and data
products).  The station signals are fed to the BG/P, which can combine
them into coherent and/or incoherent array beams, correlate them for
imaging or in fact, produce all three data products at once.

The data flow is arranged so that each input/output (I/O) node of the
BG/P handles all the data from one station, which is sent in 1~second
sections we refer to as chunks.  These 1~second chunks are then
passed, in a round-robin fashion, to one of the 16 compute nodes
attached to each of the I/O nodes (see Overview of the IBM Blue Gene/P
Project (2008)\nocite{IBM08} for a description of the BG/P
architecture).  The compute nodes must finish the full online pipeline
processing, and send the data back to the I/O node to be written out,
before the next 1~second chunk of data arrives. There are sufficient
computing resources that while keeping up with the real time
processing there is approximately 16 seconds of available computing
time for each 1~second chunk of raw data\footnote{This is because
  there are 244 (i.e. approximately 256) subbands divided over 4096
  available compute cores.}.

\subsubsection{First transpose}  

The data arrive at the BG/P such that they are arranged by station,
however in order to efficiently form correlation products and beams it
is necessary to have the data from all the stations for a particular
subband, in the 1~second chunks, present.  The process of
reordering these data is referred to as the ``First Transpose"
(Figure~\ref{fig:online_pipeline}).  Before the transpose, the memory
of a particular compute node holds multiple frequency subbands from a
single station; after the transpose, the same compute node holds
subband(s) of a particular frequency from all the stations being used.

\subsubsection{Second polyphase filter}  
\label{sppf}

Depending on the particular application, each subband can be further
split into an optional number of channels ($16-256$) via a polyphase
filter (PPF).  This is referred to as the ``2nd polyphase filter"
(2PPF, see Figure~\ref{fig:online_pipeline}), because of the preceding
PPF step which occurs at the stations themselves in order to create
subbands (see Sect. \ref{sec:LOFAR}). This further channelization serves
two purposes: it allows for the removal of narrow band interference
signals in a way which minimises the loss of bandwidth; it is also
required for incoherently dedispersing the pulsed signal.  The removal
of radio frequency interference (RFI) will be discussed in more detail
below, but typically these signals are significantly narrower than the
156/195-kHz subbands sent from the stations, so further division into
finer frequency channels is required (see Sect. 8 for details).

As discussed in Sect. \ref{sec:challenges}, any broadband pulsed signal
will be dispersed, and hence smeared in time, by the ISM between the
source and the Earth.  This can be corrected by either incoherent or
coherent dedispersion.  Coherent dedispersion will be used for
studying known pulsars, as it provides the best possible dispersive
correction along with the highest achievable time resolution for a
given bandwidth.  Since this technique is computationally very
expensive, it is unfortunately not yet feasible to use coherent
dedispersion for wide-area pulsar surveys, where it would need to be
applied over large numbers of tied-array beams and trial DMs
(thousands of trial DMs in the case of blind searches for millisecond
pulsars).  The survey processing chain will therefore operate on data
which has been through the 2PPF and use computionally efficient
incoherent dedispersion algorithms. Recently we have been able to
develop a system which can perform coherent dedispersion for up to 40
different dispersion measures at one time \citep{mr11}. This will
have application as either a hybrid option where it is combined with
incoherent dedispersion for large area surveys, or using coherent
dedispersion for searches of objects where the DM is constrained to a
limited range of values, such as in globular clusters.

\subsubsection{Beamforming}  
\label{sec:bf}

It is at this stage that the (optional) beamforming between multiple
stations is done (Figure~\ref{fig:online_pipeline}).  The stations are
combined in one or potentially several of the ways described in
Sect. \ref{sec:obsmodes}.  This is performed on a per subband or even per
channel basis depending on whether the data has been through the
2PPF. For incoherent beams, the combination is not done on the complex
samples, but instead on the Stokes parameters (see also below). The
geometrical delays between stations are computed once per second and
interpolated both in frequency and time, so that each sample is
corrected by a per-sample unique factor. The small number of possible
incoherent beams, only one per station beam, means that this is not a
computationally heavy task compared with the other calculations of the
online processing; nor is the data rate unmanageable (see Table~3).
Hence, incoherent array beams can, and often will, be produced in
parallel to the correlation products needed for imaging or any other
data-products, such as multiple tied-array beams.

In the case of tied-array beams, computational efficiency is a far
greater concern, as up-to several hundred beams must be synthesized in
parallel in order to give this mode sufficient ``survey speed" for
all-sky surveys.  Beam forming is performed by applying the
appropriate time and phase correction to the complex samples
(Sect. \ref{sec:LOFAR}).  These delays are applied in three steps: first
shifting by integer amounts of samples, then a phase correction for
the center beam (these are both shared with the imaging pipeline), and
finally a delta-phase correction between the center beam and each
tied-array beam is applied. Determining the phase corrections due to
contributions from the ionosphere and the instrument, which need to be
calculated from the data, are discussed in Sect. \ref{sec:obsmodes}.  An
efficient algorithm, partially written in assembly code, is used to
maximize the number of beams that can be calculated.  The large number
of beams also make the data rate much larger, and the 15~Gb/s designed
input limit of the current offline storage cluster plays the main
limiting role on the maximum number of beams that can be written
out. The final LOFAR offline cluster will have an increase in
throughput of about 500\%.

To form both incoherent and tied-array beams it is necessary to define
a reference phase location. It is strongly preferable for observations
where precise time tagging of events is required, such as pulsar
timing, that this location be fixed.  For LOFAR the phase center of
all observations is now by convention the geographical center of the
LBA field of station CS002, regardless of whether this particular
station is being used or not. This places the official position of the
LOFAR telescope at (x,y,z) coordinates of (3826577.462 m, 461022.624
m, 5064892.526 m) in the ETRS89 system \citep{ba01}. This reference
position will be used for barycentering, pulsar timing, and phasing-up
the array.

\subsubsection{Fly's eye mode}

For Fly's Eye mode, in which the station signals are {\it not}
combined into incoherent or tied-array beams at BG/P, the processing
proceeds largely as described above, except that individual station
beams are written out separately using the same code that stores
multiple tied-array beams.  In parallel, it is also possible to
generate the incoherent sum of the station beams if desired.  Besides
the scientific applications made possible by the huge FoV achievable
with this mode, it is particularly handy for debugging issues related
to individual station data because it allows for the comparison of
ideally identical stations. Moreover it is useful for discovering sources
of local RFI. 

\subsubsection{Stokes parameters, downsampling, and re-binning}  

Prior to beam forming, the individual time samples are in the form of
two 32-bit complex numbers, representing the amplitude and phase of
the $X$ and $Y$ polarisations of the antennas. From these, it is possible
to calculate the 4 Stokes parameters, I, Q, U, and V online, assuming
incoherent dedispersion is to be used, or to record the complex
samples for offline coherent dedispersion. The 4 Stokes values fully
encode the polarisation properties of the signal and can be used for
polarimetric studies. Calibration of the polarisation products is
discussed in Sect. \ref{sec:poln}.

To reduce the data rate one can optionally output only Stokes I, the
total intensity. This is sufficient for many observations, and is
preferable in the case of large-scale surveys where the data rate is
already a major issue.  Depending on the required sampling rate, the
data may be downsampled in time and also scaled to re-pack the 32-bit
Stokes parameters into 16-bit samples (this currently happens offline
but will later be implemented on the streaming data). Application of
RFI mitigation strategies to these streaming data may allow us to
further reduce the data to 8-bit samples or less, which will reduce
the data rate and hence potentially allow the formation of more
tied-array beams.

\subsubsection{Second transpose}

At this stage, the necessary online processing is complete but the
data products are spread across the compute nodes of BG/P in an order
that is contrary to the way they are typically, and most efficiently,
analyzed in further time-domain based scientific post-processing.  Each of the compute
nodes contains all of the synthesized beams, but only for a limited
number of subbands.  For dedispersion and other applications the data
are best organized such that all the frequency channels of a
particular beam are in one place.  Thus a ``Second Transpose"
(Figure~\ref{fig:online_pipeline}) is carried out to reorganize the
data across the various I/O nodes of BG/P before it is written to
files on the offline cluster.  This Second Transpose is critical for
handling large numbers of beams in the offline processing.  An added
advantage of the Second Transpose is that it will also make it
possible to recombine the subbands using an inverse PPF in order to
obtain a time resolution close to the original $\sim 10$~ns resolution
available before the subbanding of the data at station level.  This
mode may be of interest for detailed studies of giant pulses and for
detecting the short radio flashes created by cosmic rays entering the
atmosphere \citep{fal+07}.

\begin{figure}
\begin{center}
\includegraphics[scale=0.45]{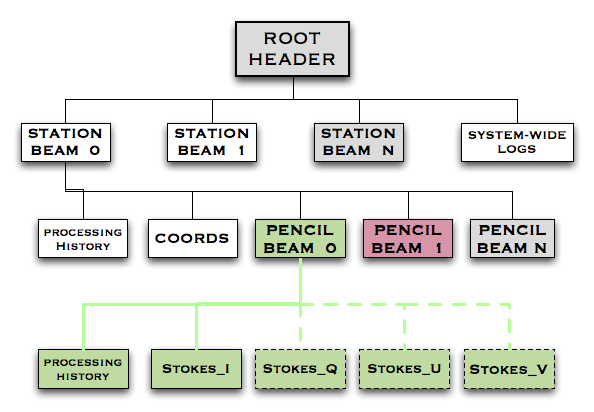}
\caption{An overview of the beamformed file structure, encapsulated
  in a hierarchical tree within an HDF5 file.  The data are split into
  separated sub-array pointing (known in this paper as station beams);
  within each sub-array pointing are pencil or tied-array Beams (we
  will use tied-array beams), along with coordinates and processing
  information; each Beam contains a data structure with either 4
  Stokes parameters (or just Stokes I for total intensity) {\it or},
  if desired, the amplitude and phase of the X and Y polarisations
  separately.  Header keywords are stored within each tree level,
  depending on their hierarchical relevance to the rest of the file,
  as the information is inherited as one steps downward through the
  tree structure. See Alexov et al. (2010a,b) for more details.}
\label{fig:hdf5} 
\end{center}
\end{figure}

\subsection{Radio frequency interference excision}

The low-frequency window, $10-240$~MHz, observed by LOFAR
contains a range of interfering sources of radio emission: from
commercial radio stations, to weather satellites, to air traffic
control communication. To mitigate the influence of this RFI on our
data we have implemented some basic strategies, which we discuss
below. We will not discuss the online RFI mitigation strategies
employed by LOFAR as these will be discussed in detail elsewhere.

Our principal mechanism for excising RFI is to simply not observe
where it is present. The FM-radio band from about 90--110~MHz is
already filtered out within the LOFAR design (van Haarlem et al. in
prep.).  The station-based frequency channelisation of LOFAR to
subbands of either 156 or 195~kHz, combined with the ability to spread
these subbands anywhere within the available 80 or 100~MHz Nyquist
zones, already provides the means to avoid commonly affected
frequencies without sacrificing observing bandwidth.  As will be shown
below (Sect. \ref{rficom}) many of the interfering sources have
bandwidths that are significantly narrower than the station-based
subband width and so by performing the 2PPF step (Sect. \ref{sppf}) to
increase the frequency resolution we are further able to excise RFI
with a low impact on the total remaining bandwidth. Moreover, the
majority of the interfering sources have a relatively low duty cycle
and so only small sections of time may need to be excised in order to
remove the interference.

Our present strategy to identify RFI in our data is such that for
every frequency channel we calculate the quasi-instantaneous
root-mean-square deviation (rms) of data chunks of duration
$\sim10$~s. The rms calculation is performed after iterating to remove
any bright data points which are likely to be RFI and biasing the
determination of the rms. If a particular time sample is larger than
$6\sigma$ above the rms then it is considered to be RFI. If there are
more than 30\% of such samples in the time series, then the whole
frequency channel is marked as RFI-corrupted and will be excluded from
further analysis. This RFI excision can be run both in manual mode and
as part of our automated pipeline. In the automated mode we exclude
only the corrupted channels, but leave the strong $>6\sigma$
individual samples if their fraction is less than 30\%. We do this to
avoid removing any very bright single pulses. This is not problematic
since some short duration sparse RFI in the data will not dramatically
impact the data quality. This is because the dispersion smearing, even
for quite low DM values, is significant at such low observing
frequencies. Thus, during the dedispersion process, the RFI, which is
at a dispersion measure of zero as it has not propagated through the
ISM, will be dispersed and thus contribute only extra noise and
will not significantly distort the pulse profile (see Figure
\ref{fig:disprfi} for an example).

\subsubsection{Final raw data products}  

The raw data streaming out of BG/P, in some cases at the eventual rate of
several GB/s, are written to multiple storage nodes of a large
offline storage and processing
cluster\footnote{\url{http://www.lofar.org/wiki/doku.php?id=public:lofar_cluster#short_lofar_cluster_layout}}.
The data are currently written as almost header-less raw binary files,
but will soon be stored using an implementation of the Hierarchical
Data Format 5 (HDF5)\footnote{HDF home page:
  http://www.hdfgroup.org/}.  The HDF5 file format was chosen because
of its flexibility and its ability to store very large, complex
datasets spread over many separate physical devices.

An overview of the beamformed file structure is shown in
Figure~\ref{fig:hdf5}; it is too complex to describe here in detail
\citep{ale10a,ale10b}.  The hierarchical tree divides the data into
the separate station beams, tied-array beams and the 4 Stokes
parameters (shown) or the X \& Y polarisation amplitude and phase
values if complex data is recorded. For each beam and polarisation
parameter, the signal as a function of time sample and frequency
channel is stored in an array.  Metadata fully describing the
observation is stored both at the root level and along the tree where
applicable.  For instance, the parameters describing individual
stations or tied-array beams are stored at the same tree level as
these data themselves. Data processing and logging information are
stored in the output beamformed HDF5 files in processing history
groups; more general information is stored at the root of the file,
while more specific information which pertains only to individual
stations or beams is stored along the tree where applicable.  Each
beam has a coordinates group which describes the pointing information
for that observation; the coordinates are analogous to the World
Coordinate System (WCS) in the FITS format definition
\citep{ale10a,ale10b}.

Though it is expected that LOFAR pulsar data will be archived in this
format, we note that we have already written a flexible data converter
in order to also write the data into several other community standard
formats like the PRESTO/SIGPROC ``filterbank'' format and the PSRFITS
format.

\subsection{Offline processing}
\label{offlineproc}

\begin{figure*}
\begin{center}
\includegraphics[width=0.9\textwidth]{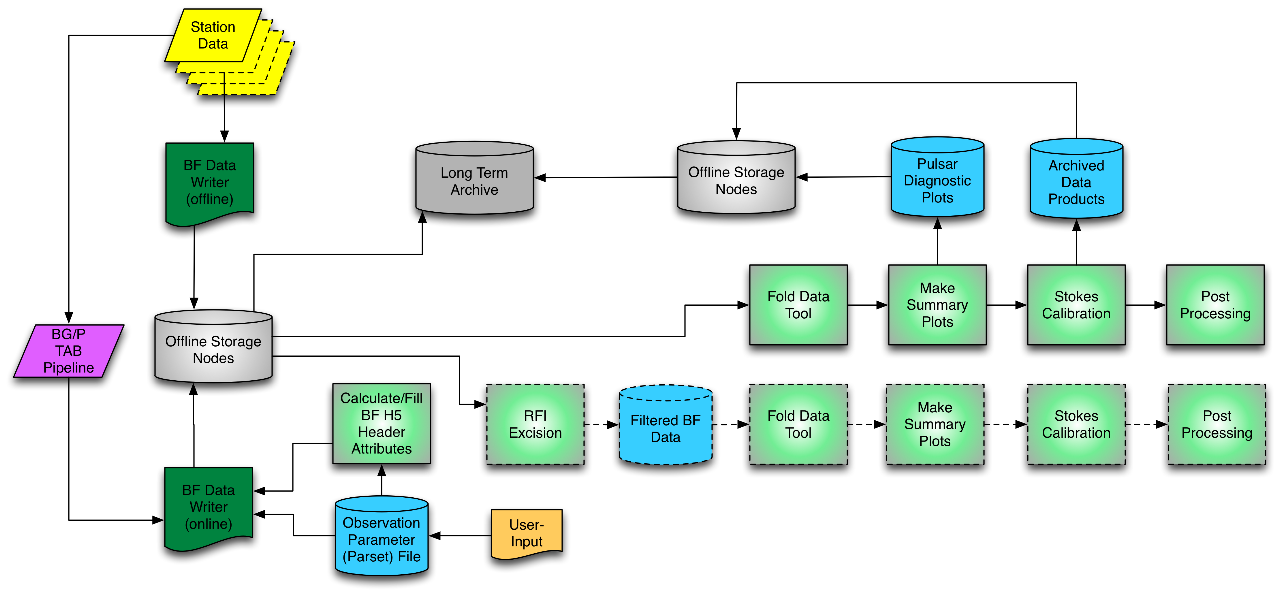}
\caption{Schematic diagram of the overall Known Pulsar Pipeline, as it
  runs ``online" on the BG/P followed by ``offline" scientific
  processing on the offline cluster.  Offline pipeline processing can
  be run on data directly out of the BG/P or on RFI-filtered data.
  Scientific tools such as de-dispersion and folding are performed,
  along with Stokes calibration to get polarisation and flux densities.  Data
  products and diagnostic plots are stored in the long term archive.}
\label{fig:known} 
\end{center}
\end{figure*}

In Figure \ref{fig:known} we show a block diagram of an example
pipeline, in this case the ``Known Pulsar Pipeline", describing the
essential steps in the process. Similar pipelines are being developed
for all the different processing streams as will be described below.

The raw data from the stations can either be written directly to the
offline storage nodes, which is useful for sub-arraying or single
station experiments, or they can be passed through the BG/P for further
channelization, beamforming, and other processing. The starting point
for all the pipelines discussed here is the beamformed data coming
out of BG/P, be that incoherent or coherent beams (though see
Sect. \ref{singlestatproc}). It is possible to take other data products,
such as imaging visibilities or data dumps from the transient buffer
boards, in parallel, but the processing of those will be discussed
elsewhere.

\subsubsection{LOFAR offline cluster}

The offline processing pipelines run on the LOFAR Offline Cluster
(LOC), which is the second main part of CEP. The data writing from the
BG/P is distributed across multiple storage nodes on the LOC, the
number of which can be chosen to match the data rate. The storage and
compute nodes are grouped into sub-clusters, each of which contains 3
storage and 9 compute nodes.  The 9 compute nodes of each sub-cluster
can access the data on the 3 storage nodes via NFS.  The current LOC
comprises 8 such sub-clusters; this will soon be expanded as LOFAR
approaches full capability.

After the data are written to the LOC it can either be passed directly
to the long term archive if that is appropriate, though that will not
normally be done due to the large volumes of the raw datasets.  More
usually the data will pass to one of the pipelines where data
processing and compression happen.

\subsubsection{Pipeline framework}

All the LOFAR processing pipelines are built to run within a generic
pipeline
framework\footnote{http://usg.lofar.org/documentation/pipeline/}. This
framework takes care of distributing the processing in parallel over
the LOC compute nodes, and provides appropriate logging and error
checking.  In the production system, the Pulsar Pipeline will start
processing automatically either during or after an observation,
depending on the data rate and availability of processing resources.
The allocation of computing and storage resources, and observing time,
will be handled by software referred to as the LOFAR Scheduler so that
commensal data-taking and processing do not collide. The Scheduler
will also be responsible for passing observation-specific metadata to the
HDF5 header. Further header information will flow from the observing
control systems SAS (Specification, Administration and Scheduling) and
MAC (Monitoring and Control) into the HDF5 file as well via the {\it
  parset} (parameter set) as shown in Figure \ref{fig:known}.

\subsubsection{Software and data access}

Wherever possible, all of the different Pulsar Pipelines are being
built around the well-tested, commonly used, open source software
packages available in the pulsar community
(e.g. PRESTO\footnote{http://www.cv.nrao.edu/$\sim$sransom/presto/},
PSRCHIVE\footnote{http://sourceforge.net/projects/psrchive/},
SIGPROC\footnote{http://sigproc.sourceforge.net/},
TEMPO\footnote{http://www.atnf.csiro.au/research/pulsar/tempo/}.  To
allow for efficient development and software management these have
been incorporated into the general LOFAR User Software
(LUS)\footnote{http://usg.lofar.org}, which uses
cmake\footnote{http://www.cmake.org/}, a cross-platform, open-source
build system, to automate the installation of these packages on a
variety of platforms (e.g. Ubuntu Linux and Mac OS 10.5.X).  This
cmake installer may be of interest to others using these same
reduction packages on other systems.

The LOFAR software interface to the HDF5 library is called the Data
Access Library (DAL).  It is also available within the LUS software
repository.  The DAL allows for high-level connectivity to HDF5 data,
and more specifically to LOFAR data structures encapsulated within an
HDF5 file.  The DAL is written in C++; it contains classes which
pertain to all the LOFAR-specific data structures, making it straight
forward to read and write LOFAR HDF5 data.  The DAL is also bound as a
Python module, called pydal, giving users access to the DAL classes
via Python. This is an important set of tools for connecting existing
or newly written processing code to the data.

\subsubsection{Radio frequency interference excision}

The Known Pulsar Pipeline in Figure \ref{fig:known} shows two parallel
processing paths in the offline reduction of LOFAR BF data.  The
standard path performs dedispersion and folding; optionally, the
pipeline can be run on the same data, but having performed
RFI-excision prior to folding and dedispersion. This process filters
the data of RFI and then runs the same tools as run for standard
processing.  The data processing time doubles since the pipeline is
essentially performing the same processing twice, once on unfiltered
data and a second time on RFI-cleaned files.  Once the RFI-cleaning
algorithms have been refined, we expect to only run the processing
once, using the RFI-cleaned data.

\begin{figure*}
\begin{center}
\subfigure{
  \includegraphics[scale=0.9]{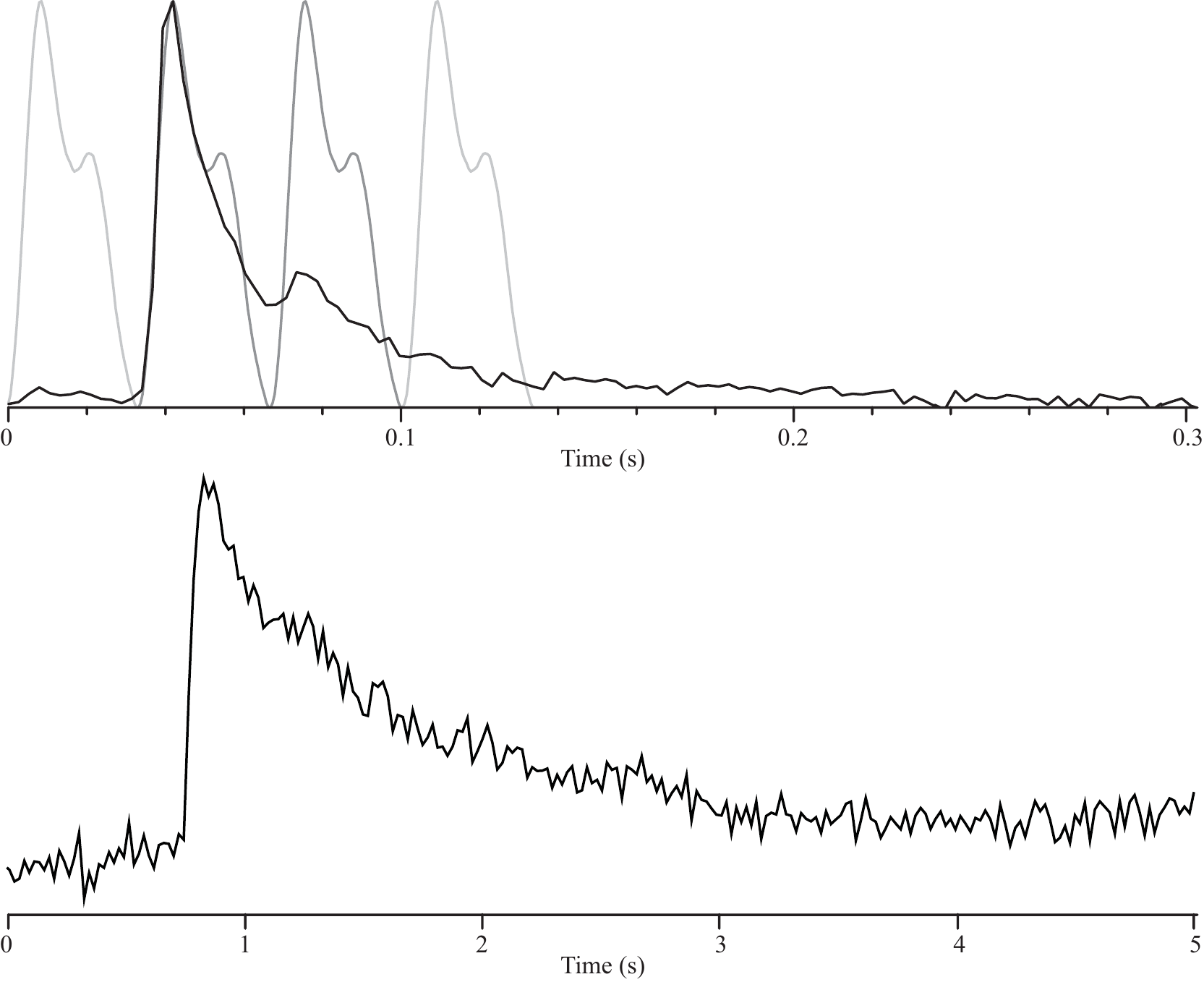}}
\caption{Observations of giant pulses from PSR B0531+21 in the Crab
  Nebula observed with both the HBA and LBA.  The top panel shows a
  ``double giant" pulse observed using the incoherent sum of the HBAs
  from 6 core stations over the frequency range $139- 187$~MHz. A
  section of the timeseries data (black line) is shown indicating a
  giant pulse which is followed by a second giant pulse in the next
  rotation of the pulsar. The fading grey lines show the average pulse
  profile for this observation repeated four times. The two giant
  pulses clearly show the influence of scattering in the ISM and from
  the Crab Nebula itself.  The lower panel shows a single giant pulse
  observed with the LBAs from 17 core stations added incoherently over
  the range $32-80$~MHz. Note the significantly different timescale
  for the scattering delay compared with the HBA observation.}
\label{fig:crab} 
\end{center}
\end{figure*}

\subsection{Known pulsar pipeline}

The Known Pulsar Pipeline processes data taken in the following modes:
Incoherent Stokes, Coherent Stokes, Coherent Complex, Fly's Eye and
Commensal Imaging. In all cases, once the data is on the LOC, the
processing done by the Known Pulsar Pipeline is very similar to
standard pulsar processing (e.g. \citealt{lk05}). Once the data have
undergone RFI excision it is then dedispersed at the known DM using
either coherent or incoherent dedispersion, depending on the
particular source and the scientific goals. The dispersion-corrected
data will then be used to generate a number of output data products
such as dedispersed time series, dynamic spectra and folded pulse
profiles. It is likely that we will simultaneously observe multiple
pulsars, and so multiple instances of the pipeline, with different
dispersion and fold parameters, will run in parallel. The processing
capability of LOFAR is such that we are in general able to produce all
of these data products for each known pulsar observation. These data
products are then flux and polarisation calibrated, stored to the
long term archive as well as being passed to post processing programs.
Multiple pipelines, from multiple separate observations, can run in
parallel depending on the available computing resources on the LOC.

As well as these scientific data products the pipeline
(Figure~\ref{fig:known}) will also produce a series of ``pulsar
diagnostic plots" e.g. signal statistics, bandpasses, RFI occupancy
and calibration information, which can be used to both calibrate the
data and assess their quality.  These will be stored in the long term
archive. In some cases raw data products with lower time and frequency
resolution will also be archived to allow further processing if
needed. This will be particularly valuable when wide FoV modes are
being used, as new sources might later be found in these same
observations.

\begin{figure*}
\begin{center}
\includegraphics[width=0.48\textwidth]{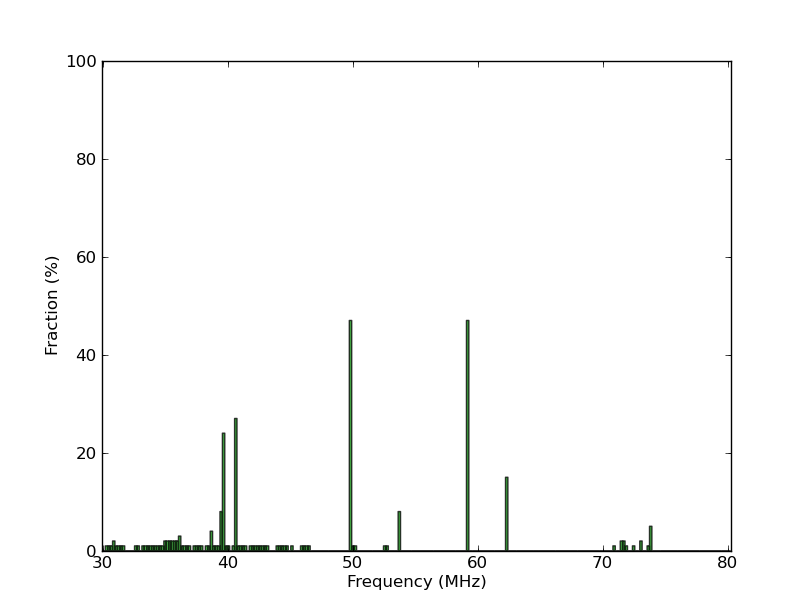}
\includegraphics[width=0.48\textwidth]{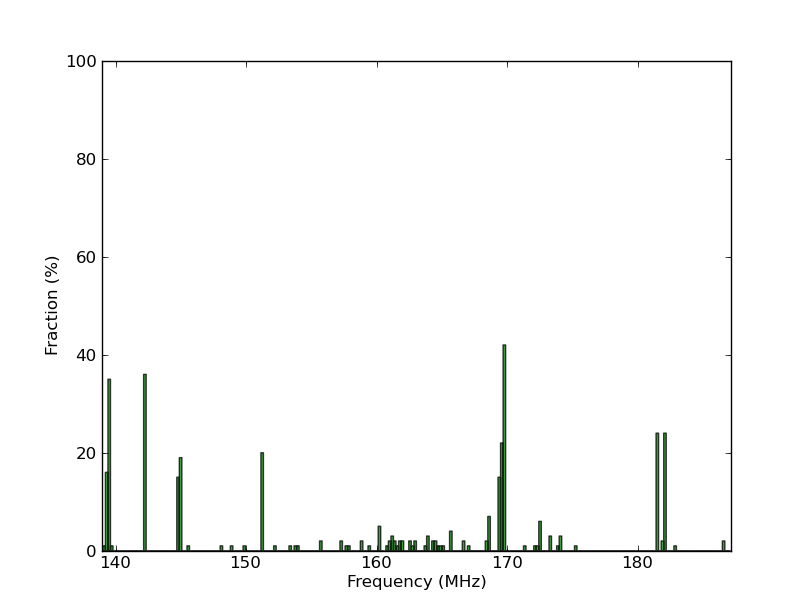}
\includegraphics[width=0.48\textwidth]{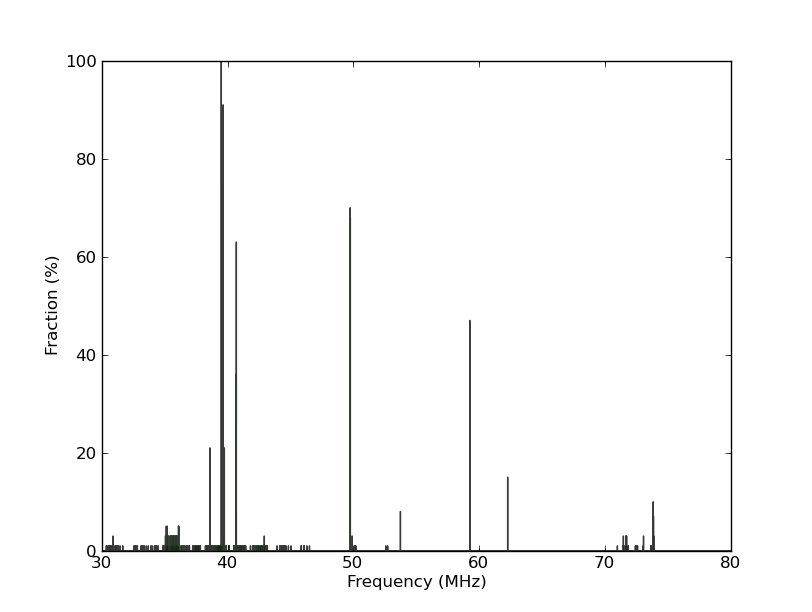}
\includegraphics[width=0.48\textwidth]{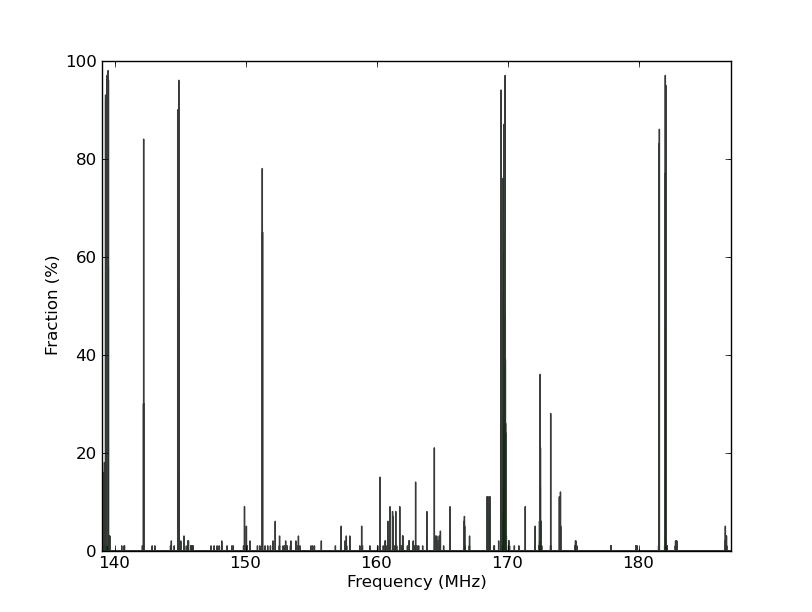}
\caption{An initial summary of the RFI situation across our most
  commonly used sections of the LBA (left-hand panels) and HBA
  (right-hand panels) frequency ranges. The upper and lower panels
  correspond to a frequency channel width of 195~kHz and 12~kHz,
  respectively. Each line in the plot corresponds to the percentage of
  observations in which the particular frequency channel was affected
  by RFI.}
\label{fig:rfireport}
\end{center}
\end{figure*}

\subsection{Periodicity and single pulse search pipeline}

As described in Sect. \ref{sec:science}, pulsar/fast-transient search 
observations are a key aspect of the LOFAR pulsar science case.  These 
can be divided into at least five types of searches, each with its own corresponding
data-reduction strategy and science goals.  Together, these complementary 
approaches span a huge range of source parameter space (e.g., source brightness, 
recurrence rate of transients, and sky location):

\begin{itemize}

\item i) Targeted searches of multi-wavelength sources using one or 
a few tied-array beams.  Here maximum sensitivity is required over a 
relatively small FoV.

\item ii) All-sky survey using hundreds of tied-array beams.  This is 
the best option for a high-instantaneous-sensitivity all-sky 
pulsar/fast-transient survey.  However, the small FoV of tied-array 
beams will limit the possible dwell time to roughly 10 minutes.

\item iii) All-sky survey using (multiple) incoherent beams.  This provides 
a shallower survey in terms of raw sensitivity, but the 
large FoV of potentially multiple incoherent beams makes 1~hour (or 
longer) dwell times feasible.

\item iv) Wide-area Fly's Eye searches for rare, bright transients.  Pointing 
the station beams in different directions allows even longer dwell times, 
with reduced sensitivity.

\item v) Piggy-back observations using a single incoherent beam.
  These intend to maximize the product of total on-sky observing time
  and FoV in order to detect the rarest transient events.  Ultimately,
  this could provide on the order of a full day's worth of integration
  time over the {\it entire} LOFAR-visible sky on the timescale of a
  few years.\\
\end{itemize}

\subsubsection{General search requirements and processing}

We currently have a basic search pipeline in place that can deal well
with searching a single incoherent beam or a few tied-array
beams. However, modes using hundreds of tied-array beams require
highly optimised processing chains and the full LOC hardware to be in
place, and so we are currently only in the planning phase of this more
complex mode\footnote{Note that this processing challenge is still
  only a fraction of what will be required for the SKA.}.  As well as
potentially dealing with large numbers of beams, the very low
frequencies and wide fractional bandwidths involved in the pulsar
searches means that in the Pulsar Search Pipeline, the number of trial
DMs will be at least 10000, which clearly makes dedispersion one of
the rate limiting steps.  We are therefore in the process of designing
an optimal mechanism to efficiently process pulsar search data within
the CEP framework. As well as considering hybrid coherent/incoherent
dedispersion modes we are looking at moving some survey processing on
to the BG/P. Distributing and parallelizing the pipeline will be done
by the Pipeline Framework; however, the mechanism itself of how data
processing will be performed and which algorithms will be used on the
LOC are still being determined.

\begin{figure}
\begin{center}
\includegraphics[scale=0.99]{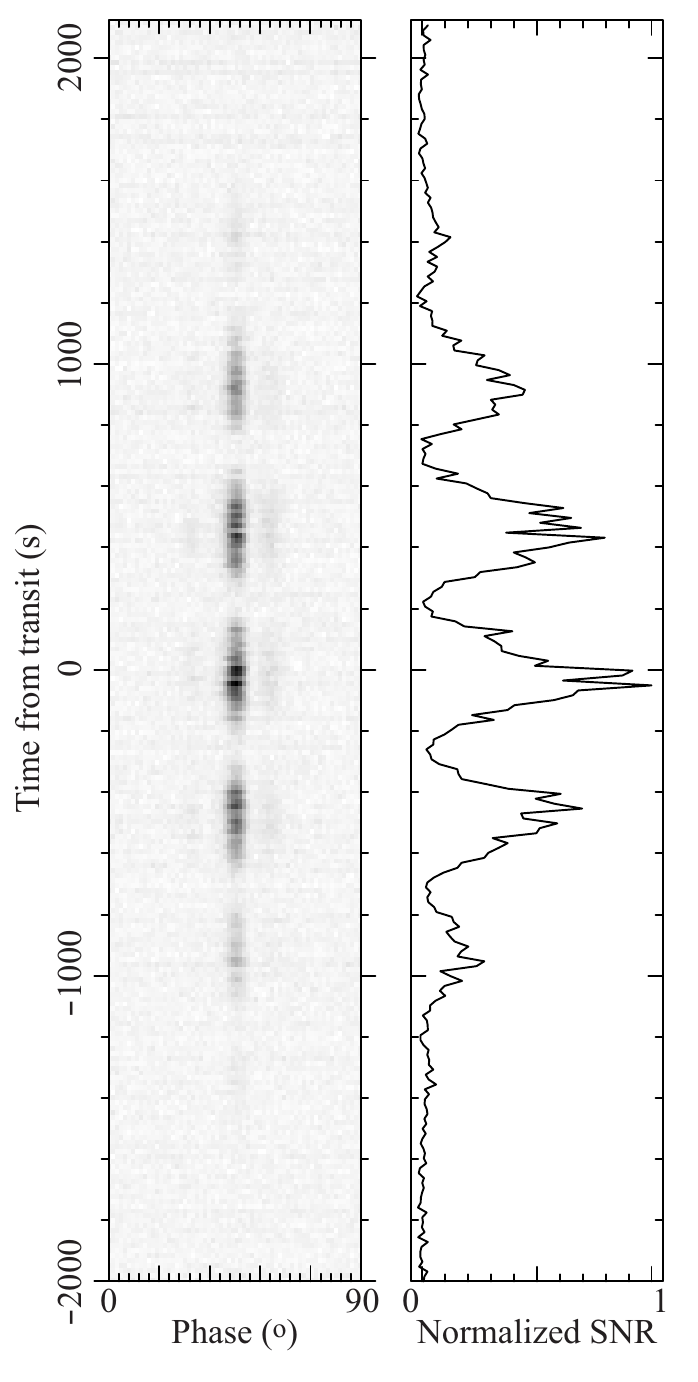}
\caption{A transit observation at a central frequency of 165~MHz
  of PSR~B0329+54 as it passes through the zenith above core station
  CS302.  The antennas were combined coherently without further
  geometrical delay in order to point directly above.  {\it Left: }The
  intensity of PSR~B0329+54 as a function of time from transit and
  rotational phase.  The variations are due to both the ``fringe"
  pattern created by the coherent addition of the two HBA sub-stations
  of station CS302 and the wider single sub-station beam width.  {\it
    Right: }The corresponding normalized S/N of the pulsar as a
  function of time from transit.}
\label{fig:hba_fringe}
\end{center}
\end{figure}

We now consider the different envisioned search modes in more detail:

\paragraph{Targeted Searches}

Targeted searches of known, multi-wavelength sources require maximum 
sensitivity over a relatively small
area of sky compared with the $\sim 20$~sq.~deg single station beam FoV. This mode will
therefore use a number of tied-array beams, the total required number
depending on the size of the source or its positional uncertainty. If
the number of beams is significantly less than needed for the all-sky
survey, then the load on BG/P and the offline cluster will be modest
and so simultaneous imaging modes are likely to be run in
parallel. For this, and other modes which use tied-array beams, one
beam may be sacrificed to point well away from the target area and be
used as a reference beam for RFI excision.  Potential targets for such searches 
include, e.g., unidentified Fermi $\gamma$-ray sources, apparently radio 
quiet $\gamma$-ray/X-ray pulsars, magnetars, and nearby galaxies or globular clusters.

\paragraph{All-Sky Coherent (Tied-Array) Survey}

The planned all-sky tied-array survey will be by far the most data and
computation intensive implementation of the pulsar pipelines. The
optimal way to perform the survey is with, at least, a couple of
hundred tied-array beams. A maximal implementation will result in a
data rate of order 23~TB~hr$^{-1}$. We are currently pursuing two options for
the survey processing, which depend greatly on the mix of computing
resources and storage capacity. If the latter is the most restrictive
then a real-time processing chain may be needed, where dedispersion,
FFT'ing, and candidate selection will be done in close to real time
with only data products related to the resultant candidates being
stored. Alternatively, relatively short observing sessions will take
place with data being stored and processed at a later date.  The
latter approach is preferable, but some hybrid process may be
required.  Such a survey provides the highest possible raw
sensitivity, but the dwell time per pointing will have to be modest
because of the limited FoV.  Certain classes of intermittent sources
may be better sampled by an all-sky survey using incoherent beams (see
below).

\begin{figure*}
\begin{center}
\includegraphics[width=0.98\textwidth]{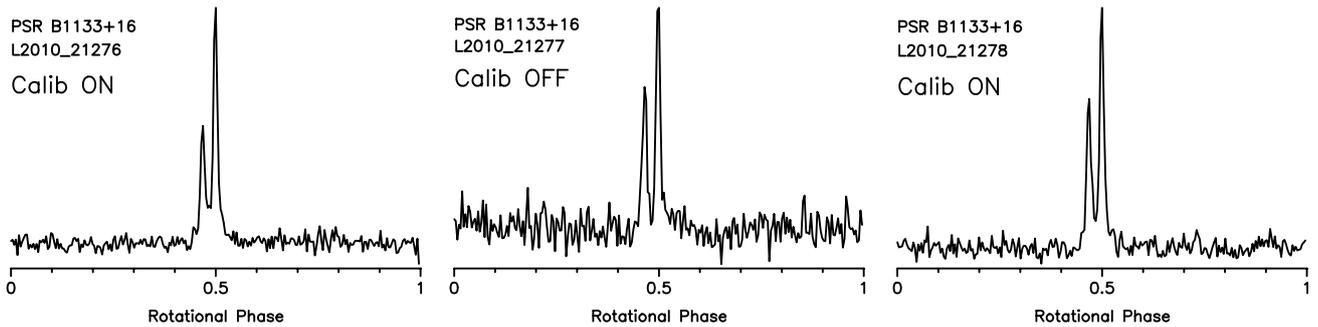}
\caption{Three consecutive 20 minute LBA observations of PSR B1133+16 
taken on 2010 Nov 5.  In between these observations, the station calibration 
was alternately turned off and then back on. The profiles are all scaled to
have the same peak height and as can be seen, the S/N of the 
cumulative pulse profile decreases by roughly a factor of 3 when the static 
calibration table is not applied.  Five Dutch stations were combined 
incoherently in these observations, and as such the S/N difference is the 
average calibration gain for all these stations.}
\label{fig:lba_calibration}
\end{center}
\end{figure*}

\paragraph{All-Sky Incoherent Survey}

The raw sensitivity of incoherent array beams (using say all 40 Dutch
stations) is roughly 3 times lower than the sensitivity of tied-array
beams made from the combination of all 24 core stations.  Nonetheless,
these incoherent beams have over 1000 times larger FoVs, making
shallow all-sky surveys in this mode an attractive possibility because
of the ability to use long dwell times.  Using multiple station beams,
the FoV can be increased even further.  For instance, using 7 station
beams provides an instantaneous FoV of $\sim 170$~sq.~deg, at a
central frequency of about 140~MHz, allowing one to search the entire
LOFAR-visible sky ($\delta > -35^{\circ}$) with only 200 pointings.
Even with a 1~hr dwell time, such a survey can be completed in roughly
1 week of observing time.  Shallow surveys such as the example
described above are not only much more tractable in terms of required
observing and processing time, but they also provide an excellent
complement to deep all-sky surveys with higher instantaneous
sensitivity.  This is because some intermittent sources, e.g. the
RRATs, may be easier to identify in surveys where the product of FoV
and dwell time is large (assuming the underlying sensitivity is also
sufficient).

\paragraph{Fly's Eye Searches}

Very bright but rare transients may be better found by surveys that sacrifice 
even more raw sensitivity in favor of increased FoV and dwell time.
Fly's Eye searches will be processed in a similar way to the all-sky surveys,
except that different data streams will correspond to individual
station beams pointing in a variety of directions, instead of
different tied-array or incoherent array beams. There could be up to a 
couple hundred such beams depending on how the observing bandwidth is 
distributed among individual station beams.  This provides, in principle, the 
ability to monitor the {\it entire} LOFAR-visible sky at once making LOFAR a
truly synoptic radio telescope.

\paragraph{Piggy-Back Searches}

Piggy-back searches aim to achieve the maximum possible on-sky time 
for transients by observing in parallel whenever possible.
This mode will be run as often as possible when the telescope is doing
standard imaging observations or even during tied-array observations. 
It will use only one incoherent or a
few tied-array beams and so the computational load is reasonable. The
simultaneous imaging data might be used when interesting
candidates are found.  These relatively shallow, but typically wide
FoV observations allow for repeated studies of the same piece of sky
in order to probe variable and transient sources.  The additional
processing requirement is small compared with that required for
imaging.

\paragraph{Single station processing}
\label{singlestatproc}

In Single Station mode, the network streams carrying the data that
would otherwise go to the BG/P need to be redirected to local
servers. This involves for the most part the 48 MHz of beamformed
data, which are streamed through 4 UDP\footnote{User Datagram
  Protocol} streams of approximately 800~Mb~s$^{-1}$ each. Members of the
LOFAR pulsar collaboration have developed a hardware and software
solution (codenamed ARTEMIS) acting as a single station backend and
capable of recording data for observations of pulsars and fast
transients. This backend has the processing power required to perform
a number of relevant operations in real time, such as RFI excision,
channelisation, and Stokes parameter generation. It is also designed
to connect to fast Graphics Processing Units for real time dispersion
measure searching on the streaming data. Besides pulsars and fast
transients ARTEMIS can be used as a general purpose single station
backend. Such an instrument allows observations to be made either in
parallel with ongoing central LOFAR observations, or allow for local
processing when the station is in local or remote control if there are
insufficient BG/P resources for the required processing. Moreover it
allows for the possibility of processing and analysing larger data
sets than can be transferred to the Netherlands.

\section{Commissioning results}
\label{sec:commissioning}

Since pulsar ``first light" in the summer of 2007, when there were
only a few test antennas available, we have been commissioning the
LOFAR telescope for pulsar observations.  In this time, LOFAR's
sensitivity and functionality have increased dramatically, as has the
data quality \citep[for the progression, see][]{slk+08,hsl09,hsa+10,sha+11}.
The ability to detect bright pulsars with just one or a few individual
HBA tiles means that pulsar observations are useful for
measuring the system performance from some of its smallest elements
up to the full, combined array.  For instance, pulsar observations can
be used to study the beam shapes of individual or beamformed elements,
verify antenna positions, and measure clock stability. Here we present
highlights from some of these system tests, followed by associated
early science results.  These demonstrate that LOFAR is beginning to
function as designed.

\begin{figure*}
\begin{center}
\includegraphics[width=0.75\textwidth]{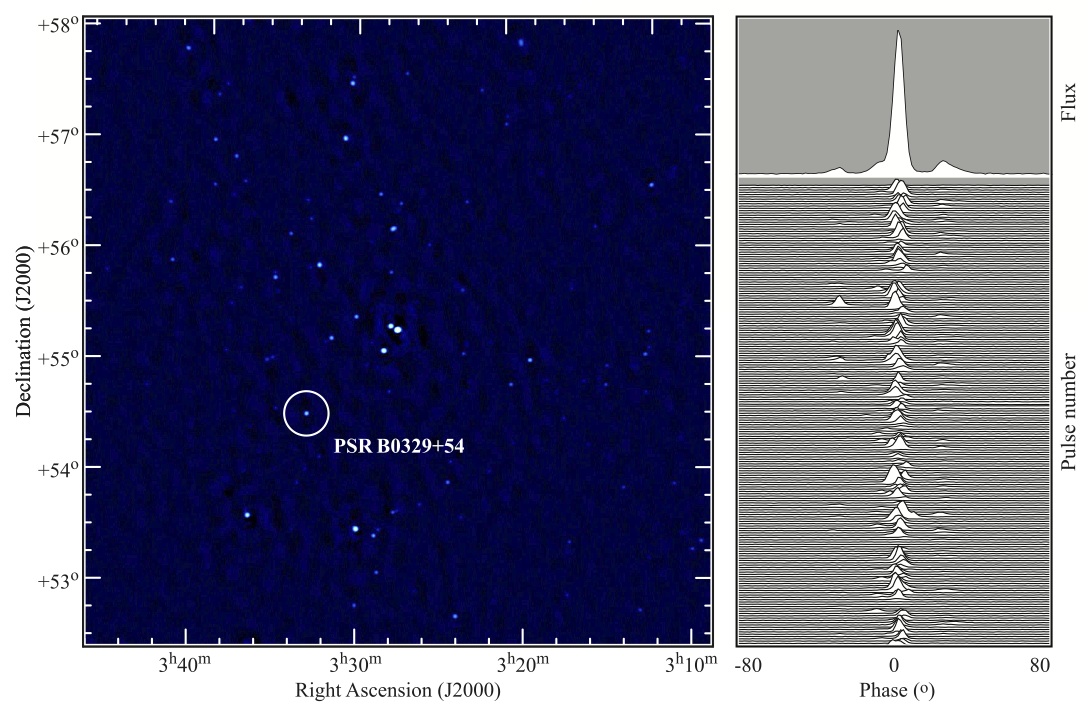}

\caption{{\it Left:} LOFAR HBA image ($139-187$~MHz) of a $5\fdg5 \times
  5\fdg5$ field including PSR B0329+54, which is circled.  The
  phase center of the image was pointed to the brightest nearby
  source, the 40~Jy: 3C~86.  The dynamic range in the image is about 1600 and the resolution about 100$\arcsec$.  {\it Right:} Pulse stack of
  roughly 200 individual pulses from PSR B0329+54, detected in
  simultaneously acquired beamformed data.  The cumulative profile of
  these pulses is shown at the top.}

\label{fig:img_psr}
\end{center}
\end{figure*}

Though already impressive, these observational results provide merely
a taste of what LOFAR will ultimately be capable of doing.  The number
of completed stations is still increasing and we have only just begun
to take data with coherently combined stations.  Also, proper phase
calibration of the LOFAR stations is currently being implemented and
will increase the raw sensitivity by up to a factor of $2-3$ (Figure
\ref{fig:lba_calibration}).  Therefore, we expect that many of the
observations presented here will be surpassed in the coming year(s) by
increasingly sensitive measurements using the full breadth of LOFAR.

\subsection{Observational tests of the system}

\subsubsection{Radio frequency interference}
\label{rficom}

In order to monitor the LOFAR RFI environment, we have recorded which
frequency channels are badly corrupted in each given observing
session. Collating the statistics from many observations, we are able
to determine which specific channels are more consistently
contaminated by RFI than others. With this information we can then
carefully select which channels to avoid when choosing the channels to
process. If whole subbands are seen to be dominated by RFI we can also
not select those when transferring data from the
stations. Figure~\ref{fig:rfireport} shows the histograms of corrupted
channels as a percentile of the total number of observations that
included that particular channel. These data do not show the full
LOFAR observing range, as they reflect the frequency range most
commonly observed in these commissioning observations (which was
chosen to optimize sensitivity for pulsar observations).

We have used more than 350 observations in the HBAs and more than 50
observations with the LBAs, spread over about a 3 month period and
occurring during both day and night-time. We compare the statistics
for subbands, the 195~kHz channels delivered directly by the stations,
with those of the typically 12~kHz channels created by the second
polyphase filter on BG/P. One can see that in the majority of
frequency channels the fraction of observations affected by RFI is
less than a few percent, with a higher total percentage of the band
being affected at the low end of the LBA range. It is also noticeable
that in the majority of cases the RFI is unresolved in the broader
195-kHz subbands, meaning that by further channelising the data we
lose a smaller percentage of the available band\footnote{Tests from
  imaging observations indicate that a good fraction of the
  narrow-band RFI is only resolved at the 1-kHz level.}. These results
are extremely encouraging and indicate that in general, RFI, if they
stay at these levels, will have a limited detrimental effect on pulsar
observations.

\begin{figure*}
\begin{center}
\includegraphics[width=0.9\textwidth]{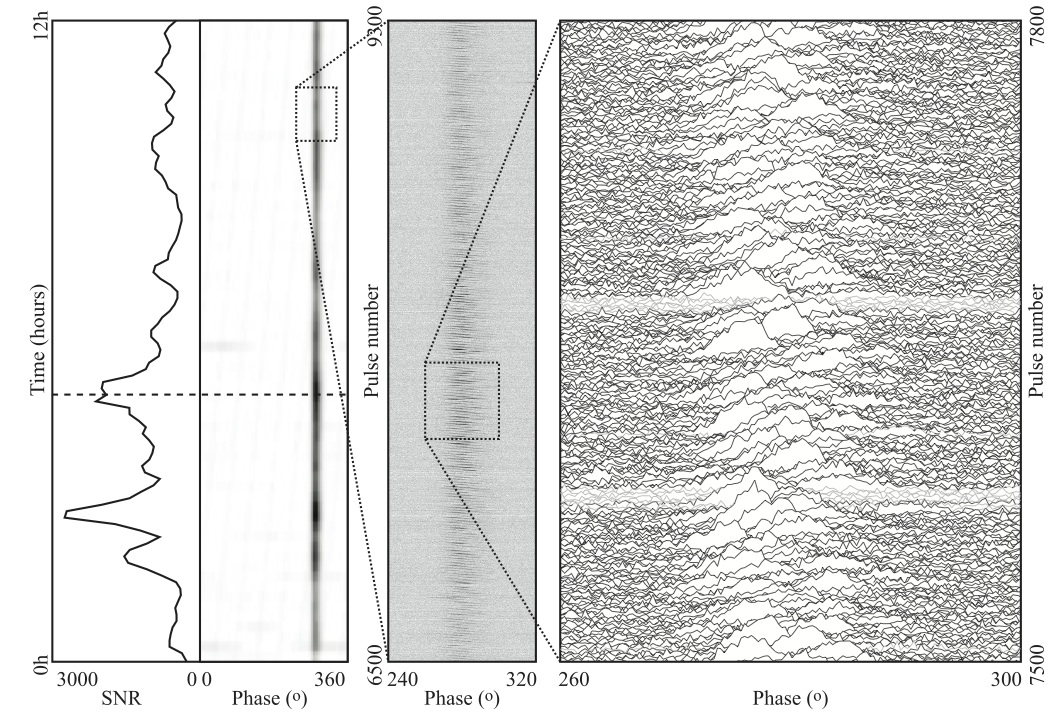}
\caption{A 12 hr observation of PSR~B0809+74 made with 4 incoherently
  added HBA stations in the frequency range $140-164$~MHz. The
  left-most plots highlight how the S/N varies throughout the
  observation, mostly due to scintillation but also from the changing
  telescope sensitivity with zenith angle. The dashed line marks the
  time when the pulsar crosses the meridian. The S/N value shown in
  the left hand plot is calculated for each 7.2 min integration. The
  middle plot shows an hour of data when the pulsar most frequently
  nulled and exhibited mode changing. The right-most plot zooms in on
  two long nulls (grey), that are easily distinguished from the
  on-pulses (black). Note the quality of the data even in
  this region when the pulsar was not particularly bright, and at a high
  zenith angle, where the sensitivity of LOFAR is reduced by a factor of roughly two.}
\label{fig:drift} 
\end{center}
\end{figure*}

\subsubsection{HBA sub-station beam shape comparison}

The two 24-tile HBA sub-stations associated with each core station can be
used separately or combined coherently at station level.  In order to
verify that a truly coherent addition of the two sub-stations was taking
place, and to confirm that the resulting beam shape agreed with
theoretical expectations, we performed an experiment in which the two
sub-stations were pointed at the Zenith and did not track a celestial
position.  Data were then recorded as the very bright pulsar B0329+54
transited through the beam\footnote{The latitude of the core of LOFAR
  is $52\fdg9$ and the $5\fdg5$ single-sub-station beam is wide enough
  to encompass the transit of this source.}.  The resulting ``fringe''
pattern is shown in Figure~\ref{fig:hba_fringe}.  As the pulsar enters
the single-sub-station beam, it becomes visible and the detected intensity
then varies as a function of time as the pulsar passes through the
fan-beam pattern that results from the coherent addition of the two
sub-stations.  A simple comparison of the time between maxima, 500~s, agrees
with what is expected for two sub-stations separated by 120~m, at a central
observing frequency of 165~MHz.  This conclusively verifies that the
sub-stations are indeed being added coherently.

\subsubsection{LBA station calibration}
\label{sec:lbacalib}

Beam-server software has been implemented in order to apply static
calibration tables that correct for the remaining phase differences
between the individual elements at each of the stations
(e.g. uncorrected differences in cable lengths).  This is necessary in
order to make a true coherent addition of the station antennas, which
maximizes station sensitivity and produces a cleaner station beam
shape.  These calibration tables are calculated empirically from
24~hour calibration runs, whose long length aims to reduce the
influence of ionospheric turbulence and RFI present in individual
observations. These long term values will only need to be updated
infrequently and will be augmented with a few minutes of observations
on bright calibration sources. As can be seen in
Figure~\ref{fig:lba_calibration}, station calibration of the LBA
antennas is now in place and is providing a sizable increase in raw
sensitivity.  This is particularly important as LBA pulsar
observations are strongly sensitivity limited. We note that station
calibration for the HBAs has also recently been implemented.

\subsubsection{Simultaneous imaging and pulsar observations}
\label{sec:simult}

As discussed earlier, the ability to simultaneously image the sky {\it
  and} record high-time-resolution, beamformed data increases the
observing efficiency of the telescope and affords new scientific
opportunities.  We have performed a number of simultaneous imaging and
pulsar observations to test the functionality of this mode.  Shown in
Figure~\ref{fig:img_psr} is a 12 hr observation of a field including
the bright pulsar B0329+54 (with a flux density at 100~MHz of $S_{100}
\sim 1$~Jy).  These data were taken on 2010 April 19 and included 7
core and 3 remote stations.  An initial calibration was made using the
three brightest sources in the field, the brightest of which was
placed at the phase center.  These observations have an odd point
spread function (PSF), which is the result of there being only short
and long baselines, with little uv-coverage on intermediate baselines.
Note however that the image here has been ``cleaned'', largely
removing this effect.  Current observations are already using
significantly more stations and far better $(u,v)$-coverage.

\subsubsection{Multiple station beams}

To demonstrate the ability of LOFAR stations to create multiple beams
on the sky (at the expense of total bandwidth per beam), we performed
an observation in which two station beams, of $\sim 24$~MHz bandwidth,
each simultaneously tracked the pulsars B0329+54 and B0450+55 for
0.5~hr.  The system was configured such that the $\sim
20^{\circ}$-wide element beam of the HBA tiles was pointed half-way
between the two pulsars, which are separated by about 12$^{\circ}$ on
the sky.  The resulting pulse profiles can be seen in Figure
\ref{fig:profiles} and an illustrative graphic can be seen in
\cite{hsa+10}. It is worth noting that this angular separation is far
larger than what is possible to observe with a multi-beam receiver or
a focal plane array.  Though the HBA system, using the full
sensitivity, is limited to creating multiple station beams within the
20$^{\circ}$-wide tile beams, the LBA dipoles do not have this extra
level of beamforming and each has a roughly 120$^{\circ}$-wide beam,
making it possible to simultaneously observe sources {\it almost
  anywhere} above the local horizon. In fact, such observations have
recently been made in which 6 pulsars distributed across the
intantaneous LBA FoV were simultaneously observed.

\subsubsection{Observing with multiple stations.}  

As demonstrated below, individual LOFAR stations are sensitive
telescopes in their own right.  Combining the 24 stations in the LOFAR
core will increase sensitivity over that of a single station by up to
a factor of about 5 or 24 depending on whether the addition is
incoherent or coherent respectively (see Sect. \ref{sec:LOFAR}).  We have
performed a series of observations comparing the measured S/N of the
pulsar B1508+55 for different numbers of stations added.  The increase
in S/N seen in these profiles agrees well with the theoretical
expectation that the S/N should increase with the square-root of the
number of stations which have been incoherently combined (see
\citealt{hsa+10}).  We will continue to test this relation as more
LOFAR stations come online, especially to verify that coherently added
stations are delivering the expected sensitivity increase (see below).
It is possible that RFI local to individual stations will also affect
the summed signal, although less dramatically if they are more distant
and we are forming the coherent sum, and so we are exploring options to
flag these stations online and dynamically remove them from the
combined beam.

\subsubsection{Early (Superterp) tied-array observations}

\begin{figure}
\begin{center}
\includegraphics[scale=0.33]{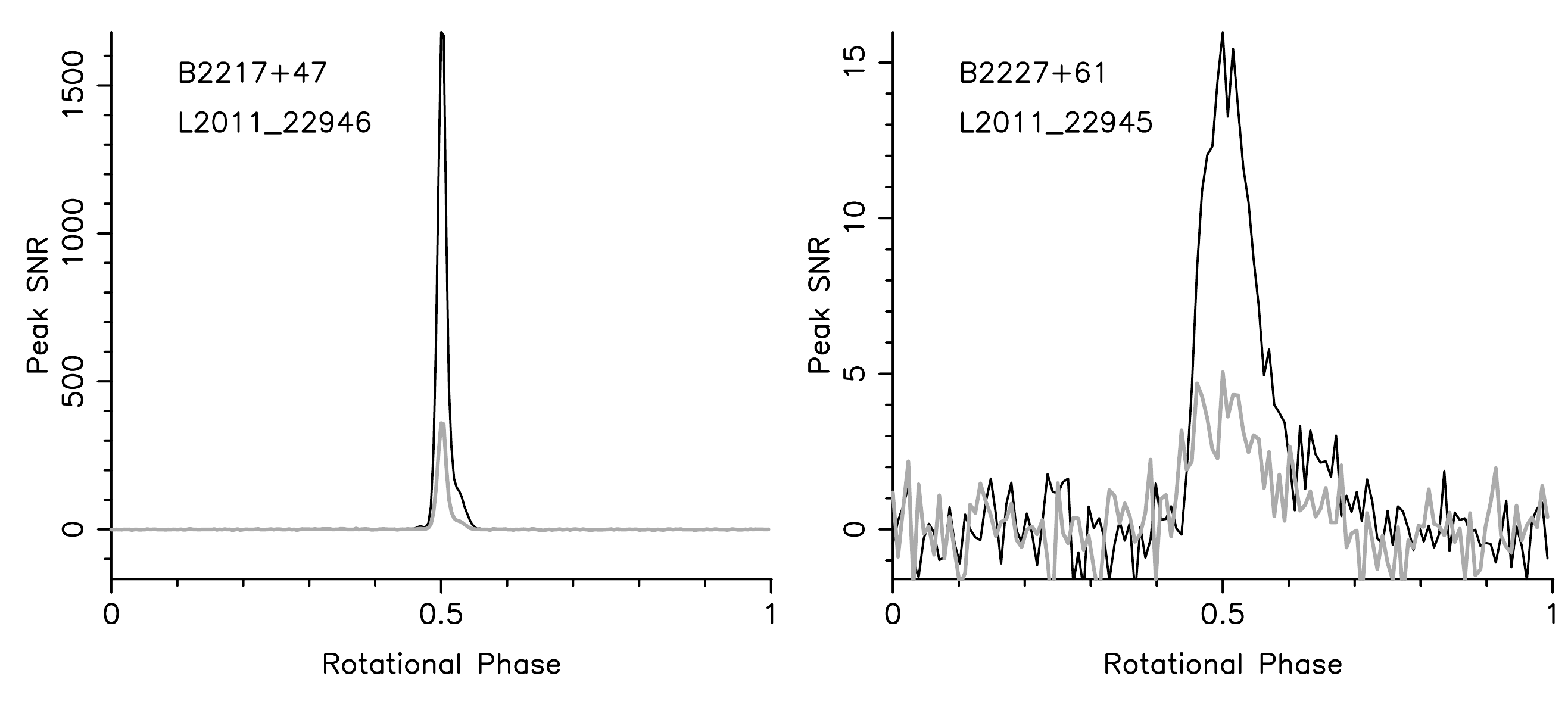}
\caption{The large sensitivity gain given by coherently adding the station
beams is illustrated in these two sample observations.  In each
observation, both the incoherent {\it and} the coherent sum of the
station beams for all 12 Superterp HBA sub-stations were recorded
simultaneously. The resulting profiles were then scaled so that the
off-pulse noise has a standard deviation of 1. This comparison shows
the expected increase in sensitivity due to coherent summation
and smaller beam size (see text for details).}
\label{fig:coh} 
\end{center}
\end{figure}

With the installation of a system to provide a single clock signal to
all of the stations located on the Superterp (6 LBA core fields and 12
HBA sub-stations), the task of calibrating tied-array beams between these
stations was greatly simplified.  By observing a bright calibrator
source in imaging mode, it is possible to solve for the phase offsets
between stations and then apply these corrections to future
observations.  Experience shows that the phase offsets between
Superterp stations, which should most of the time see the same
ionospheric patch, for the majority of the frequencies we are
considering here are sufficiently constant on timescales of hours to
ensure that the gain in the direction of the pulsar remains
optimal. Our commissioning observations so far have shown that
coherent addition produces an improvement in the S/N compared with the
incoherent sum by a factor of $\sqrt{n_{\rm stations}} = \sqrt{12} =
3.5$ as would be expected for tied-array addition to be working (see Figure \ref{fig:coh}).

\subsection{Early science observations}

\subsubsection{Pulsars with the LOFAR HBAs and LBAs.}  
\label{sec:lba-hba}
 
 We have detected many of the known, bright ($S_{400} > 50$~mJy)
 northern-hemisphere pulsars using both the HBAs and LBAs
 \citep{sha+11}.  Some pulsars (e.g., PSRs B0329+54, B0809+74,
 B1508+55, and B1919+21) are even bright enough to be visible in 1 hr
 integrations with {\it individual} HBA tiles.  As mentioned above,
 this proved quite useful for testing the beam shape, phasing, and the
 tracking accuracy of single, and later summed, HBA dipole elements.
 In the case of the LBAs, roughly a whole Dutch station (48 active
 dipoles) is necessary to achieve reasonable S/N within a 1 hr
 integration on these same bright pulsars.  We note that single pulses
 are seen with the LBAs as expected (Sect. \ref{sec:aips}).

As described earlier in Sect. \ref{sec:singlepulses}, even individual
LOFAR stations have sufficient sensitivity to be interesting for a
variety of scientific applications.  For example,
Figure~\ref{fig:simult} shows the average profiles resulting from the
simultaneous detection of PSR B1133+16 using 96 active LBA dipoles
with the Effelsberg station (called DE601) and using the 48 HBA tiles
of the Dutch core station (called CS302).

\begin{figure*}
\begin{center}
\includegraphics[scale=0.6]{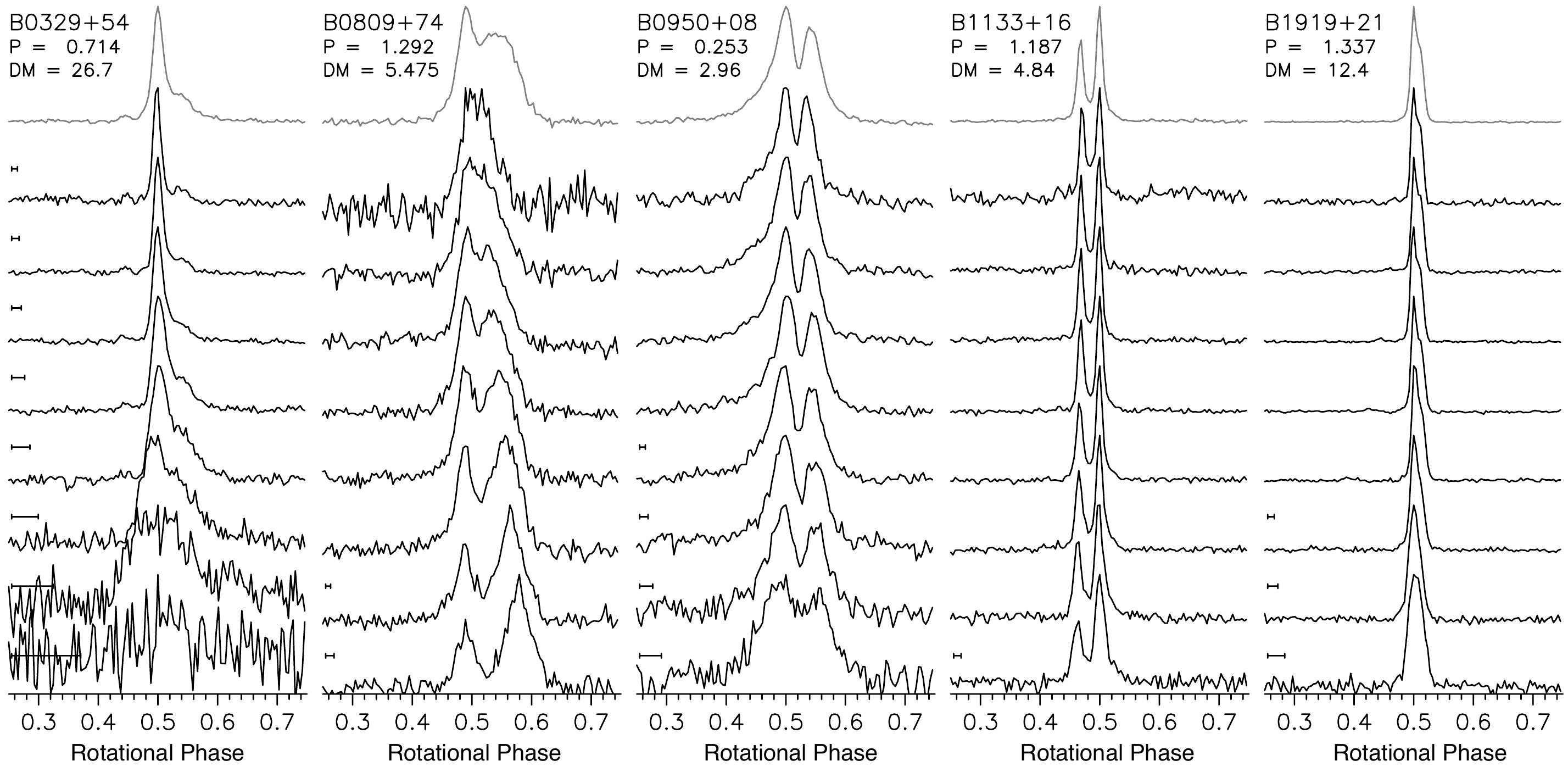}
\caption{A sample of average profiles of five pulsars observed with the
  LBAs using a total of 48~MHz with a channel bandwidth of 12.2~kHz
  and sampled at a rate of 1.3~ms. In all cases 17 stations were
  combined incoherently and the observation duration was 2~hr,
  except for PSR B1919+21 which was observed for 1~hr. The lower 8
  profiles correspond to observations with 6~MHz of bandwidth
  centered at frequencies, from lowest to highest, of 35, 41, 47, 53,
  59, 65, 71 and 77~MHz respectively. The top profile shows the
  summed profile from all 8 bands. At present there is no flux
  calibration for the LOFAR data hence no flux density scale is shown and
  profiles have been normalised to the peak intensity. The bar on the
  left hand side indicates the smearing due to dispersion across the
  12.2~kHz channels and is shown only when it exceeds 3~ms. The
  periods are given in seconds and the dispersion measure in
  pc~cm$^{-3}$.}
\label{fig:lbaprofiles}
\end{center}
\end{figure*}

The recent advent of station calibration for the LBAs discussed in
Sect. \ref{sec:lbacalib} has led to a significant improvement in
sensitivity and this is demonstrated by the observations of PSRs
B0329+54, B0809+74, B0950+08, B1133+16 and B1919+21 shown in Figure
\ref{fig:lbaprofiles}. These observations cover 32$--$80~MHz, likely
making them the widest contiguous bandwidth observations ever made of
radio pulsars at these frequencies. We have even been able to detect
pulsars down to below 16~MHz \citep{sha+11}. This provides us with an
unprecedented view into the evolution of the pulse profile as a
function of frequency, allowing any ambiguities, for example in
aligning pulse components, due to dispersive effects, to be
resolved. The contributions from changing geometry, spectral index
variations, scattering in the interstellar medium, and new components
can be separated and studied with data like these. These data were
obtained with the incoherent sum of 17 LBA stations; once we can form
the coherent addition of all the stations in the LOFAR core the S/N of
these observations will be improved by a further factor of about
six. However we already have data of sufficient quality to start such
a study. We note that dispersive effects are affecting the profiles at
the lowest frequencies shown in Figure \ref{fig:lbaprofiles}; however
observations with more than an order of magnitude improvement in
effective time resolution are already possible. Moreover, coherent
dedispersion has now been implemented which further increases the time
resolution we can achieve for the least scattered pulsars.

\subsubsection{Giant pulses from the Crab pulsar.}  

The Crab pulsar, B0531+21, is particularly interesting, in part
because it emits extremely bright ``giant pulses'' \citep{sr68,han71}.
These giant pulses have the highest brightness temperature of any
observed astronomical phenomenon and have been seen at frequencies
from a few 10's of GHz all the way down to a few 10's of MHz
\citep{pku+06}. LOFAR will be especially useful for studying the
Crab pulsar, and other young pulsars in supernova remnants, because it
will be possible to form small (arc-second to arc-minute depending on
what fraction of the array is used) tied-array beams which can
potentially partially resolve out these nebulae. This will greatly
increase sensitivity to the pulsations over the nebular
background. There are significant variations in the scattering
timescale with time, as seen by Kuz'min et al. (2008)\nocite{kljs08},
and the possibility of frequently monitoring these over a wide
bandwidth with LOFAR allows one to study the changes in the nebular
scattering properties on timescales ranging from the 33~ms rotation
period up to years.

Even before having the ability to form tied-array beams we have
detected giant pulses from the Crab pulsar using the LOFAR LBAs and
HBAs.  Figure~\ref{fig:crab} shows a ``double giant pulse'', where two
consecutive giant pulses are separated by only one rotation period.
Note that at the observing frequency of $\sim 150$~MHz the scattering
tail of each pulse is longer than the 33-ms pulse period.  We have
observed the Crab on a number of occassions and have already seen
significant variations in this scattering timescale, indicating
changes in the nebula along our line of sight. As shown in
Figure~\ref{fig:crab} we have also detected giant pulses with the LBAs
in the frequency range of 32--80~MHz that exhibits a scattering tail
that extends to many seconds. Studying the evolution of the scattering
of these giant pulses over this sort of frequency range will be a
useful probe of the frequency scaling laws appropriate for this type
of scattering.

\subsubsection{Multi-day observations of PSR B0809+74}

Previous low-frequency monitoring of pulsars has been hampered by the
limited observing times achievable by transit instruments or
telescopes with equatorial mounts.  In stark contrast to this, we have
observed several circumpolar sources for up to 64~hr continuously,
using LOFAR's full tracking ability.  In Figure \ref{fig:drift} we
show a 12 hr segment of one of these observations of PSR B0809+74.
This corresponds to 33000 pulses and the full data set to a remarkable
178000 pulses. These observations used the incoherent combination of
just 4 core HBA stations and yet show high S/N: the individual
pulses from this pulsar are clearly visible, and the pulsar's
occasional sudden turn-off, the so-called nulls, can be clearly
distinguished above the low noise floor. Interspersed by nulls, the
individual pulses form a drift pattern in the time versus rotational
phase plane (middle and right-most panels of Figure \ref{fig:drift})
$-$ a well known phenomenon that provides important insight into the
pulsar emission mechanism.  For understanding the interaction between
nulling and drifting, long integrations have been instrumental,
together with the occasional, fortuitous boost in pulsar brightness
through scintillation (e.g. \citealt{vsrr03}).  Never before has a
data set been gathered on a source like this with such a large number
of pulses, over such a wide bandwidth and at this time
resolution. These data, and more like it, will provide a unique view
of the pulse emission physics continuously over timescale of
milliseconds to days and we have already begun such studies.

\subsubsection{Observation of the Galactic centre and PSR B1749$-$28}
We have easily detected the bright ($S_{400} = 1.1$~Jy) pulsar
B1749$-$28, which is only $1^{\circ}$ away from the direction of the
Galactic Centre (see Figure~\ref{fig:profiles} for pulse profile).
Three aspects of this detection are noteworthy.  First, LOFAR has
adequate sensitivity to pick out a 1.1 Jy pulsed source against the
high sky background in the direction of the Galactic Centre
(especially considering that the single-HBA-sub-station beam is very
wide at these low elevations).  Second, LOFAR is able to observe
bright pulsars at a zenith angle (ZA) of at least $80^{\circ}$.  At
such a high ZA, the sensitivity of the dipoles is reduced by close to
a factor of 6, simply because of projection. Third, these
  observations were made with the incoherent sum of the stations,
  subsequent observations will be able to use the coherent sum of the
  stations reducing the size of the beam and thus resolving out part
  of the bright Galactic plane and further improving the sensitivity.
Furthermore, RFI may be more pernicious for observations at higher ZA.
Nonetheless, this detection clearly demonstrates that it will be
possible to monitor the Galactic Center for bright fast transients
(though scattering will remain a major limitation).

\begin{figure}
\begin{center}
\includegraphics[width=\columnwidth]{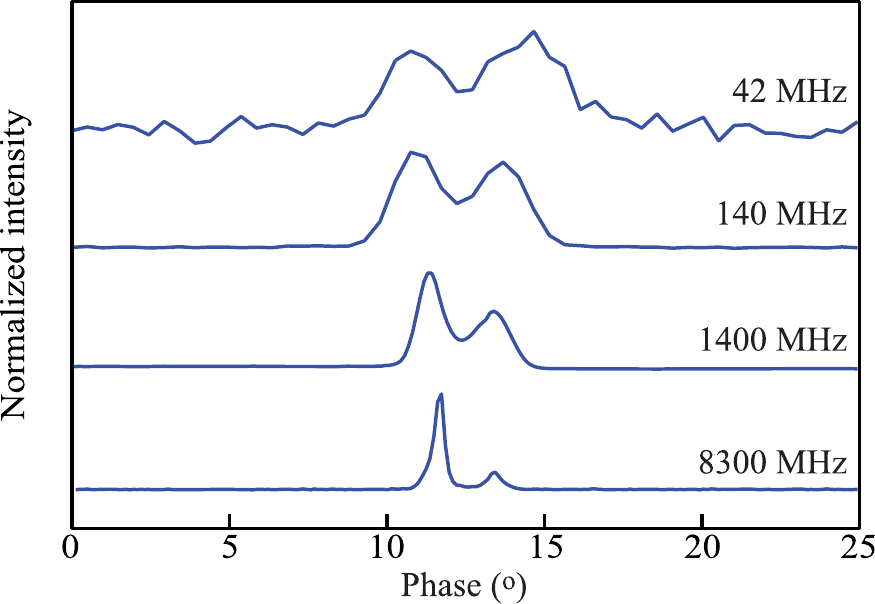}
\caption{PSR B1133+16's average pulse profile observed simultaneously
  over nearly 8 octaves in frequency.  LOFAR observations were
  acquired in both the LBA (42~MHz, station DE601) and HBA (140~MHz,
  station CS302) bands and were supplemented by contemporaneous
  observations with the Lovell (1524~MHz) and Effelsberg
  (8300~MHz) telescopes.}
\label{fig:simult}
\end{center}
\end{figure}

\begin{figure}
\begin{center}
\includegraphics[scale=0.45]{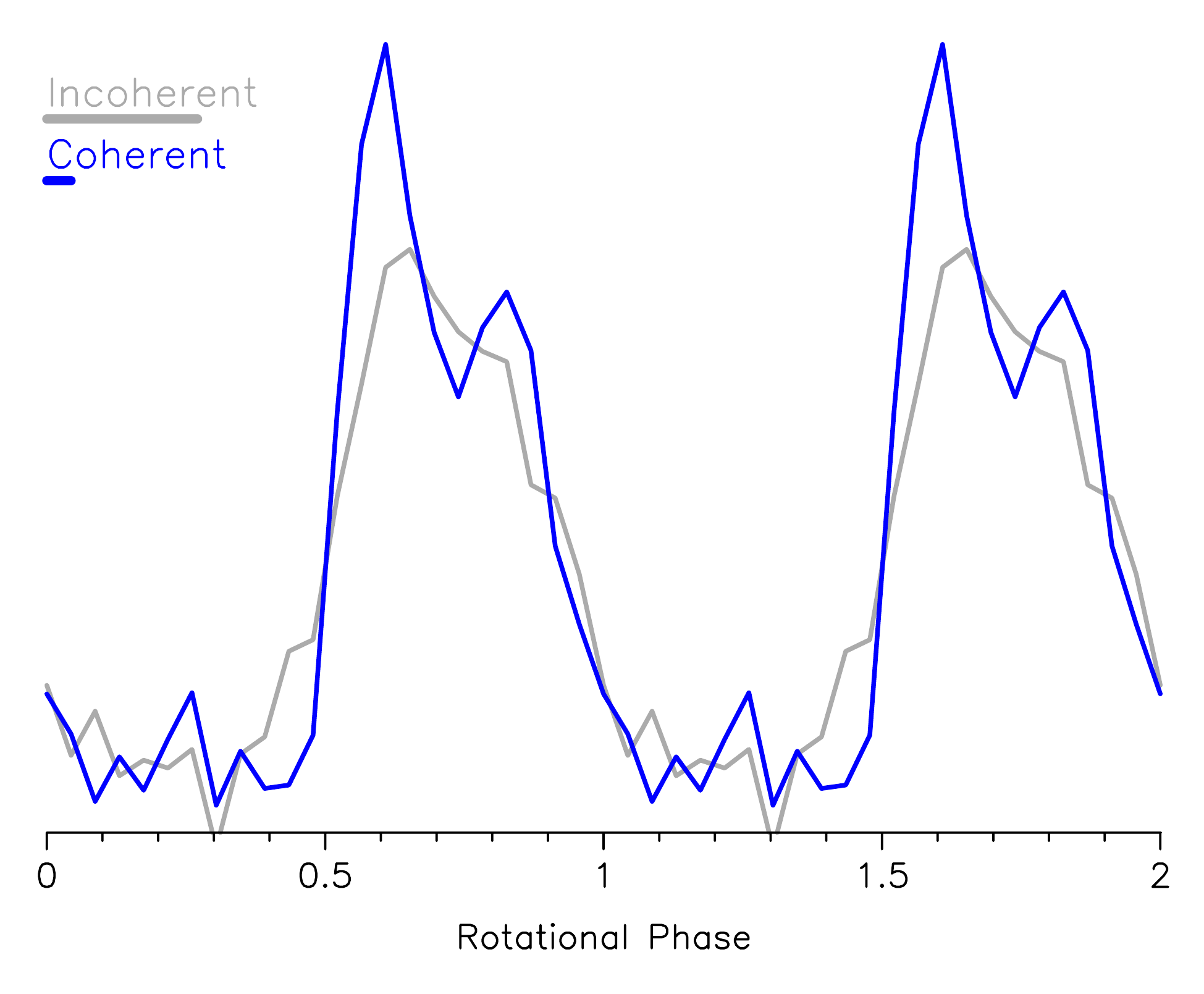}
\caption{For comparison, two consecutive 20-minute observations of the
  1.88-ms pulsar J0034-0534 at a central observing frequency of 143
  MHz and with a spectral resolution of 12kHz (corresponding time
  resolution of 82$\mu$s) are shown.  The data for the ``Incoherent"
  profile has only been corrected for dispersion by applying a
  frequency dependent time delay to the spectral channels in the
  offline post-processing.  In contrast, the data for the ``Coherent"
  profile has also been coherently dedispersed online in order to
  remove the intra-channel dispersive smearing.  The increased
  effective time resolution of the coherently dedispersed data is
  evident from comparing the two profile morphologies; e.g., two
  profile components are visible in the ``Coherent" profile which are
  washed out in the ``Incoherent" profile.  The off-pulse noise has
  been scaled to the same level to facilitate comparison of the S/N.
  The effective time resolution is represented by the two horizontal
  bars on the left-hand side, and reflects the intra-channel smearing
  in the case of the incoherently dedispersed profile.}
\label{fig:cohdd}
\end{center}
\end{figure}

\subsubsection{Millisecond pulsars}

Though scattering strongly limits the effective time resolution
achievable at low observing frequencies, LOFAR is still highly capable
of observing nearby MSPs, which are only mildly
scattered\footnote{MSPs that are further away may also have low
  scattering measures in some cases, as the general correlation
  between distance, dispersion measure, and scattering shows large
  deviations from the general trend, see \citet{bcc+04}.}.
Figure~\ref{fig:profiles} shows the detections of the 16-ms pulsar
J2145$-$0750 and the 6.2-ms pulsar B1257+12.  These pulsars both have
low DMs ($\sim 10$~pc cm$^{-3}$), which made these detections possible
without the need for coherent dedispersion.  Despite the low DM
however, one can still see that the profiles are broadened by
dispersive smearing within the channels.  As discussed earlier,
creating narrower channels is possible but will come at the expense of
the time resolution of the samples, such that the narrow pulse profile
will still be poorly sampled.  To circumvent this problem, we have
implemented coherent dedispersion on the BG/P itself to remove
intra-channel dispersive smearing and to open the possibility of
observing MSPs with much larger DMs. An example of the results of
observations in this mode is shown in Figure \ref{fig:cohdd}. We will soon be
able to do this over larger bandwidths and LOFAR will then be able to
provide measurements that eclipse those achieved in \cite{skh08}.

\begin{figure*}
\begin{center}
\includegraphics[scale=0.68]{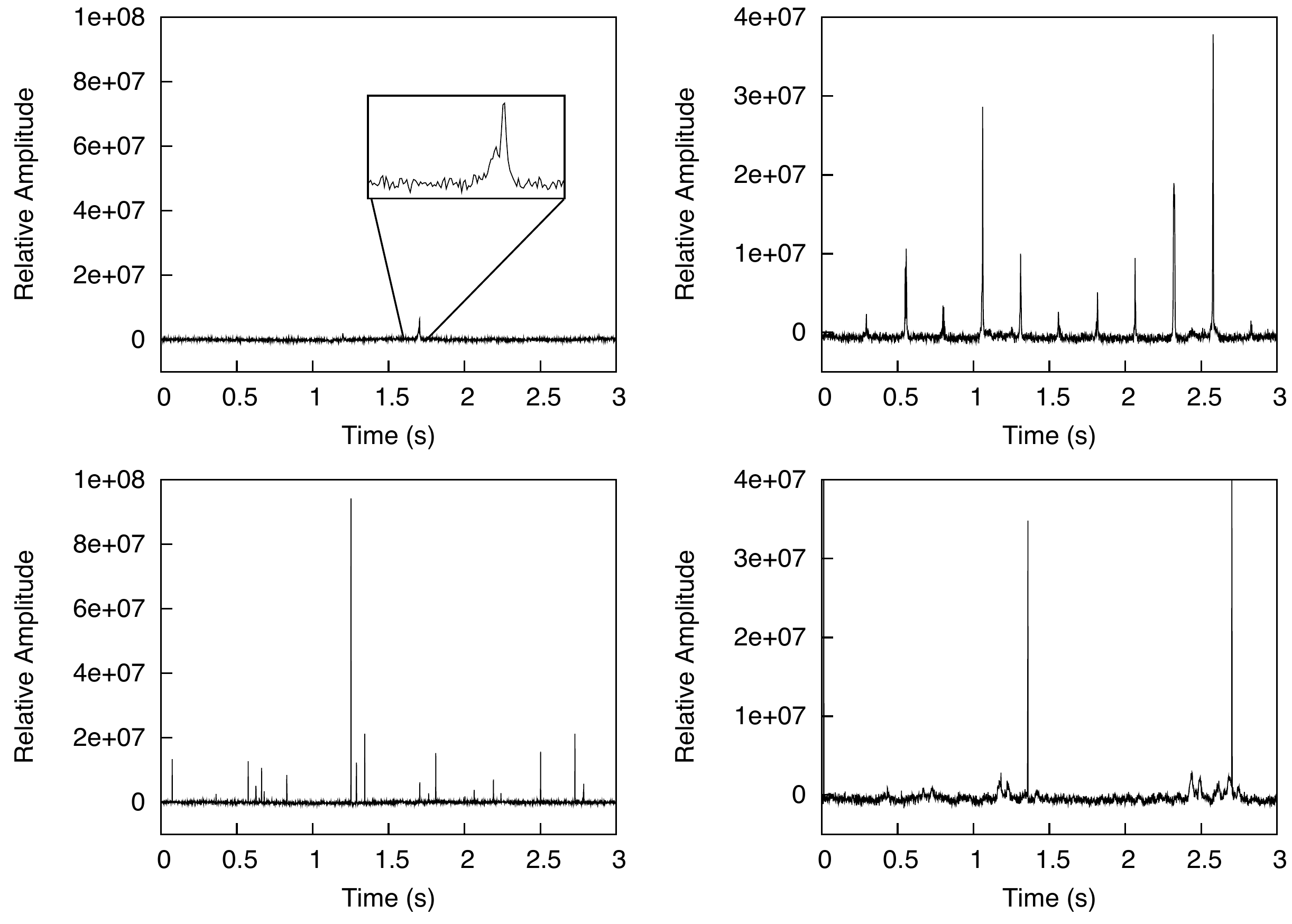}
\caption{Observations of PSR B0950+08 made using the LBAs (left) and
  HBAs (right).  The top panels show sections of the timeseries after
  dedispersion to the pulsar DM of 2.96~pc~cm$^{-3}$. We easily detect
  single pulses and one can see the wide range in intensity of the
  pulses, in particular in the low band. In the lower panels the same
  timeseries are shown but without dedipsersion, the so-called DM zero
  timeseries. In both the LBAs and HBAs we can see evidence for narrow
  spikes of RFI at zero DM, however these are clearly broadband in
  nature and have been completely dispersed beyond a detectable level
  at the dispersion measure of the pulsar. We note that the broad
  features seen in the DM zero timeseries of the HBAs (e.g. near 2.5~s
  in the lower-right panel) are due to the pulsar itself. The ordinate
  axes corresponds to an arbitrary amplitude but is preserved in the
  top and bottom plots.}
\label{fig:disprfi}
\end{center}
\end{figure*}

\subsubsection{Simultaneous LBA/HBA observations}

Sub-arraying and {\it independent} simultaneous observations make it
possible to use multiple LOFAR stations to cover a larger frequency
band than the maximum 48~MHz instantaneously available from any one
station.  In this way, it is possible to almost\footnote{One can in
  principle observe in the FM band, but the data are very strongly
  affected by the interference from radio stations.} completely cover
the $10-240$~MHz radio band (see Sect. \ref{sec:LOFAR}).  In 2009
December we performed the first LOFAR observations in which the LBA
and HBA systems were used simultaneously.  This was achieved by
setting the Dutch core station CS302 in the HBA mode and the German
station at Effelsberg (DE601) in the LBA mode and running two
independent observations in parallel on the BG/P.  In addition, the
76~m Lovell and 100~m Effelsberg single dish telescopes were used
concurrently to record at 1.4~GHz and 8.3~GHz respectively.
Figure~\ref{fig:simult} shows the cumulative pulse profile of PSR
B1133+16 observed in four frequency bands centered at 42, 140, 1400,
and 8300~MHz. In these initial test observations, only one station
could be used for each of the LBA/HBA bands; in the future it will be
possible to combine multiple stations to boost sensitivity. These data
show how we can further extend the studies of the ISM and pulsar
emission physics discussed in Sect. \ref{sec:lba-hba}.

\subsubsection{LBA detections of ``Anomalously Intensive Pulses''}
\label{sec:aips}

Another example of a phenomenon which may only be visible at low
frequencies are the ``anomalously intensive pulses" (AIPs) reported by
Ulyanov et al (2006)\nocite{uzk+06} from five pulsars with low DMs, 
at decameter wavelengths with the UTR-2 radio
telescope. Currently, 6 pulsars are known to emit such strong sporadic
pulses at frequencies below 35 MHz \citep{uzk+06,udz+07}. These pulses
are 10$-$15~ms wide with energy exceeding 10$-$100 times the energy of the
average profile. The emission is seen to be quite narrow band with the
emission typically $\sim 1$~MHz wide or seen in a few narrow 1.5--5~MHz
frequency channels.  These pulses are very intriguing but are still
not carefully studied over a broad bandwidth. Such a study can shed
light on a possible link between AIPs and similar phenomena observed
at higher frequencies, such as giant pulses (e.g.,
\citealt{pku+06,spb+04b}) giant micropulses
(e.g. \citealt{cjd04,smi06}), and spiky emission \citep{wwsr06}.
LOFAR can excel in this area, and Figure~\ref{fig:disprfi} shows
several examples of bright single pulses detected from PSR B0950+08.

\subsubsection{Interstellar medium}

As well as dispersion, several propagation effects have been proposed
which could cause a frequency dependent delay in the arrival time of
pulses. Many of these effects have not, as yet, been directly
detected or studied. They include refractive delays, DM variations,
delays associated with pulse broadening from scattering \citep{fc90},
propagation effects from within the pulsar magnetosphere (Michel 1992),
and super dispersion (Shitov and Malofeev 1985; Kuz'min 1986; Shitov
et al 1988; Kuz'min et al 2008). Many of these effects are strongly
frequency dependent, with scaling indices between $\nu^{-3}$ and 
$\nu^{-5}$. LOFAR is ideally suited for studying these types of effects
and Figure \ref{fig:0329scatt} shows a frequency-phase greyscale of
simultaneous observations with the HBAs and LBAs which were
dedispersed to a DM of 26.768~pc~cm$^{-3}$. This
observation corresponds to observing over more than two octaves
simultaneously and so any deviation from the $\nu^{-2}$ law which is
greater than $\sim100$~ms should be visible over this large
bandwidth and at these low frequencies. Once the effects have been
identified we should be able to constrain how much of an impact they
have on pulsar timing at higher frequencies and also extract
information about the composition of the ISM.

\section{Conclusions and future prospects}
\label{conclusions}

\begin{figure}
\begin{center}
\includegraphics[width=0.48\textwidth]{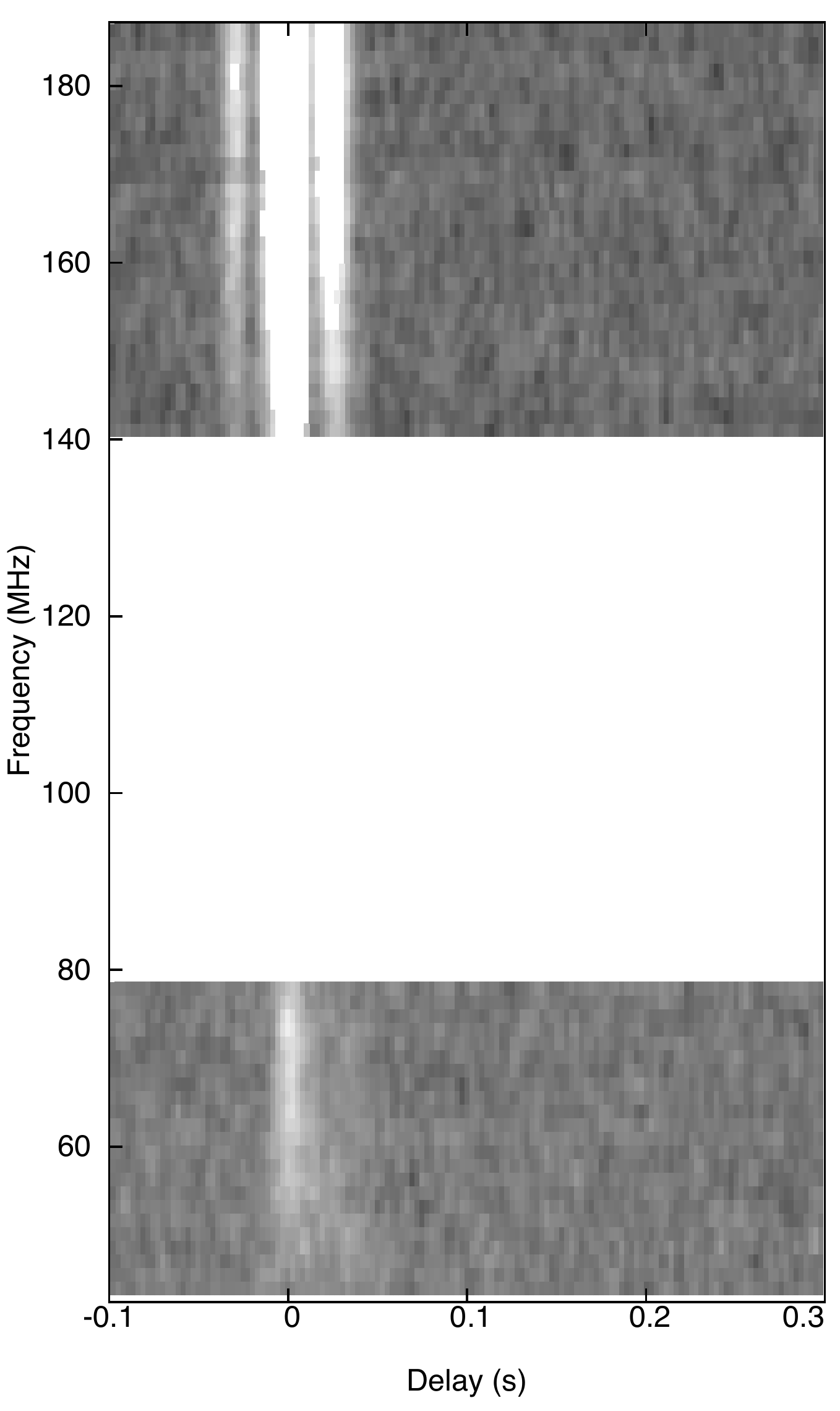}
\caption{An observation of PSR B0329+54 from December 2009. Data were taken
using the HBAs and LBAs simultaneously. The low band data was taken
using the Effelsberg international station (DE601) with 36.328125 MHz
of bandwidth between 42.1875 MHz and 78.3203125 MHz. The high band
data was taken using a single core station (CS302) with a bandwidth of
48.4375 MHz between 139.0625 MHz and 187.5 MHz. The figure shows no
significant deviation from the $\nu^{-2}$ cold dispersion law.}
\label{fig:0329scatt}
\end{center}
\end{figure}

We have shown that LOFAR will provide a massive improvement in our
ability to study pulsars and fast transients in the lowest frequency
range observable from Earth ($10-240$\,MHz). Many of these modes will
also find application for observations of solar and extra-solar
planets, flare stars and other time variable sources. In particular,
we have discussed how wide-band, low-frequency pulsar observations
will address the nature of the still enigmatic pulsar emission
mechanism (Figure~\ref{fig:simult}) and will provide a useful probe
of the ISM (Figure~\ref{fig:0329scatt}).  LOFAR also promises to
discover many new pulsars and fast transients through the combined
power of its large collecting area and (multiple) large FoVs.  Through
a dedicated all-sky pulsar/fast-transient survey, a dedicated
``radio-sky monitor'', {\it and} regular piggy-backing on imaging
observations, LOFAR can revolutionize our understanding of the
population of rare/weak radio transients by charting parts of
parameter space effectively inaccesible to traditional radio
telescopes.

Though the LOFAR antennas are still in the process of being deployed
in the field, the commissioning of pulsar observing modes has been
underway for over 3 years, with a steady increase in the level of
activity and results.  Many of the desired observing modes are now
functional, and have resulted in the observations presented here.
While LOFAR is still in its operational infancy, it has already
demonstrated several of the capabilities and technologies that make it
an important precursor to the Square Kilometre Array (SKA).  This
includes the use of multiple digital beams to track sources widely
separated on the sky (Figure~\ref{fig:profiles}), simultaneous
high-spatial and high-time-resolution observations
(Figure~\ref{fig:img_psr}), and large instantaneous fractional
observing bandwidth (Figure~\ref{fig:simult}).

With these successes in hand, we are focussed on continuing
development of the pulsar pipelines in order to take full advantage of
LOFAR's flexible data acquisition capabilities.  Along with these
opportunities come several challenges.

The combination of many thousands of antennas, each with ``all-sky''
FoV, is key to LOFAR's design.  Applying proper phase corrections to
add these elements and stations in phase remains one of our biggest
challenges.  In particular, the lack of ionospheric calibration has
limited us to using stations within the Superterp, or adding stations
incoherently.  We are tackling this problem as part of a larger effort
to provide real-time inonospheric calibration for the purposes of
generating short-timescale images that can be fed into a transient
detection pipeline.

Even with coherently summed, ``tied-array'' beams in hand, the beam
pattern itself will remain very complex and requires careful study.
As we have already seen from blind searches of our data, the beam
response of the combined array will often allow bright pulsars to
creep into observations nominally pointed far away from their
position.  While this might have some interesting applications, this
effect is a particularly pernicious form of interference for a blind
pulsar survey and requires careful treatment so that weak pulsars can
still be found in the same data.  This issue becomes even more
delicate in the case of one-off transient events.  The imaging
capability of LOFAR will certainly help alleviate this problem, but
care is nonetheless needed in ascribing positions to such events.

To maximize data quality and time resolution, studies of known pulsars
should use coherent dedispersion.  For LOFAR, this is true even for
slowly rotating pulsars because the number of channels required for
incoherent dedispersion quickly becomes so large as to degrade the
time resolution of the data to an unacceptable level.  It is already
possible to apply coherent dedispersion offline to LOFAR data that is
recorded as complex samples, but this mode requires a very large data
rate and large amounts of offline processing.  We have now also developed
online coherent dedispersion (on BG/P) in order to circumvent these issues
for a range of pulsars and DMs.

An all-sky pilot imaging survey with LOFAR is planned in order to
address many of the outstanding calibration issues, such as beam
pattern and flux density scale.  This survey has as primary purpose
the creation of a low-frequency sky model for future LOFAR imaging and
calibration.  In addition to the obvious benefits that better
calibration will have for pulsar observations, this survey will also
provide a platform for a commensal all-sky shallow survey using an
incoherent sum of the station beams.  This will be an opportunity for
us to further automate and refine our data acquisition and reduction
pipelines and to get the first real glimpse at what an all-sky LOFAR
survey for pulsars and fast transients will offer.

\section{Acknowledgements}

LOFAR, the Low Frequency Array designed and constructed by ASTRON, has
facilities in several countries, that are owned by various parties
(each with their own funding sources), and that are collectively
operated by the International LOFAR Telescope (ILT) foundation under a
joint scientific policy. The data in Figure \ref{fig:simult} based on
observations with the 100-m telescope of the MPIfR
(Max-Planck-Institut f\"ur Radioastronomie) at Effelsberg and the
Lovell Telescope at Jodrell Bank Observatory. Ben Stappers, Patrick
Weltevrede and the Lovell observations are supported through the STFC
RG R108020.  Jason Hessels is a Veni Fellow of The Netherlands
Organisation for Scientific Research (NWO). Joeri van Leeuwen and
Thijs Coenen are supported by the Netherlands Research School for
Astronomy (Grant NOVA3-NW3-2.3.1) and by the European Commission
(Grant FP7-PEOPLE-2007-4-3-IRG \#224838). Aris Karastergiou is
grateful to the Leverhulme Trust for financial support. Ralph Wijers
 acknowledges support from the European Research Council via
Advanced Investigator Grant no. 247295. LVEK is supported by an ERC ST Grant.


\begin{thebibliography}{167}
\expandafter\ifx\csname natexlab\endcsname\relax\def\natexlab#1{#1}\fi

\bibitem[{{Abbott} {et~al.}(2010){Abbott}, {Abbott}, {Acernese}, {Adhikari},
  {Ajith}, {Allen}, {Allen}, {Alshourbagy}, {Amin}, {Anderson}, \&
  et~al.}]{aaa+10b}
{Abbott}, B.~P., {Abbott}, R., {Acernese}, F., {et~al.} 2010, \apj, 713, 671

\bibitem[{{Abdo} {et~al.}(2010{\natexlab{a}}){Abdo}, {Ackermann}, {Ajello},
  {Allafort}, {Antolini}, {Atwood}, {Axelsson}, {Baldini}, {Ballet},
  {Barbiellini}, \& et~al.}]{aaa+10c}
{Abdo}, A.~A., {Ackermann}, M., {Ajello}, M., {et~al.} 2010{\natexlab{a}},
  \apjs, 188, 405

\bibitem[{{Abdo} {et~al.}(2009){Abdo}, {Ackermann}, {Ajello}, {Anderson},
  {Atwood}, {Axelsson}, {Baldini}, {Ballet}, {Barbiellini}, {Baring},
  {Bastieri}, {Baughman}, {Bechtol}, {Bellazzini}, {Berenji}, {Bignami},
  {Blandford}, {Bloom}, {Bonamente}, {Borgland}, {Bregeon}, {Brez}, {Brigida},
  {Bruel}, {Burnett}, {Caliandro}, {Cameron}, {Caraveo}, {Casandjian},
  {Cecchi}, {{\c C}elik}, {Chekhtman}, {Cheung}, {Chiang}, {Ciprini}, {Claus},
  {Cohen-Tanugi}, {Conrad}, {Cutini}, {Dermer}, {de Angelis}, {de Luca}, {de
  Palma}, {Digel}, {Dormody}, {do Couto e Silva}, {Drell}, {Dubois}, {Dumora},
  {Farnier}, {Favuzzi}, {Fegan}, {Fukazawa}, {Funk}, {Fusco}, {Gargano},
  {Gasparrini}, {Gehrels}, {Germani}, {Giebels}, {Giglietto}, {Giommi},
  {Giordano}, {Glanzman}, {Godfrey}, {Grenier}, {Grondin}, {Grove},
  {Guillemot}, {Guiriec}, {Gwon}, {Hanabata}, {Harding}, {Hayashida}, {Hays},
  {Hughes}, {J{\'o}hannesson}, {Johnson}, {Johnson}, {Johnson}, {Kamae},
  {Katagiri}, {Kataoka}, {Kawai}, {Kerr}, {Kn{\"o}dlseder}, {Kocian}, {Kuss},
  {Lande}, {Latronico}, {Lemoine-Goumard}, {Longo}, {Loparco}, {Lott},
  {Lovellette}, {Lubrano}, {Madejski}, {Makeev}, {Marelli}, {Mazziotta},
  {McConville}, {McEnery}, {Meurer}, {Michelson}, {Mitthumsiri}, {Mizuno},
  {Monte}, {Monzani}, {Morselli}, {Moskalenko}, {Murgia}, {Nolan}, {Norris},
  {Nuss}, {Ohsugi}, {Omodei}, {Orlando}, {Ormes}, {Paneque}, {Parent},
  {Pelassa}, {Pepe}, {Pesce-Rollins}, {Pierbattista}, {Piron}, {Porter},
  {Primack}, {Rain{\`o}}, {Rando}, {Ray}, {Razzano}, {Rea}, {Reimer}, {Reimer},
  {Reposeur}, {Ritz}, {Rochester}, {Rodriguez}, {Romani}, {Ryde},
  {Sadrozinski}, {Sanchez}, {Sander}, {Parkinson}, {Scargle}, {Sgr{\`o}},
  {Siskind}, {Smith}, {Smith}, {Spandre}, {Spinelli}, {Starck}, {Strickman},
  {Suson}, {Tajima}, {Takahashi}, {Takahashi}, {Tanaka}, {Thayer}, {Thompson},
  {Tibaldo}, {Tibolla}, {Torres}, {Tosti}, {Tramacere}, {Uchiyama}, {Usher},
  {Van Etten}, {Vasileiou}, {Vilchez}, {Vitale}, {Waite}, {Wang}, {Watters},
  {Winer}, {Wolff}, {Wood}, {Ylinen}, \& {Ziegler}}]{aaa+09a}
{Abdo}, A.~A., {Ackermann}, M., {Ajello}, M., {et~al.} 2009, Science, 325, 840

\bibitem[{{Abdo} {et~al.}(2010{\natexlab{b}}){Abdo}, {Ackermann}, {Ajello},
  {Atwood}, {Axelsson}, {Baldini}, {Ballet}, {Barbiellini}, {Baring},
  {Bastieri}, \& et~al.}]{aaa+10a}
{Abdo}, A.~A., {Ackermann}, M., {Ajello}, M., {et~al.} 2010{\natexlab{b}},
  \apjs, 187, 460

\bibitem[{{Abranin} {et~al.}(2001){Abranin}, {Bruk}, {Zakharenko}, \&
  {Konovalenko}}]{abzk01}
{Abranin}, E.~P., {Bruk}, I.~M., {Zakharenko}, V.~V., \& {Konovalenko}, A.~A.
  2001, Experimental Astronomy, 11, 85

\bibitem[{{Aharonian} {et~al.}(2007){Aharonian}, {Akhperjanian}, {Bazer-Bachi},
  {Beilicke}, {Benbow}, {Berge}, {Bernl{\"o}hr}, {Boisson}, {Bolz}, {Borrel},
  {Braun}, {Brion}, {Brown}, {B{\"u}hler}, {B{\"u}sching}, {Boutelier},
  {Carrigan}, {Chadwick}, {Chounet}, {Coignet}, {Cornils}, {Costamante},
  {Degrange}, {Dickinson}, {Djannati-Ata{\"i}}, {O'C.~Drury}, {Dubus},
  {Egberts}, {Eifert}, {Emmanoulopoulos}, {Espigat}, {Farnier}, {Feinstein},
  {Ferrero}, {Fiasson}, {Fontaine}, {Funk}, {Funk}, {F{\"u}{\ss}ling},
  {Gallant}, {Giebels}, {Glicenstein}, {Gl{\"u}ck}, {Goret}, {Hadjichristidis},
  {Hauser}, {Hauser}, {Heinzelmann}, {Henri}, {Hermann}, {Hinton}, {Hoffmann},
  {Hofmann}, {Holleran}, {Hoppe}, {Horns}, {Jacholkowska}, {de Jager},
  {Kendziorra}, {Kerschhaggl}, {Kh{\'e}lifi}, {Komin}, {Kosack}, {Lamanna},
  {Latham}, {Le Gallou}, {Lemi{\`e}re}, {Lemoine-Goumard}, {Lohse},
  {Manchester}, {Martin}, {Martineau-Huynh}, {Marcowith}, {Masterson},
  {Maurin}, {McComb}, {Moulin}, {de Naurois}, {Nedbal}, {Nolan}, {Noutsos},
  {Olive}, {Orford}, {Osborne}, {Panter}, {Pelletier}, {Petrucci}, {Pita},
  {P{\"u}hlhofer}, {Punch}, {Ranchon}, {Raubenheimer}, {Raue}, {Rayner},
  {Reimer}, {Ripken}, {Rob}, {Rolland}, {Rosier-Lees}, {Rowell}, {Sahakian},
  {Santangelo}, {Saug{\'e}}, {Schlenker}, {Schlickeiser}, {Schmidt},
  {Schr{\"o}der}, {Schwanke}, {Schwarzburg}, {Schwemmer}, {Shalchi}, {Sol},
  {Spangler}, {Spanier}, {Steenkamp}, {Stegmann}, {Superina}, {Tam},
  {Tavernet}, {Terrier}, {Tluczykont}, {van Eldik}, {Vasileiadis}, {Venter},
  {Vialle}, {Vincent}, {V{\"o}lk}, {Wagner}, \& {Ward}}]{aab+07}
{Aharonian}, F., {Akhperjanian}, A.~G., {Bazer-Bachi}, A.~R., {et~al.} 2007,
  \aap, 466, 543

\bibitem[{Alexov {et~al.}(2010{\natexlab{a}})Alexov, Anderson, B\"ahren,
  Grie{\ss}meier, Hessels, Masters, \& Stappers}]{ale10a}
Alexov, A., Anderson, K., B\"ahren, L., {et~al.} 2010{\natexlab{a}}, LOFAR Data
  Format ICD: Beam-Formed Data, Tech. rep., LOFAR-USG

\bibitem[{Alexov {et~al.}(2010{\natexlab{b}})Alexov, Anderson, L.~B\"ahren,
  Holties, \& Wise}]{ale10b}
Alexov, A., Anderson, K., L.~B\"ahren, A.~G., Holties, H., \& Wise, M.
  2010{\natexlab{b}}, LOFAR Data Format ICD: File Naming Conventions, Tech.
  rep., LOFAR-USG

\bibitem[{{Aliu} {et~al.}(2008){Aliu}, {Anderhub}, {Antonelli}, {Antoranz},
  {Backes}, {Baixeras}, {Barrio}, {Bartko}, {Bastieri}, {Becker}, {Bednarek},
  {Berger}, {Bernardini}, {Bigongiari}, {Biland}, {Bock}, {Bonnoli}, {Bordas},
  {Bosch-Ramon}, {Bretz}, {Britvitch}, {Camara}, {Carmona}, {Chilingarian},
  {Commichau}, {Contreras}, {Cortina}, {Costado}, {Covino}, {Curtef}, {Dazzi},
  {De Angelis}, {De Cea del Pozo}, {de los Reyes}, {De Lotto}, {De Maria}, {De
  Sabata}, {Delgado Mendez}, {Dominguez}, {Dorner}, {Doro}, {Els{\"a}sser},
  {Errando}, {Fagiolini}, {Ferenc}, {Fernandez}, {Firpo}, {Fonseca}, {Font},
  {Galante}, {Garcia Lopez}, {Garczarczyk}, {Gaug}, {Goebel}, {Hadasch},
  {Hayashida}, {Herrero}, {H{\"o}hne}, {Hose}, {Hsu}, {Huber}, {Jogler},
  {Kranich}, {La Barbera}, {Laille}, {Leonardo}, {Lindfors}, {Lombardi},
  {Longo}, {Lopez}, {Lorenz}, {Majumdar}, {Maneva}, {Mankuzhiyil}, {Mannheim},
  {Maraschi}, {Mariotti}, {Martinez}, {Mazin}, {Meucci}, {Meyer}, {Miranda},
  {Mirzoyan}, {Moles}, {Moralejo}, {Nieto}, {Nilsson}, {Ninkovic}, {Otte},
  {Oya}, {Paoletti}, {Paredes}, {Pasanen}, {Pascoli}, {Pauss}, {Pegna},
  {Perez-Torres}, {Persic}, {Peruzzo}, {Piccioli}, {Prada}, {Prandini},
  {Puchades}, {Raymers}, {Rhode}, {Rib{\'o}}, {Rico}, {Rissi}, {Robert},
  {R{\"u}gamer}, {Saggion}, {Saito}, {Salvati}, {Sanchez-Conde}, {Sartori},
  {Satalecka}, {Scalzotto}, {Scapin}, {Schweizer}, {Shayduk}, {Shinozaki},
  {Shore}, {Sidro}, {Sierpowska-Bartosik}, {Sillanp{\"a}{\"a}}, {Sobczynska},
  {Spanier}, {Stamerra}, {Stark}, {Takalo}, {Tavecchio}, {Temnikov}, {Tescaro},
  {Teshima}, {Tluczykont}, {Torres}, {Turini}, {Vankov}, {Venturini}, {Vitale},
  {Wagner}, {Wittek}, {Zabalza}, {Zandanel}, {Zanin}, {Zapatero}, {de Jager},
  {de Ona Wilhelmi}, \& {MAGIC Collaboration}}]{aaa+08}
{Aliu}, E., {Anderhub}, H., {Antonelli}, L.~A., {et~al.} 2008, Science, 322,
  1221

\bibitem[{{Archibald} {et~al.}(2009){Archibald}, {Stairs}, {Ransom}, {Kaspi},
  {Kondratiev}, {Lorimer}, {McLaughlin}, {Boyles}, {Hessels}, {Lynch}, {van
  Leeuwen}, {Roberts}, {Jenet}, {Champion}, {Rosen}, {Barlow}, {Dunlap}, \&
  {Remillard}}]{asr+09}
{Archibald}, A.~M., {Stairs}, I.~H., {Ransom}, S.~M., {et~al.} 2009, Science,
  324, 1411

\bibitem[{Armstrong {et~al.}(1981)Armstrong, Cordes, \& Rickett}]{acr81}
Armstrong, J.~W., Cordes, J.~M., \& Rickett, B.~J. 1981, Nature, 291, 561

\bibitem[{Armstrong {et~al.}(1995)Armstrong, Rickett, \& Spangler}]{ars95}
Armstrong, J.~W., Rickett, B.~J., \& Spangler, S.~R. 1995, ApJ, 443, 209

\bibitem[{{Asgekar} \& {Deshpande}(2005)}]{ad05}
{Asgekar}, A. \& {Deshpande}, A.~A. 2005, MNRAS, 357, 1105

\bibitem[{{Beck}(2007)}]{bbb+07}
{Beck}, R. 2007, Studying Cosmic Magnetism by Polarization Observations with
  LOFAR

\bibitem[{{Bennett} {et~al.}(2010){Bennett}, {van Eysden}, \&
  {Melatos}}]{bvm10}
{Bennett}, M.~F., {van Eysden}, C.~A., \& {Melatos}, A. 2010, \mnras, 1370

\bibitem[{{Bhat} {et~al.}(2004){Bhat}, {Cordes}, {Camilo}, {Nice}, \&
  {Lorimer}}]{bcc+04}
{Bhat}, N.~D.~R., {Cordes}, J.~M., {Camilo}, F., {Nice}, D.~J., \& {Lorimer},
  D.~R. 2004, ApJ, 605, 759

\bibitem[{Boucher \& Altamimi(2001)}]{ba01}
Boucher, C. \& Altamimi, Z. 2001, http://etrs89.ensg.ign.fr/memo-V7.pdf

\bibitem[{{Bruk} \& {Ustimenko}(1976)}]{bu76}
{Bruk}, I.~M. \& {Ustimenko}, B.~I. 1976, \nat, 260, 766

\bibitem[{{Bruk} \& {Ustimenko}(1977)}]{bu77}
{Bruk}, I.~M. \& {Ustimenko}, B.~I. 1977, \apss, 49, 349

\bibitem[{{Cairns} {et~al.}(2004){Cairns}, {Johnston}, \& {Das}}]{cjd04}
{Cairns}, I.~H., {Johnston}, S., \& {Das}, P. 2004, MNRAS, 353, 270

\bibitem[{{Camilo} {et~al.}(2007){Camilo}, {Ransom}, {Halpern}, \&
  {Reynolds}}]{crhr07}
{Camilo}, F., {Ransom}, S.~M., {Halpern}, J.~P., \& {Reynolds}, J. 2007, ApJ,
  666, L93

\bibitem[{{Camilo} {et~al.}(2006){Camilo}, {Ransom}, {Halpern}, {Reynolds},
  {Helfand}, {Zimmerman}, \& {Sarkissian}}]{crh+06}
{Camilo}, F., {Ransom}, S.~M., {Halpern}, J.~P., {et~al.} 2006, Nature, 442,
  892

\bibitem[{{Carilli} \& {Rawlings}(2004)}]{cr04}
{Carilli}, C. \& {Rawlings}, S. 2004, astro-ph/0409274

\bibitem[{{Cohen} {et~al.}(2007){Cohen}, {Lane}, {Cotton}, {Kassim}, {Lazio},
  {Perley}, {Condon}, \& {Erickson}}]{clc+07}
{Cohen}, A.~S., {Lane}, W.~M., {Cotton}, W.~D., {et~al.} 2007, \aj, 134, 1245

\bibitem[{{Cole}(1969)}]{col69}
{Cole}, T.~W. 1969, \nat, 223, 487

\bibitem[{Cordes(1978)}]{cor78}
Cordes, J.~M. 1978, ApJ, 222, 1006

\bibitem[{{Cordes}(2008)}]{cor08}
{Cordes}, J.~M. 2008, in Astronomical Society of the Pacific Conference Series,
  Vol. 395, Frontiers of Astrophysics: A Celebration of NRAO's 50th
  Anniversary, ed. {A.~H.~Bridle, J.~J.~Condon, \& G.~C.~Hunt}, 225

\bibitem[{Cordes {et~al.}(2004)Cordes, Kramer, Lazio, Stappers, Backer, \&
  Johnston}]{ckl+04}
Cordes, J.~M., Kramer, M., Lazio, T. J.~W., {et~al.} 2004, New Astr., 48, 1413

\bibitem[{{Cordes} \& {Lazio}(2002)}]{cl02}
{Cordes}, J.~M. \& {Lazio}, T.~J.~W. 2002, ArXiv e-prints, preprint
  (arXiv:astro-ph/0207156)

\bibitem[{{Cordes} {et~al.}(2004){Cordes}, {Lazio}, \& {McLaughlin}}]{clm04}
{Cordes}, J.~M., {Lazio}, T.~J.~W., \& {McLaughlin}, M.~A. 2004, New Astronomy
  Review, 48, 1459

\bibitem[{{Cordes} \& {Shannon}(2010)}]{cs10}
{Cordes}, J.~M. \& {Shannon}, R.~M. 2010, \apj, submitted, arXiv: 1010.3785

\bibitem[{Cordes {et~al.}(1990)Cordes, Weisberg, \& Hankins}]{cwh90}
Cordes, J.~M., Weisberg, J.~M., \& Hankins, T.~H. 1990, AJ, 100, 1882

\bibitem[{{D'Alessandro}(1996)}]{dal96}
{D'Alessandro}, F. 1996, \apss, 246, 73

\bibitem[{{de Vos} {et~al.}(2009){de Vos}, {Gunst}, \& {Nijboer}}]{dgn09}
{de Vos}, M., {Gunst}, A.~W., \& {Nijboer}, R. 2009, IEEE Proceedings, 97, 1431

\bibitem[{{Deshpande} \& {Radhakrishnan}(1994)}]{dr94}
{Deshpande}, A.~A. \& {Radhakrishnan}, V. 1994, Journal of Astrophysics and
  Astronomy, 15, 329

\bibitem[{Deshpande {et~al.}(1989)Deshpande, Shevgaonkar, \& Shastry}]{dss89}
Deshpande, A.~A., Shevgaonkar, R.~K., \& Shastry, C.~V. 1989, J. Inst.
  Electron. Telecommunications Eng.

\bibitem[{{Edwards} \& {Stappers}(2003)}]{es03}
{Edwards}, R.~T. \& {Stappers}, B.~W. 2003, A\&A, 407, 273

\bibitem[{{Ellingson} {et~al.}(2009){Ellingson}, {Clarke}, {Cohen}, {Craig},
  {Kassim}, {Pihlstrom}, {Rickard}, \& {Taylor}}]{ecc+09}
{Ellingson}, S.~W., {Clarke}, T.~E., {Cohen}, A., {et~al.} 2009, IEEE
  Proceedings, 97, 1421

\bibitem[{{Falcke} \& {LOFAR Cosmic Ray Key Science Project}(2007)}]{fal+07}
{Falcke}, H. \& {LOFAR Cosmic Ray Key Science Project}. 2007, Astronomische
  Nachrichten, 328, 593

\bibitem[{{Faucher-Gigu{\`e}re} \& {Kaspi}(2006)}]{fk06}
{Faucher-Gigu{\`e}re}, C.-A. \& {Kaspi}, V.~M. 2006, ApJ, 643, 332

\bibitem[{{Fender} {et~al.}(2008){Fender}, {Wijers}, {Stappers}, \& {LOFAR
  Transients Key Science Project}}]{fws+08}
{Fender}, R., {Wijers}, R., {Stappers}, B., \& {LOFAR Transients Key Science
  Project}, t. 2008, ArXiv e-prints, arXiv:0805.4349

\bibitem[{{Fender} {et~al.}(2006){Fender}, {Wijers}, {Stappers}, {Braun},
  {Wise}, {Coenen}, {Falcke}, {Griessmeier}, {Van Haarlem}, {Jonker}, {Law},
  {Markoff}, {Masters}, {Miller-Jones}, {Osten}, {Scheers}, {Spreeuw},
  {Swinbank}, {Vogt}, {Wijnands}, \& {Zarka}}]{fws+06}
{Fender}, R.~P., {Wijers}, R.~A.~M.~J., {Stappers}, B., {et~al.} 2006, in VI
  Microquasar Workshop: Microquasars and Beyond

\bibitem[{Foster \& Cordes(1990)}]{fc90}
Foster, R.~S. \& Cordes, J.~M. 1990, ApJ, 364, 123

\bibitem[{Fruchter {et~al.}(1988)Fruchter, Stinebring, \& Taylor}]{fst88}
Fruchter, A.~S., Stinebring, D.~R., \& Taylor, J.~H. 1988, Nature, 333, 237

\bibitem[{{Golap} {et~al.}(1998){Golap}, {Shankar}, {Sachdev}, {Dodson}, \&
  {Sastry}}]{gss+98}
{Golap}, K., {Shankar}, N.~U., {Sachdev}, S., {Dodson}, R., \& {Sastry}, C.~V.
  1998, Journal of Astrophysics and Astronomy, 19, 35

\bibitem[{Gould \& Lyne(1998)}]{gl98}
Gould, D.~M. \& Lyne, A.~G. 1998, MNRAS, 301, 235

\bibitem[{{Gupta} {et~al.}(2000){Gupta}, {Gothoskar}, {Joshi}, {Vivekanand},
  {Swain}, {Sirothia}, \& {Bhat}}]{ggj+00}
{Gupta}, Y., {Gothoskar}, P., {Joshi}, B.~C., {et~al.} 2000, in Astronomical
  Society of the Pacific Conference Series, Vol. 202, IAU Colloq. 177: Pulsar
  Astronomy - 2000 and Beyond, ed. {M.~Kramer, N.~Wex, \& R.~Wielebinski},
  277

\bibitem[{{Gurevich} {et~al.}(1993){Gurevich}, {Beskin}, \& {Istomin}}]{gbi93}
{Gurevich}, A., {Beskin}, V., \& {Istomin}, Y. 1993, {Physics of the Pulsar
  Magnetosphere} (Physics of the Pulsar Magnetosphere, by Alexandr Gurevich and
  Vassily Beskin and Yakov Istomin, pp.~432.~ISBN 0521417465.~Cambridge, UK:
  Cambridge University Press, August 1993.)

\bibitem[{{Hamaker} {et~al.}(1996){Hamaker}, {Bregman}, \& {Sault}}]{hbs96}
{Hamaker}, J.~P., {Bregman}, J.~D., \& {Sault}, R.~J. 1996, A\&AS, 117, 137

\bibitem[{{Han}(2009)}]{han09}
{Han}, J. 2009, in IAU Symposium, Vol. 259, IAU Symposium, 455--466

\bibitem[{Hankins(1971)}]{han71}
Hankins, T.~H. 1971, ApJ, 169, 487

\bibitem[{{Heald} {et~al.}(2010){Heald}, {McKean}, {Pizzo}, {van Diepen}, {van
  Zwieten}, {van Weeren}, {Rafferty}, {van der Tol}, {Birzan}, {Shulevski},
  {Swinbank}, {Orru}, {De Gasperin}, {Ker}, {Bonafede}, {Macario}, {Ferrari},
  \& {on behalf of the LOFAR Collaboration}}]{hmp+10}
{Heald}, G., {McKean}, J., {Pizzo}, R., {et~al.} 2010, ArXiv e-prints,
  arXiv:1008.4693

\bibitem[{{Hessels} {et~al.}(2010){Hessels}, {Stappers}, {Alexov}, {Coenen},
  {Hassall}, {Karastergiou}, {Kondratiev}, {Kramer}, {van Leeuwen}, {Mol},
  {Noutsos}, {Weltevrede}, \& {the LOFAR Collaboration}}]{hsa+10}
{Hessels}, J., {Stappers}, B., {Alexov}, A., {et~al.} 2010, ArXiv e-prints,
  arXiv:1009.1758

\bibitem[{{Hessels} {et~al.}(2011){Hessels}, {Roberts}, \&
  {McLaughlin}}]{hrm+11}
{Hessels}, J.~W.~T., {Roberts}, M.~S.~E., \& {McLaughlin}, M. 2011, in American
  Institute of Physics Proceedings, Vol. 999, Radio Pulsars: a key to unlock
  the secrets of the Universe, ed. {A. Possenti}

\bibitem[{{Hessels} {et~al.}(2009){Hessels}, {Stappers}, {van Leeuwen}, \& {The
  LOFAR Transients Key Science Project}}]{hsl09}
{Hessels}, J.~W.~T., {Stappers}, B.~W., {van Leeuwen}, J., \& {The LOFAR
  Transients Key Science Project}. 2009, in Astronomical Society of the Pacific
  Conference Series, Vol. 407, Astronomical Society of the Pacific Conference
  Series, ed. {D.~J.~Saikia, D.~A.~Green, Y.~Gupta, \& T.~Venturi}, 318

\bibitem[{Hewish {et~al.}(1968)Hewish, Bell, Pilkington, Scott, \&
  Collins}]{hbp+68}
Hewish, A., Bell, S.~J., Pilkington, J. D.~H., Scott, P.~F., \& Collins, R.~A.
  1968, Nature, 217, 709

\bibitem[{{Intema}(2009)}]{int09}
{Intema}, H.~T. 2009, PhD thesis, Leiden Observatory, Leiden University,
  P.O.~Box 9513, 2300 RA Leiden, The Netherlands

\bibitem[{Izvekova {et~al.}(1994)Izvekova, Jessner, Kuzmin, Malofeev, Sieber,
  \& Wielebinski}]{ijk+94}
Izvekova, V.~A., Jessner, A., Kuzmin, A.~D., {et~al.} 1994, A\&AS, 105, 235

\bibitem[{{Jenet} {et~al.}(2001){Jenet}, {Anderson}, \& {Prince}}]{jap01}
{Jenet}, F.~A., {Anderson}, S.~B., \& {Prince}, T.~A. 2001, ApJ, 546, 394

\bibitem[{Johnston \& Kulkarni(1991)}]{jk91}
Johnston, H.~M. \& Kulkarni, S.~R. 1991, ApJ, 368, 504

\bibitem[{Karastergiou {et~al.}(2003)Karastergiou, Johnston, \& Kramer}]{kjk03}
Karastergiou, A., Johnston, S., \& Kramer, M. 2003, A\&A, 404, 325

\bibitem[{Karastergiou {et~al.}(2001)Karastergiou, {von Hoensbroech}, Kramer,
  Lorimer, Lyne, Doroshenko, Jessner, Jordan, \& Wielebinski}]{khk+01}
Karastergiou, A., {von Hoensbroech}, A., Kramer, M., {et~al.} 2001, A\&A, 379,
  270

\bibitem[{{Karuppusamy} {et~al.}(2008){Karuppusamy}, {Stappers}, \& {van
  Straten}}]{ksv08}
{Karuppusamy}, R., {Stappers}, B., \& {van Straten}, W. 2008, \pasp, 120, 191

\bibitem[{{Karuppusamy} {et~al.}(2011){Karuppusamy}, {Stappers}, \&
  {Serylak}}]{kss10}
{Karuppusamy}, R., {Stappers}, B.~W., \& {Serylak}, M. 2011, \aap, 525, A55+

\bibitem[{{Keane} \& {Kramer}(2008)}]{kk08}
{Keane}, E.~F. \& {Kramer}, M. 2008, MNRAS, 391, 2009

\bibitem[{{Keith} {et~al.}(2010){Keith}, {Jameson}, {van Straten}, {Bailes},
  {Johnston}, {Kramer}, {Possenti}, {Bates}, {Bhat}, {Burgay}, {Burke-Spolaor},
  {D'Amico}, {Levin}, {McMahon}, {Milia}, \& {Stappers}}]{kjv+10}
{Keith}, M.~J., {Jameson}, A., {van Straten}, W., {et~al.} 2010, \mnras, 1356

\bibitem[{{Kondratiev} {et~al.}(2009){Kondratiev}, {McLaughlin}, {Lorimer},
  {Burgay}, {Possenti}, {Turolla}, {Popov}, \& {Zane}}]{kml+09}
{Kondratiev}, V.~I., {McLaughlin}, M.~A., {Lorimer}, D.~R., {et~al.} 2009,
  \apj, 702, 692

\bibitem[{{Kramer} {et~al.}(2003){Kramer}, {Karastergiou}, {Gupta}, {Johnston},
  {Bhat}, \& {Lyne}}]{kkg+03}
{Kramer}, M., {Karastergiou}, A., {Gupta}, Y., {et~al.} 2003, A\&A, 407, 655

\bibitem[{Kramer {et~al.}(1999)Kramer, Lange, Lorimer, Backer, Xilouris,
  Jessner, \& Wielebinski}]{kll+99}
Kramer, M., Lange, C., Lorimer, D.~R., {et~al.} 1999, ApJ, 526, 957

\bibitem[{{Kramer} {et~al.}(2006){Kramer}, {Lyne}, {O'Brien}, {Jordan}, \&
  {Lorimer}}]{klo+06}
{Kramer}, M., {Lyne}, A.~G., {O'Brien}, J.~T., {Jordan}, C.~A., \& {Lorimer},
  D.~R. 2006, Science, 312, 549

\bibitem[{Kramer {et~al.}(1994)Kramer, Wielebinski, Jessner, Gil, \&
  Seiradakis}]{kwj+94}
Kramer, M., Wielebinski, R., Jessner, A., Gil, J.~A., \& Seiradakis, J.~H.
  1994, A\&AS, 107, 515

\bibitem[{Kramer {et~al.}(1998)Kramer, Xilouris, Lorimer, Doroshenko, Jessner,
  Wielebinski, Wolszczan, \& Camilo}]{kxl+98}
Kramer, M., Xilouris, K.~M., Lorimer, D.~R., {et~al.} 1998, ApJ, 501, 270

\bibitem[{{Kuzmin} {et~al.}(2008){Kuzmin}, {Losovsky}, {Jordan}, \&
  {Smith}}]{kljs08}
{Kuzmin}, A., {Losovsky}, B.~Y., {Jordan}, C.~A., \& {Smith}, F.~G. 2008, \aap,
  483, 13

\bibitem[{Kuzmin \& Losovskii(1997)}]{kl97}
Kuzmin, A.~D. \& Losovskii, B.~Y. 1997, Astron. Lett., 23, 283

\bibitem[{{Kuzmin} \& {Losovsky}(2001)}]{kl01}
{Kuzmin}, A.~D. \& {Losovsky}, B.~Y. 2001, \aap, 368, 230

\bibitem[{Kuzmin {et~al.}(1978)Kuzmin, Malofeev, Shitov, Davies, Lyne, \&
  Rowson}]{kms+78}
Kuzmin, A.~D., Malofeev, V.~M., Shitov, Y.~P., {et~al.} 1978, MNRAS, 185, 441

\bibitem[{Lawson {et~al.}(1987)Lawson, Mayer, Osborne, \& Parkinson}]{lmop87}
Lawson, K.~D., Mayer, C.~J., Osborne, J.~L., \& Parkinson, M.~L. 1987, MNRAS,
  225, 307

\bibitem[{{Levin} {et~al.}(2010){Levin}, {Bailes}, {Bates}, {Bhat}, {Burgay},
  {Burke-Spolaor}, {D'Amico}, {Johnston}, {Keith}, {Kramer}, {Milia},
  {Possenti}, {Rea}, {Stappers}, \& {van Straten}}]{lbb+10}
{Levin}, L., {Bailes}, M., {Bates}, S., {et~al.} 2010, ApJ, 721, L33

\bibitem[{L{\"o}hmer {et~al.}(2004)L{\"o}hmer, Mitra, Gupta, Kramer, \&
  Ahuja}]{lmg+04}
L{\"o}hmer, O., Mitra, D., Gupta, Y., Kramer, M., \& Ahuja, A. 2004, A\&A, 425,
  569

\bibitem[{{Lonsdale} {et~al.}(2009){Lonsdale}, {Cappallo}, {Morales}, {Briggs},
  {Benkevitch}, {Bowman}, {Bunton}, {Burns}, {Corey}, {Desouza}, {Doeleman},
  {Derome}, {Deshpande}, {Gopala}, {Greenhill}, {Herne}, {Hewitt}, {Kamini},
  {Kasper}, {Kincaid}, {Kocz}, {Kowald}, {Kratzenberg}, {Kumar}, {Lynch},
  {Madhavi}, {Matejek}, {Mitchell}, {Morgan}, {Oberoi}, {Ord},
  {Pathikulangara}, {Prabu}, {Rogers}, {Roshi}, {Salah}, {Sault}, {Shankar},
  {Srivani}, {Stevens}, {Tingay}, {Vaccarella}, {Waterson}, {Wayth}, {Webster},
  {Whitney}, {Williams}, \& {Williams}}]{lcm+09}
{Lonsdale}, C.~J., {Cappallo}, R.~J., {Morales}, M.~F., {et~al.} 2009, IEEE
  Proceedings, 97, 1497

\bibitem[{{Lorimer} {et~al.}(2007){Lorimer}, {Bailes}, {McLaughlin},
  {Narkevic}, \& {Crawford}}]{lbm+07}
{Lorimer}, D.~R., {Bailes}, M., {McLaughlin}, M.~A., {Narkevic}, D.~J., \&
  {Crawford}, F. 2007, Science, 318, 777

\bibitem[{{Lorimer} {et~al.}(2006){Lorimer}, {Faulkner}, {Lyne}, {Manchester},
  {Kramer}, {McLaughlin}, {Hobbs}, {Possenti}, {Stairs}, {Camilo}, {Burgay},
  {D'Amico}, {Corongiu}, \& {Crawford}}]{lfl+06}
{Lorimer}, D.~R., {Faulkner}, A.~J., {Lyne}, A.~G., {et~al.} 2006, MNRAS, 372,
  777

\bibitem[{Lorimer \& Kramer(2005)}]{lk05}
Lorimer, D.~R. \& Kramer, M. 2005, Handbook of Pulsar Astronomy (Cambridge
  University Press)

\bibitem[{{Lyne} {et~al.}(2010){Lyne}, {Hobbs}, {Kramer}, {Stairs}, \&
  {Stappers}}]{lhk+10}
{Lyne}, A., {Hobbs}, G., {Kramer}, M., {Stairs}, I., \& {Stappers}, B. 2010,
  Science, 329, 408

\bibitem[{Lyne \& Smith(2004)}]{ls04}
Lyne, A.~G. \& Smith, F.~G. 2004, Pulsar Astronomy, 3rd ed. (Cambridge:
  Cambridge University Press)

\bibitem[{Lyutikov(2002)}]{lyu02}
Lyutikov, M. 2002, ApJ, 580, L65

\bibitem[{{Malofeev}(2000)}]{mal00}
{Malofeev}, V.~M. 2000, in Astronomical Society of the Pacific Conference
  Series, Vol. 202, IAU Colloq. 177: Pulsar Astronomy - 2000 and Beyond, ed.
  {M.~Kramer, N.~Wex, \& R.~Wielebinski}, 221

\bibitem[{Malofeev {et~al.}(1994)Malofeev, Gil, Jessner, Malov, Seiradakis,
  Sieber, \& Wielebinski}]{mgj+94}
Malofeev, V.~M., Gil, J.~A., Jessner, A., {et~al.} 1994, A\&A, 285, 201

\bibitem[{Malofeev \& Malov(1997)}]{mm97}
Malofeev, V.~M. \& Malov, O.~I. 1997, Nature, 389, 697

\bibitem[{{Malofeev} {et~al.}(2000){Malofeev}, {Malov}, \&
  {Shchegoleva}}]{mms00}
{Malofeev}, V.~M., {Malov}, O.~I., \& {Shchegoleva}, N.~V. 2000, Astronomy
  Reports, 44, 436

\bibitem[{{Malofeev} {et~al.}(2006){Malofeev}, {Malov}, \& {Teplykh}}]{mmt06}
{Malofeev}, V.~M., {Malov}, O.~I., \& {Teplykh}, D.~A. 2006, Chin. J. Astron.
  Astrophys. Suppl. 2, 6, 68

\bibitem[{{Malofeev} {et~al.}(2007){Malofeev}, {Malov}, \& {Teplykh}}]{mmt07}
{Malofeev}, V.~M., {Malov}, O.~I., \& {Teplykh}, D.~A. 2007, \apss, 308, 211

\bibitem[{{Malofeev} {et~al.}(2005){Malofeev}, {Malov}, {Teplykh},
  {Tyul'Bashev}, \& {Tyul'Basheva}}]{mmt+05}
{Malofeev}, V.~M., {Malov}, O.~I., {Teplykh}, D.~A., {Tyul'Bashev}, S.~A., \&
  {Tyul'Basheva}, G.~E. 2005, Astronomy Reports, 49, 242

\bibitem[{{Malov} \& {Malofeev}(2010)}]{mm10}
{Malov}, O.~I. \& {Malofeev}, V.~M. 2010, Astronomy Reports, 54, 210

\bibitem[{{Manchester} {et~al.}(2005){Manchester}, {Hobbs}, {Teoh}, \&
  {Hobbs}}]{mhth05}
{Manchester}, R.~N., {Hobbs}, G.~B., {Teoh}, A., \& {Hobbs}, M. 2005, AJ, 129,
  1993

\bibitem[{{Maron} {et~al.}(2000){Maron}, {Kijak}, {Kramer}, \&
  {Wielebinski}}]{mkkw00a}
{Maron}, O., {Kijak}, J., {Kramer}, M., \& {Wielebinski}, R. 2000, A\&AS, 147,
  195

\bibitem[{{McLaughlin} \& {Cordes}(2003)}]{mc03}
{McLaughlin}, M.~A. \& {Cordes}, J.~M. 2003, ApJ, 596, 982

\bibitem[{{McLaughlin} {et~al.}(2006){McLaughlin}, {Lyne}, {Lorimer}, {Kramer},
  {Faulkner}, {Manchester}, {Cordes}, {Camilo}, {Possenti}, {Stairs}, {Hobbs},
  {D'Amico}, {Burgay}, \& {O'Brien}}]{mll+06}
{McLaughlin}, M.~A., {Lyne}, A.~G., {Lorimer}, D.~R., {et~al.} 2006, Nature,
  439, 817

\bibitem[{{McLaughlin} {et~al.}(2009){McLaughlin}, {Lyne}, {Keane}, {Kramer},
  {Miller}, {Lorimer}, \& {Manchester}}]{mlk+09}
{McLaughlin}, M.~A., {Lyne}, A.~G., {Keane}, E.~F., {et~al.} 2009, MNRAS, 400, 1431

\bibitem[{{Melrose}(2004)}]{mel04}
{Melrose}, D. 2004, in Young Neutron Stars and Their Environments, {IAU}
  Symposium 218, ed. F.~Camilo \& B.~M. Gaensler (San Francisco: Astronomical
  Society of the Pacific), 349--356

\bibitem[{Mitra \& Rankin(2002)}]{mr02a}
Mitra, D. \& Rankin, J.~M. 2002, ApJ, 577, 322

\bibitem[{Mol \& Romein(2011)}]{mr11}
Mol, J.~D. \& Romein, J.~W. 2011, {The LOFAR Beam Former: Implementation and
  Performance Analysis}, under review

\bibitem[{Morris {et~al.}(1997)Morris, Kramer, Thum, \& et~al.}]{mkt+97}
Morris, D., Kramer, M., Thum, C., \& et~al. 1997, A\&A, 322, L17

\bibitem[{Narayan(1992)}]{nar92a}
Narayan, R. 1992, Philos. Trans. Roy. Soc. London A, 341, 151

\bibitem[{Navarro {et~al.}(1995)Navarro, de~Bruyn, Frail, Kulkarni, \&
  Lyne}]{nbf+95}
Navarro, J., de~Bruyn, G., Frail, D., Kulkarni, S.~R., \& Lyne, A.~G. 1995,
  ApJ, 455, L55

\bibitem[{{Nijboer} \& {Noordam}(2007)}]{nn07}
{Nijboer}, R.~J. \& {Noordam}, J.~E. 2007, in Astronomical Society of the
  Pacific Conference Series, Vol. 376, Astronomical Data Analysis Software and
  Systems XVI, ed. {R.~A.~Shaw, F.~Hill, \& D.~J.~Bell}, 237

\bibitem[{{Noutsos} {et~al.}(2008){Noutsos}, {Johnston}, {Kramer}, \&
  {Karastergiou}}]{njkk08}
{Noutsos}, A., {Johnston}, S., {Kramer}, M., \& {Karastergiou}, A. 2008,
  \mnras, 386, 1881

\bibitem[{{Petrova}(2006)}]{pet06}
{Petrova}, S.~A. 2006, \mnras, 368, 1764

\bibitem[{{Petrova}(2008)}]{pet08}
{Petrova}, S.~A. 2008, \mnras, 383, 1413

\bibitem[{{Popov} {et~al.}(2006{\natexlab{a}}){Popov}, {Kuzmin}, {Ulyanov},
  {Deshpande}, {Ershov}, {Kondratiev}, {Kostyuk}, {Losovsky}, {Soglasnov}, \&
  {Zakharenko}}]{pku+06}
{Popov}, M.~V., {Kuzmin}, A.~D., {Ulyanov}, O.~M., {et~al.} 2006{\natexlab{a}},
  On the Present and Future of Pulsar Astronomy, 26th meeting of the IAU, Joint
  Discussion 2, 16-17 August, 2006, Prague, Czech Republic, JD02, \#19, 2

\bibitem[{{Popov} {et~al.}(2006{\natexlab{b}}){Popov}, {Kuz'min}, {Ul'yanov},
  {Deshpande}, {Ershov}, {Zakharenko}, {Kondrat'ev}, {Kostyuk},
  {Losovski{\"a}{shy}}, \& {Soglasnov}}]{pku+06b}
{Popov}, M.~V., {Kuz'min}, A.~D., {Ul'yanov}, O.~M., {et~al.}
  2006{\natexlab{b}}, Astronomy Reports, 50, 562

\bibitem[{Rankin(1993)}]{ran93}
Rankin, J.~M. 1993, ApJ, 405, 285

\bibitem[{Rankin {et~al.}(1970)Rankin, Comella, Craft, Richards, Campbell, \&
  Counselman}]{rcc+70}
Rankin, J.~M., Comella, J.~M., Craft, H.~D., {et~al.} 1970, ApJ, 162, 707

\bibitem[{{Ransom} {et~al.}(2003){Ransom}, {Cordes}, \& {Eikenberry}}]{rce03}
{Ransom}, S.~M., {Cordes}, J.~M., \& {Eikenberry}, S.~S. 2003, ApJ, 589, 911

\bibitem[{{Ransom} {et~al.}(2011){Ransom}, {Ray}, {Camilo}, {Roberts}, {{\c
  C}elik}, {Wolff}, {Cheung}, {Kerr}, {Pennucci}, {DeCesar}, {Cognard}, {Lyne},
  {Stappers}, {Freire}, {Grove}, {Abdo}, {Desvignes}, {Donato}, {Ferrara},
  {Gehrels}, {Guillemot}, {Gwon}, {Harding}, {Johnston}, {Keith}, {Kramer},
  {Michelson}, {Parent}, {Saz Parkinson}, {Romani}, {Smith}, {Theureau},
  {Thompson}, {Weltevrede}, {Wood}, \& {Ziegler}}]{rrc+11}
{Ransom}, S.~M., {Ray}, P.~S., {Camilo}, F., {et~al.} 2011, \apjl, 727, L16+

\bibitem[{{Reich} \& {Reich}(1988)}]{rr88}
{Reich}, P. \& {Reich}, W. 1988, A\&AS, 74, 7

\bibitem[{{Rickett} {et~al.}(2009){Rickett}, {Johnston}, {Tomlinson}, \&
  {Reynolds}}]{rjt+09}
{Rickett}, B., {Johnston}, S., {Tomlinson}, T., \& {Reynolds}, J. 2009, \mnras,
  395, 1391

\bibitem[{Rickett(1970)}]{ric70}
Rickett, B.~J. 1970, MNRAS, 150, 67

\bibitem[{Rickett(1990)}]{ric90}
Rickett, B.~J. 1990, Ann. Rev. Astr. Ap., 28, 561

\bibitem[{{Roger} {et~al.}(1999){Roger}, {Costain}, {Landecker}, \&
  {Swerdlyk}}]{rcls99}
{Roger}, R.~S., {Costain}, C.~H., {Landecker}, T.~L., \& {Swerdlyk}, C.~M.
  1999, A\&AS, 137, 7

\bibitem[{Romein {et~al.}(2010)Romein, Broekema, Mol, \& van
  Nieuwpoort}]{rbmn10}
Romein, J.~W., Broekema, P.~C., Mol, J.~D., \& van Nieuwpoort, R.~V. 2010, in
  ACM Symposium on Principles and Practice of Parallel Programming (PPoPP'10),
  Bangalore, India, 169--178

\bibitem[{{Ryabov} {et~al.}(2010){Ryabov}, {Vavriv}, {Zarka}, {Ryabov},
  {Kozhin}, {Vinogradov}, \& {Denis}}]{rvz+10}
{Ryabov}, V.~B., {Vavriv}, D.~M., {Zarka}, P., {et~al.} 2010, \aap, 510, A16

\bibitem[{Shishov \& Smirnova(2002)}]{ss02}
Shishov, V.~I. \& Smirnova, T.~V. 2002, Astron.\ Rep., 46, 731

\bibitem[{Shitov {et~al.}(1988)Shitov, Malofeev, \& Izvekova}]{smi88}
Shitov, Y.~P., Malofeev, V.~M., \& Izvekova, V.~A. 1988, Sov. Astron. Lett.,
  14, 181

\bibitem[{Shitov \& Pugachev(1997)}]{sp97b}
Shitov, Y.~P. \& Pugachev, V.~D. 1997, New Astr., 3, 101

\bibitem[{Shrauner {et~al.}(1998)Shrauner, Taylor, \& Woan}]{stw98}
Shrauner, J.~A., Taylor, J.~H., \& Woan, G. 1998, ApJ, 509, 785

\bibitem[{{Singh} {et~al.}(2008){Singh}, {B{\"a}hren}, {Falcke}, \& {et
  al.}}]{sbf+08}
{Singh}, K., {B{\"a}hren}, L., {Falcke}, H., \& {et al.} 2008, in International
  Cosmic Ray Conference, Vol.~4, International Cosmic Ray Conference, 429--432

\bibitem[{Slee {et~al.}(1986)Slee, Alurkar, \& Bobra}]{sab86}
Slee, O.~B., Alurkar, S.~K., \& Bobra, A.~D. 1986, Aust. J. Phys., 39, 103

\bibitem[{{Smirnova}(2006)}]{smi06}
{Smirnova}, T.~V. 2006, Astronomy Reports, 50, 915

\bibitem[{{Smirnova} \& {Shishov}(2010)}]{ss10}
{Smirnova}, T.~V. \& {Shishov}, V.~I. 2010, Astronomy Reports, 54, 139

\bibitem[{{Smirnova} {et~al.}(1994){Smirnova}, {Tul'bashev}, \&
  {Boriakoff}}]{stb94}
{Smirnova}, T.~V., {Tul'bashev}, S.~A., \& {Boriakoff}, V. 1994, \aap, 286, 807

\bibitem[{{Soglasnov} {et~al.}(2004){Soglasnov}, {Popov}, {Bartel}, {Cannon},
  {Novikov}, {Kondratiev}, \& {Altunin}}]{spb+04b}
{Soglasnov}, V.~A., {Popov}, M.~V., {Bartel}, N., {et~al.} 2004, \apj, 616, 439

\bibitem[{{Soglasnov} {et~al.}(1983){Soglasnov}, {Popov}, \& {Kuz'min}}]{spk83}
{Soglasnov}, V.~A., {Popov}, M.~V., \& {Kuz'min}, O.~A. 1983, \azh, 60, 293

\bibitem[{Staelin \& Reifenstein(1968)}]{sr68}
Staelin, D.~H. \& Reifenstein, {III}, E.~C. 1968, Science, 162, 1481

\bibitem[{Stappers {et~al.}(1996)Stappers, Bailes, Lyne, Manchester, D'Amico,
  Tauris, Lorimer, Johnston, \& Sandhu}]{sbl+96}
Stappers, B.~W., Bailes, M., Lyne, A.~G., {et~al.} 1996, ApJ, 465, L119

\bibitem[{{Stappers} {et~al.}(2011){Stappers}, {Hessels}, {Alexov}, {Coenen},
  {Hassall}, {Karastergiou}, {Kondratiev}, {Kramer}, {van Leeuwen}, {Mol},
  {Noutsos}, {Weltevrede}, \& {the LOFAR Collaboration}}]{sha+11}
{Stappers}, B.~W., {Hessels}, J., {Alexov}, A., {et~al.} 2011, in Radio
  Pulsars: a key to unlock the secrets of the Universe (American Institute of
  Physics Conference Series)

\bibitem[{{Stappers} {et~al.}(2008){Stappers}, {Karappusamy}, \&
  {Hessels}}]{skh08}
{Stappers}, B.~W., {Karappusamy}, R., \& {Hessels}, J.~W.~T. 2008, in American
  Institute of Physics Conference Series, Vol. 983, 40 Years of Pulsars:
  Millisecond Pulsars, Magnetars and More, ed. {C.~Bassa, Z.~Wang, A.~Cumming,
  \& V.~M.~Kaspi}, 593--597

\bibitem[{{Stappers} {et~al.}(2007){Stappers}, {van Leeuwen}, {Kramer},
  {Stinebring}, \& {Hessels}}]{slk+08}
{Stappers}, B.~W., {van Leeuwen}, A.~G.~J., {Kramer}, M., {Stinebring}, D., \&
  {Hessels}, J. 2007, in WE-Heraeus Seminar on Neutron Stars and Pulsars 40
  years after the Discovery, ed. {W.~Becker \& H.~H.~Huang}, 100

\bibitem[{{Stinebring} {et~al.}(2001){Stinebring}, {McLaughlin}, {Cordes},
  {Becker}, {Goodman}, {Kramer}, {Sheckard}, \& {Smith}}]{smc+01}
{Stinebring}, D.~R., {McLaughlin}, M.~A., {Cordes}, J.~M., {et~al.} 2001, ApJ,
  549, L97

\bibitem[{Swarup {et~al.}(1991)Swarup, Ananthakrishnan, Kapahi, Rao,
  Subrahmanya, \& Kulkarni}]{sak+91}
Swarup, G., Ananthakrishnan, S., Kapahi, V.~K., {et~al.} 1991, Current Science,
  60, 95

\bibitem[{Tang \& Cheng(2001)}]{tc01}
Tang, A.~P.~S. \& Cheng, K.~S. 2001, ApJ, 549, 1039

\bibitem[{{Tavani} {et~al.}(2009){Tavani}, {Barbiellini}, {Argan}, {Boffelli},
  {Bulgarelli}, {Caraveo}, {Cattaneo}, {Chen}, {Cocco}, {Costa}, {D'Ammando},
  {Del Monte}, {de Paris}, {Di Cocco}, {di Persio}, {Donnarumma},
  {Evangelista}, {Feroci}, {Ferrari}, {Fiorini}, {Fornari}, {Fuschino},
  {Froysland}, {Frutti}, {Galli}, {Gianotti}, {Giuliani}, {Labanti}, {Lapshov},
  {Lazzarotto}, {Liello}, {Lipari}, {Longo}, {Mattaini}, {Marisaldi},
  {Mastropietro}, {Mauri}, {Mauri}, {Mereghetti}, {Morelli}, {Morselli},
  {Pacciani}, {Pellizzoni}, {Perotti}, {Piano}, {Picozza}, {Pontoni},
  {Porrovecchio}, {Prest}, {Pucella}, {Rapisarda}, {Rappoldi}, {Rossi},
  {Rubini}, {Soffitta}, {Traci}, {Trifoglio}, {Trois}, {Vallazza},
  {Vercellone}, {Vittorini}, {Zambra}, {Zanello}, {Pittori}, {Preger},
  {Santolamazza}, {Verrecchia}, {Giommi}, {Colafrancesco}, {Antonelli},
  {Cutini}, {Gasparrini}, {Stellato}, {Fanari}, {Primavera}, {Tamburelli},
  {Viola}, {Guarrera}, {Salotti}, {D'Amico}, {Marchetti}, {Crisconio},
  {Sabatini}, {Annoni}, {Alia}, {Longoni}, {Sanquerin}, {Battilana}, {Concari},
  {Dessimone}, {Grossi}, {Parise}, {Monzani}, {Artina}, {Pavesi},
  {Marseguerra}, {Nicolini}, {Scandelli}, {Soli}, {Vettorello}, {Zardetto},
  {Bonati}, {Maltecca}, {D'Alba}, {Patan{\'e}}, {Babini}, {Onorati},
  {Acquaroli}, {Angelucci}, {Morelli}, {Agostara}, {Cerone}, {Michetti},
  {Tempesta}, {D'Eramo}, {Rocca}, {Giannini}, {Borghi}, {Garavelli}, {Conte},
  {Balasini}, {Ferrario}, {Vanotti}, {Collavo}, \& {Giacomazzo}}]{tba+09}
{Tavani}, M., {Barbiellini}, G., {Argan}, A., {et~al.} 2009, \aap, 502, 995

\bibitem[{team(2008)}]{IBM08}
team, I. B.~G. 2008, IBM Journal of Research and Development, 52

\bibitem[{{Thompson}(2000)}]{tho00b}
{Thompson}, D.~J. 2000, Advances in Space Research, 25, 659

\bibitem[{Tinbergen(1996)}]{tin96}
Tinbergen, J. 1996, Astronomical Polarimetry (Cambridge University Press),
  iSBN: 0521475317

\bibitem[{{Ulyanov} {et~al.}(2007){Ulyanov}, {Deshpande}, {Zakharenko},
  {Asgekar}, \& {Shankar}}]{udz+07}
{Ulyanov}, O.~M., {Deshpande}, A.~A., {Zakharenko}, V.~V., {Asgekar}, A., \&
  {Shankar}, U. 2007, Radiofizika and Radioastronomia, 12, 5

\bibitem[{{Ul'Yanov} {et~al.}(2008){Ul'Yanov}, {Zakharenko}, \& {Bruk}}]{uzb08}
{Ul'Yanov}, O.~M., {Zakharenko}, V.~V., \& {Bruk}, I.~M. 2008, Astronomy
  Reports, 52, 917

\bibitem[{{Ulyanov} {et~al.}(2006){Ulyanov}, {Zakharenko}, {Konovalenko},
  {Lecacheux}, {Rosolen}, \& {Rucker}}]{uzk+06}
{Ulyanov}, O.~M., {Zakharenko}, V.~V., {Konovalenko}, A.~A., {et~al.} 2006,
  Radiofizika and Radioastronomia, 11, 113

\bibitem[{{van der Horst} {et~al.}(2008){van der Horst}, {Kamble}, {Resmi},
  {Wijers}, {Bhattacharya}, {Scheers}, {Rol}, {Strom}, {Kouveliotou},
  {Oosterloo}, \& {Ishwara-Chandra}}]{vkr+08}
{van der Horst}, A.~J., {Kamble}, A., {Resmi}, L., {et~al.} 2008, \aap, 480, 35

\bibitem[{{van Eysden} \& {Melatos}(2010)}]{vm10}
{van Eysden}, C.~A. \& {Melatos}, A. 2010, \mnras, 1459

\bibitem[{{van Leeuwen} {et~al.}(2003){van Leeuwen}, {Stappers},
  {Ramachandran}, \& {Rankin}}]{vsrr03}
{van Leeuwen}, A.~G.~J., {Stappers}, B.~W., {Ramachandran}, R., \& {Rankin},
  J.~M. 2003, A\&A, 399, 223

\bibitem[{{van Leeuwen} \& {Stappers}(2010)}]{ls10}
{van Leeuwen}, J. \& {Stappers}, B.~W. 2010, \aap, 509, A7+

\bibitem[{Van~Riper {et~al.}(1991)Van~Riper, Epstein, \& Miller}]{rem91}
Van~Riper, K.~A., Epstein, R.~I., \& Miller, G.~S. 1991, ApJ, 381, L17

\bibitem[{{van Straten}(2004)}]{van04d}
{van Straten}, W. 2004, \apjs, 152, 129

\bibitem[{{van Straten} {et~al.}(2010){van Straten}, {Manchester}, {Johnston},
  \& {Reynolds}}]{vmj+10}
{van Straten}, W., {Manchester}, R.~N., {Johnston}, S., \& {Reynolds}, J.~E.
  2010, \pasa, 27, 104

\bibitem[{von Hoensbroech \& Xilouris(1997)}]{hx97}
von Hoensbroech, A. \& Xilouris, K.~M. 1997, A\&AS, 126, 121

\bibitem[{{Vranesevic} {et~al.}(2004){Vranesevic}, {Manchester}, {Lorimer},
  {Hobbs}, {Lyne}, {Kramer}, {Camilo}, {Stairs}, {Kaspi}, {D'Amico},
  {Possenti}, {Crawford}, {Faulkner}, \& {McLaughlin}}]{vml+04}
{Vranesevic}, N., {Manchester}, R.~N., {Lorimer}, D.~R., {et~al.} 2004, ApJ,
  617, L139

\bibitem[{{Welch} {et~al.}(2009){Welch}, {Backer}, {Blitz}, {Bock}, {Bower},
  {Cheng}, {Croft}, {Dexter}, {Engargiola}, {Fields}, {Forster},
  {Gutierrez-Kraybill}, {Heiles}, {Helfer}, {Jorgensen}, {Keating}, {Lugten},
  {MacMahon}, {Milgrome}, {Thornton}, {Urry}, {van Leeuwen}, {Werthimer},
  {Williams}, {Wright}, {Tarter}, {Ackermann}, {Atkinson}, {Backus}, {Barott},
  {Bradford}, {Davis}, {Deboer}, {Dreher}, {Harp}, {Jordan}, {Kilsdonk},
  {Pierson}, {Randall}, {Ross}, {Shostak}, {Fleming}, {Cork}, {Vitouchkine},
  {Wadefalk}, \& {Weinreb}}]{wbb+09}
{Welch}, J., {Backer}, D., {Blitz}, L., {et~al.} 2009, IEEE Proceedings, 97,
  1438

\bibitem[{{Weltevrede} {et~al.}(2007){Weltevrede}, {Stappers}, \&
  {Edwards}}]{wse07}
{Weltevrede}, P., {Stappers}, B.~W., \& {Edwards}, R.~T. 2007, \aap, 469, 607

\bibitem[{{Weltevrede} {et~al.}(2003){Weltevrede}, {Stappers}, {van den Horn},
  \& {Edwards}}]{wsve03}
{Weltevrede}, P., {Stappers}, B.~W., {van den Horn}, L.~J., \& {Edwards}, R.~T.
  2003, \aap, 412, 473

\bibitem[{{Weltevrede} {et~al.}(2006){Weltevrede}, {Wright}, {Stappers}, \&
  {Rankin}}]{wwsr06}
{Weltevrede}, P., {Wright}, G.~A.~E., {Stappers}, B.~W., \& {Rankin}, J.~M.
  2006, \aap, 458, 269

\bibitem[{{Wijnholds} {et~al.}(2010){Wijnholds}, {van der Tol}, {Nijboer}, \&
  {van der Veen}}]{wtnv10}
{Wijnholds}, S., {van der Tol}, S., {Nijboer}, R., \& {van der Veen}, A. 2010,
  IEEE Signal Processing Magazine, 27, 30

\bibitem[{Wijnholds \& van~der Veen(2009{\natexlab{a}})}]{wv09a}
Wijnholds, S.~J. \& van~der Veen, A. 2009{\natexlab{a}}, IEEE Transactions on
  Signal Processing, 57, 3512

\bibitem[{Wijnholds \& van~der Veen(2009{\natexlab{b}})}]{wv09b}
Wijnholds, S.~J. \& van~der Veen, A. 2009{\natexlab{b}}, in Proceedings of the
  17th European Signal Processing Conference (EuSiPCo), arXiv:1003.2497v1

\bibitem[{{Wolleben} {et~al.}(2010){Wolleben}, {Fletcher}, {Landecker},
  {Carretti}, {Dickey}, {Gaensler}, {Haverkorn}, {McClure-Griffiths}, {Reich},
  \& {Taylor}}]{wfl+10}
{Wolleben}, M., {Fletcher}, A., {Landecker}, T.~L., {et~al.} 2010, \apjl, 724,
  L48

\bibitem[{{You} {et~al.}(2007){You}, {Hobbs}, {Coles}, {Manchester}, {Edwards},
  {Bailes}, {Sarkissian}, {Verbiest}, {van Straten}, {Hotan}, {Ord}, {Jenet},
  {Bhat}, \& {Teoh}}]{yhc+07}
{You}, X.~P., {Hobbs}, G., {Coles}, W.~A., {et~al.} 2007, MNRAS, 378, 493

\end{thebibliography}
\end{document}